\documentclass[a4paper,11pt]{article}
\pdfoutput=1
\usepackage{jcappub}

\usepackage[T1]{fontenc}

\usepackage[english]{babel}
\usepackage{amsfonts}
\usepackage{txfonts}
\usepackage{amssymb}
\usepackage{ae,aecompl}
\usepackage{fancyhdr}
\usepackage{multicol}
\usepackage{layout}
\usepackage{multirow}
\usepackage{color}
\usepackage{times}
\usepackage{textcomp}
\usepackage{cprotect}

\usepackage[outdir=./figures/]{epstopdf}
\usepackage[export]{adjustbox}
    
\graphicspath{{./figures/}}

\newif\ifAMStwofonts
\AMStwofontstrue

\title{Dark sector evolution in Horndeski models}

\author[a,1]{Francesco Pace,\note{Corresponding author.}}
\author[a]{Richard A. Battye,}
\author[a]{Boris Bolliet}
\author[a]{and Damien Trinh}

\emailAdd{francesco.pace@manchester.ac.uk}
\emailAdd{richard.battye@manchester.ac.uk}
\emailAdd{boris.bolliet@manchester.ac.uk}
\emailAdd{damien.trinh@manchester.ac.uk}

\affiliation[a]{%
Jodrell Bank Centre for Astrophysics, School of Natural Sciences, Department of Physics and Astronomy, The University 
of Manchester, Manchester, M13 9PL, U.K.
}

\abstract{
We use the Equation of State (EoS) approach to study the evolution of the dark sector in Horndeski models, the most 
general scalar-tensor theories with second order equations of motion. By including the effects of the dark sector into 
our code EoS\_class, we demonstrate the numerical stability of the formalism and excellent agreement with results 
from other publicly available codes for a range of parameters describing the evolution of the function characterising 
the perturbations for Horndeski models, $\alpha_{\rm x}$, with ${\rm x}=\{{\rm K}, {\rm B}, {\rm M}, {\rm T}\}$. 
After demonstrating that on sub-horizon scales ($k\gtrsim 10^{-3}~{\rm Mpc}^{-1}$ at $z=0$) velocity perturbations in 
both the matter and the dark sector are typically subdominant with respect to density perturbations in the equation of 
state for perturbations, we find an attractor solution for the dark sector gauge-invariant density perturbation 
$\Delta_{\rm ds}$ by neglecting its time derivatives in the equation describing its time evolution, as commonly done in 
the well-known quasi-static approximation. 
Using this result, we provide simplified expressions for the equation-of-state functions: the dark sector entropy 
perturbations $w_{\rm ds}\Gamma_{\rm ds}$ and anisotropic stress $w_{\rm ds}\Pi_{\rm ds}$. 
From this we derive a growth factor-like equation for both matter and dark sector and are able to capture the relevant 
physics for several observables with great accuracy. We finally present new analytical expressions for the well-known 
modified gravity phenomenological functions $\mu$, $\eta$ and $\Sigma$ for a generic Horndeski model as functions of 
$\alpha_{\rm x}$. 
We show that on small scales they reproduce expressions presented in previous works, but on large scales, we find 
differences with respect to other works.
}


\arxivnumber{1905.06795}

\begin{document}

\label{firstpage}

\maketitle
\flushbottom

\section{Introduction}
There is a solid amount of observational data suggesting that our Universe is currently undergoing accelerated 
expansion. All current data, ranging from the Cosmic Microwave Background (CMB) radiation 
\citep{Planck2018_VI,Planck2018_VIII,Planck2018_X}, Type Ia supernovae \citep{Riess1998,Perlmutter1999,Riess2007} and 
Large Scale Structure (LSS) \citep{Cole2005,Troxel2017,DES_Y1_2018,Gruen2018} can be well described within the 
standard cosmological model, known as $\Lambda$CDM, where the cosmological constant $\Lambda$ is responsible for the 
accelerated expansion and the cold dark matter (CDM) component determines the evolution of cosmic structures. 
While the former contributes to about 68\% of the present cosmic energy budget and the latter to about 28\%, the 
remaining 4\% is ordinary baryonic matter with small but non-negligible contributions from radiation and relativistic 
species, e.g.\ neutrinos \citep{Planck2018_VI}. 
In this context, the cosmological constant is generally interpreted as the energy of the vacuum and gravitational laws 
are those of General Relativity, assumed to be valid on all scales.

Despite the very good agreement between theoretical predictions and observational results, the cosmological constant 
is far from being a satisfactory explanation, especially when interpreted in the context of quantum field theory 
\citep{Weinberg1989}. 
This has opened the door to an intensive theoretical effort to explore different explanations for the accelerated 
expansion and to investigate dark energy and modified gravity models 
\citep{Joyce2015,Joyce2016,Koyama2016,Sami2016,BeltranJimenez2018,Ishak2018,Heisenberg2019}, in particular 
scalar-tensor theories.

The most general scalar-tensor model with second order equations of motion is described by the Horndeski Lagrangian 
\citep{Horndeski1974,Deffayet2011,Kobayashi2011}. Most dark energy and modified gravity models (quintessence 
\citep{Ford1987,Peebles1988,Ratra1988a}, $k$-essence \citep{ArmendarizPicon1999,Mukhanov2006}, KGB 
\citep{Deffayet2010,Pujolas2011,Kobayashi2010,Kimura2011}, $f(R)$ \citep{Silvestri2009,Sotiriou2010,DeFelice2010}, 
Brans-Dicke \citep{Brans1961}, Galileon cosmology \citep{Chow2009,Nicolis2009,Deffayet2009}, Gauss-Bonnet models 
\citep{Carroll2005}) are subclasses of the Horndeski Lagrangian. 
In this work we study the Horndeski Lagrangian and analyse specific subclasses.

Understanding the evolution of cosmological perturbations and their effects on observables such as the CMB and the 
matter power spectrum is important to fully characterise cosmological models. To do so, one relies on 
Einstein-Boltzmann codes which solve the linearised Einstein and Boltzmann equations on an expanding background. These 
codes have usually been designed to study perturbations in a standard $\Lambda$CDM model, for instance \verb|CMBFAST| 
\cite{Seljak1996}, \verb|DASh| \cite{Kaplinghat2002}, \verb|CMBEASY| \cite{Doran2005a}, \verb|CAMB| \cite{Lewis2000} 
and \verb|CLASS| \cite{Lesgourgues2011a,Blas2011}.

Recently, the \verb|CAMB| and \verb|CLASS| codes have been extended to include models beyond the $\Lambda$CDM 
paradigm with \verb|EFTCAMB| \cite{Hu2014a,Raveri2014}, \verb|hi_class| \cite{Zumalacarregui2017} and our 
previous code \verb|CLASS_EOS_FR| \citep{Battye2018a}, respectively, for example. 
Two independent implementations, \verb|COOP| \cite{Huang2012}, and a recent extension of \verb|hi_class| 
\citep{Traykova2019}, allow to even study models beyond Horndeski. All these codes are based on the Effective Field 
Theory (EFT) formalism in its different flavours \cite{Gubitosi2013,Gleyzes2013,Bellini2014,Gleyzes2014} and their 
predictions were recently shown to agree at the sub-percent level \cite{Bellini2018}. \verb|EFCLASS| 
\cite{Arjona2019a,Arjona2019b} implements an effective fluid approach for $f(R)$ models in the subhorizon 
approximation and for designing Horndeski. 
Here we extend our previous code and implement the full Horndeski dynamics into \verb|CLASS|, using the Equation of 
State (EoS) approach \cite{Battye2007,Battye2012,Battye2013,Battye2013a,Battye2014,Battye2016a}.\footnote{The code 
will be made publicly available when this paper is accepted for publication.}

The EoS approach is a powerful formalism based on the identification of all modifications to General Relativity with 
an effective fluid described by a non-trivial stress-energy tensor $U_{\mu\nu}$. At background order this is 
completely specified by the choice of an equation of state $P_{\rm ds}=w_{\rm ds}\rho_{\rm ds}$, where ${\rm ds}$ 
denotes the dark sector (the scalar field and its effective fluid representation), while at linear perturbation 
order two new (perturbed) gauge-invariant equations of state are introduced, the entropy perturbations 
$w_{\rm ds}\Gamma_{\rm ds}$ and the anisotropic stress $w_{\rm ds}\Pi_{\rm ds}$. 
In this approach one studies the evolution of the density ($\Delta_{\rm ds}$) and velocity ($\Theta_{\rm ds}$) 
perturbations of the dark fluid and by knowing these quantities it is possible, as we will show later, to describe and 
compute the phenomenology of the model under consideration in a relatively simple way.

The paper is organised as follows. In Sect.~\ref{sect:Horndeski} we briefly discuss the background and perturbation 
dynamics of the Horndeski models and introduce the particular subclasses analysed in this work. 
In Sect.~\ref{sect:EoS} we review the theoretical foundations of the EoS formalism and describe our numerical 
implementation into our code \verb|EoS_class|. 
In Sect.~\ref{sect:validation} we validate our code  with results from \verb|hi_class|, similarly to what was done in 
\cite{Bellini2018}. In Sect.~\ref{sect:ds} we present our numerical results on the evolution of the dark sector and of 
the phenomenological functions parameterizing modified gravity models. Using the fundamental equations of the EoS 
formalism and following the discussion in \cite{Battye2018a}, we prove the existence of an attractor solution whose 
validity is established by comparing it with the numerical results of the code, under the assumption that velocity 
perturbations in the EoS are subdominant with respect to density perturbations. These considerations allow us to 
reproduce and extend analytical results available in the literature and provide expressions describing the 
phenomenology of modified gravity models. 
In Sect.~\ref{sect:phenomenology} we study the phenomenology of modified gravity models. We conclude in 
Sect.~\ref{sect:conclusions}.

In Appendix~\ref{sect:background} we present the background expressions of the full Horndeski models, while in 
Appendix~\ref{sect:perturbations} we show the expressions for the four functions $\alpha_{\rm x}$, 
${\rm x}=\{{\rm K}, {\rm B}, {\rm M}, {\rm T}\}$, modelling linear perturbations according to the EFT formalism. In 
Appendix~\ref{sect:coefficientsEoS} we present the full expression of the EoS coefficients. 
Finally, in Appendix~\ref{sect:parameters} we report the precision parameters used for the numerical comparisons, 
similar to those in Appendix~C of \cite{Bellini2018}.

In the following we use natural units with $c=\hbar=1$ and a metric with positive signature. We choose the same 
fiducial cosmology as in the \verb|hi_class| paper \citep{Zumalacarregui2017}: the CMB temperature 
$T_{\rm CMB}=2.725{\rm K}$, the dimensionless Hubble parameter $h=0.675$, flat spatial geometry $\Omega_{\rm k}=0$, 
baryon density parameter today $\omega_{\rm b}=\Omega_{\rm b}h^2=0.022$, cold dark matter density parameter today 
$\omega_{\rm CDM}=\Omega_{\rm CDM}h^2=0.12$, effective number of neutrino species $N_{\rm eff}=3.046$, dark sector 
density parameter today, as inferred by the closure relation ($\sum_i\Omega_i=1$), $\Omega_{\rm ds}=0.688$. For all 
the models, the normalisation of the amplitude of perturbations is $A_{\rm s}=2.215\times 10^{-9}$, the slope of the 
primordial power spectrum is $n_{\rm s}=0.962$ and the reionization redshift is $z_{\rm reio}=11.36$. The background 
equation of state for the dark sector is $w_{\rm ds}=-1$ and it is kept fixed in the numerical analysis.

\section{Horndeski gravity models}\label{sect:Horndeski}

\subsection{Background}
Horndeski models \cite{Horndeski1974,Deffayet2011,Kobayashi2011} represent the most general Lagrangian for 
scalar-tensor theories leading to second order equations of motion in space and time. 
Their Lagrangian is given by the sum of the four following Lagrangians which encode the dynamics of the scalar field 
in the Jordan frame with metric $g_{\mu\nu}$ \citep{Gleyzes2014}:
\begin{align}
 {\cal L}_2 & \equiv G_2(\phi,X)\,,\label{eqn:G2}\\
 {\cal L}_3 & \equiv G_3(\phi,X)\Box\phi\,,\label{eqn:G3}\\
 {\cal L}_4 & \equiv G_4(\phi,X)R - 2G_{4,X}(\phi,X)
                     \left[\left(\Box\phi\right)^2 - \left(\nabla_{\mu}\nabla_{\nu}\phi\right)^2\right]\,,
                     \label{eqn:G4}\\
 {\cal L}_5 & \equiv G_5(\phi,X)G_{\mu\nu}\nabla^{\mu\nu}\phi + \frac{1}{3}G_{5,X}(\phi,X)
                     \left[
                       \left(\Box\phi\right)^3-
                       3\Box\phi\left(\nabla_{\mu}\nabla_{\nu}\phi\right)^2+
                       2\left(\nabla_{\mu}\nabla_{\nu}\phi\right)^3                    
                       \right]\,,\label{eqn:G5}
\end{align}
where $\phi$ and $X=g^{\mu\nu}\nabla_{\mu}\phi\nabla_{\nu}\phi$ are the scalar field and its canonical kinetic term, 
respectively, $R$ is the Ricci scalar and $G_{\mu\nu}$ the Einstein tensor. 
The subscript $X$ denotes the derivative with respect to the canonical kinetic term $X$, i.e.\ 
$G_{i,X} \equiv \partial G_i/\partial X$. We also have the following short-hand for compactness: 
$\left(\nabla_{\mu}\nabla_{\nu}\phi\right)^2=
\left(\nabla_{\mu}\nabla_{\nu}\phi\right)\left(\nabla^{\mu}\nabla^{\nu}\phi\right)$ and 
$\left(\nabla_{\mu}\nabla_{\nu}\phi\right)^3=
\left(\nabla_{\mu}\nabla_{\nu}\phi\right)
\left(\nabla^{\sigma}\nabla^{\nu}\phi\right)\left(\nabla_{\sigma}\nabla^{\mu}\phi\right)$.

The functions $G_i$ are arbitrary functions of both $\phi$ and $X$ and it is thanks to this freedom that Horndeski 
models enjoy a rich phenomenology. General Relativity plus the cosmological constant is recovered when 
$G_2=-M_{\rm pl}^2\Lambda$, $G_4=M_{\rm pl}^2/2$ and $G_3=G_5=0$, with $M_{\rm pl}=1/\sqrt{8\pi G}$ the reduced Planck 
mass. Other models, such as quintessence \citep{Wetterich2002}, $k$-essence 
\cite{ArmendarizPicon1999,ArmendarizPicon2001}, kinetic gravity braiding (KGB) \citep{Deffayet2010,Pujolas2011}, 
Brans-Dicke \citep{Brans1961}, covariant Galileons \citep{Nicolis2009,Deffayet2009}, $f(R)$ \citep{Carroll2004} and 
$f(G)$ \citep{Carroll2005} can be recovered by a suitable choice of the free functions, see for example Table~1 of 
\cite{Bellini2014}.

Let us now consider the total action of the system, including the matter sector described by the action $S_{\rm m}$,
\begin{equation}\label{eqn:action}
 S = \int\mathrm{d}^4x\,\sqrt{-g}\,{\cal L} + S_{\rm m}\,,
\end{equation}
where $g$ is the determinant of the metric and the scalar field Lagrangian ${\cal L}$ is 
${\cal L} = \sum_{i=2}^{5} {\cal L}_i$, where the functions ${\cal L}_i$ are defined in 
Eqs.~(\ref{eqn:G2})--(\ref{eqn:G5}). By \textit{matter sector}, we mean all the species other than the scalar field, 
i.e.\ CDM, baryons, photons and neutrinos. By \textit{dark sector} we mean the scalar field and its effective fluid 
representation.

By varying Eqn.~(\ref{eqn:action}) with respect to the metric and the scalar field, we obtain the field equations and 
the equation of motion of the scalar field, respectively. In a compact way, the gravitational- and scalar-field 
equations read
\begin{equation}
 -2M_{\rm pl}^2\sum_{i=2}^{5}{\cal G}_{\mu\nu}^{i} = -T_{\mu\nu}^{\rm m}\,, \quad 
 \sum_{i=2}^{5}\left(\nabla^{\mu}J_{\mu}^{i} - P_{\phi}^i\right)=0\,,
\end{equation}
respectively, where $i=1$ is not written because it corresponds to a simple cosmological constant and is included in 
the $i=2$ term in this formalism. In the previous expression, $T_{\mu\nu}^{\rm m}$ represents the stress-energy tensor 
for the matter components. Since the precise expressions for ${\cal G}_{\mu\nu}^{i}$, $J_{\mu}^{i}$ and $P_{\phi}^i$ 
are rather cumbersome, we do not report them here, but we refer the reader to \cite{Kobayashi2011,Gao2011}\footnote{
In this work we use a different notation for the canonical kinetic term and the sign of the function $G_3$ with 
respect to \cite{Kobayashi2011,Gao2011,DeFelice2011,Bellini2014,Zumalacarregui2017}.}.

In full generality, the stress-energy tensor for the Horndeski Lagrangian describes an imperfect fluid, whose 
components are its energy density $\rho$, pressure $P$, energy flow $q_{\mu}$ and anisotropic stress $\pi_{\mu\nu}$.

Assuming a flat FLRW metric at the background level, $\mathrm{d}s^2=-\mathrm{d}t^2+a^2(t)\mathrm{d}\mathbf{x}^2$, the 
expressions above simplify considerably. Taking into account that the velocity component for the scalar field is 
defined as $u_{\mu}=(\nabla_{\mu}\phi)/\dot{\phi}=\delta_{\mu}^{0}$ to satisfy the metric symmetries, it is easy to 
find the corresponding expressions for the density $\rho_{\rm ds}$ and the pressure $P_{\rm ds}$ of the dark sector 
fluid associated to the scalar field \cite{DeFelice2011,Nishi2014,Kase2018c}. We report the full expressions in 
Appendix~\ref{sect:background}, Eqs.~(\ref{eqn:rho}) and (\ref{eqn:P}), respectively. Then the background quantities 
satisfy the standard continuity equation: 
\begin{equation}\label{eqn:ce}
 \dot{\rho}_{\rm ds}+3H(\rho_{\rm ds}+P_{\rm ds})=0\,.
\end{equation}

The equation of motion for the scalar field is now \cite{DeFelice2011,Nishi2014}
\begin{equation}\label{eqn:eom}
 \frac{1}{a^3}\frac{\mathrm{d}}{\mathrm{d}t}(a^3J) = P_{\phi}\,,
\end{equation}
where the expressions for $J$ [Eqn.~(\ref{eqn:J})] and $P_{\phi}$ [Eqn.~(\ref{eqn:Pphi})] are given in 
Appendix~\ref{sect:background}. Due to the Bianchi identities, Eqn.~(\ref{eqn:eom}) is equivalent to the continuity 
equation (\ref{eqn:ce}).

\subsection{Effective Field Theory parametrisation of Horndeski models}
To study the evolution of the perturbations in dark energy and modified gravity models, several methods have been 
proposed. Among them, we recall the Effective Field Theory (EFT) approach 
\citep{Bellini2014,Gleyzes2013,Gubitosi2013,Gleyzes2014} which will be useful to describe linear perturbations in 
Horndeski models.

The EFT approach identifies a few functions, only depending on time, which are consistent with the background 
symmetries and which act as multiplying factors to the operators encoding the scale dependence of the system, see 
e.g.\ Eqn.~(86) in \cite{Gleyzes2014}. Since the equations of motion are at most second order, in the Fourier space we 
expect terms of order $k^2$, where $k$ is the wavenumber of a mode. For Horndeski models, it turns out that four 
time-dependent functions are sufficient to describe the entirety of modifications to perturbations, while the 
background evolution is fully described by the equation of state $P_{\rm ds}=w_{\rm ds}\rho_{\rm ds}$. 
These four functions only depend on background quantities, such as the scalar field $\phi$, the Hubble parameter $H$, 
and their respective derivatives. Hence, once the Lagrangian is specified, it is easy to explicitly write down the 
four time-dependent functions, as well as the Hubble parameter, in terms of the $G_i$ functions in 
Eqs.~(\ref{eqn:G2})--(\ref{eqn:G5}). 
Here we follow \cite{Bellini2014} and use four functions \{$\alpha_{\rm K}$, $\alpha_{\rm B}$, $\alpha_{\rm M}$, 
$\alpha_{\rm T}$\}. The expressions linking the $\alpha_{\rm x}$ functions, with 
${\rm x}=\{{\rm K}, {\rm B}, {\rm M}, {\rm T}\}$, to the $G_i$ functions are shown in 
Appendix~\ref{sect:perturbations}. 
Let us now recall the main properties associated with the $\alpha_{\rm x}$ functions.

The \textit{kineticity}, $\alpha_{\rm K}$, affects only scalar perturbations and receives contributions from all of 
the $G_i$ functions. It is the only term describing perfect fluid (no energy flow and anisotropic stress) dark energy 
models. 
When $\alpha_{\rm K}$ increases, the sound speed in the dark sector fluid decreases and eventually leads to a sound 
horizon smaller than the cosmological horizon. As a consequence, there is dark energy clustering on scales larger than 
the sound horizon. For constraints on $\alpha_{\rm K}$, see \cite{Sawicki2015,Gleyzes2016,Alonso2017}.

The \textit{braiding}, $\alpha_{\rm B}$, also only affects scalar perturbations and receives contributions from the 
functions $G_3$, $G_4$ and $G_5$ in Eqs.~(\ref{eqn:G3})--(\ref{eqn:G5})\footnote{In this work, we follow the convention 
used by \cite{Gleyzes2014}, rather than that of \cite{Bellini2014} and \cite{Zumalacarregui2017} for the braiding 
function $\alpha_{\rm B}$. The two definitions differ by a factor $-2$, which is taken care of automatically into our 
code.}. 
The braiding refers to the mixing of the kinetic terms of the scalar field and of the metric as can be appreciated in 
Eqn.~(86) of \cite{Gleyzes2014}. 
It modifies the coupling between matter and curvature, giving rise to an additional fifth force which is usually 
described as a modification of the effective Newton's constant in the equations of motion of the perturbations 
\cite{Battye2018a}.

The \textit{rate of running of the Planck mass}, $\alpha_{\rm M}$, affects both the scalar and the tensor 
perturbations and gets contributions from $G_4$ and $G_5$ only. 
A value different from zero leads to anisotropic stress, ultimately yielding differences between the gravitational 
potentials $\phi$ and $\psi$.

Finally, the \textit{tensor speed excess}, $\alpha_{\rm T}$, represents the deviations of the speed of gravitational 
waves $c_{\rm T}$ from that of light: $c_{\rm T}^2=1+\alpha_{\rm T}$. It receives contributions from $G_4$ and $G_5$ 
and leads to anisotropic stress. This quantity was extensively discussed after the detection of the gravitational 
wave signal GW170817, due to binary neutron star merger. Indeed, the almost simultaneous detection of GW170817 and 
its electromagnetic counterpart excludes models with $\alpha_{\rm T}\neq 0$ to high significance  
\cite{LigoVirgo2017,LigoVirgoIntegral2017,LigoVirgo2017a}, presuming that the Horndeski model applies on all scales. 
This implies $G_5=0$ and $G_4=G_4(\phi)$. See \cite{Ezquiaga2017,Ezquiaga2018,Kase2018c} for a review of Horndeski 
models surviving constraints from GW170817 and \cite{Lombriser2016a,McManus2016,Lombriser2017} for earlier works on 
the implications of a luminal propagation for GW. 
We note that some models can evade these constraints provided they have a scale-dependent tensor speed excess, or are 
Lorentz-violating or are non-local \citep{Battye2018b}.

A generic model might suffer from instabilities leading to exponentially unstable perturbations. Horndeski models can 
be unstable for several reasons, for example when specifying the $\alpha_{\rm x}$s arbitrarily without linking 
them to the actual scalar theory or when a given background equation of state $w_{\rm ds}$ leads to the wrong sign of 
the kinetic term. This is, for example, the case for $f(R)$ models, where, despite allowing a background with 
$w_{\rm ds}<-1$, the corresponding perturbation sector is unstable, as discussed in \cite{Battye2018a}, where it was 
shown that a vast sector of $f(R)$ theories with $w_{\rm ds}\neq -1$ can be ruled out by current cosmological data.

Perturbations can face three specific varieties of instabilities: ghost instabilities arise when the kinetic term is 
negative; gradient instabilities arise when the sound speed squared is negative, affecting small scale modes most 
strongly; tachyon instabilities arise when the mass squared is negative.

In Horndeski models, the stability of perturbations requires the following quantities to be positive 
\citep{DeFelice2012,Bellini2014,Gleyzes2014}:
\begin{itemize}
 \item $\alpha\equiv\alpha_{\rm K}+6\alpha_{\rm B}^2$, which represents the kinetic coefficient and hence the absence 
       of ghosts;
 \item the sound speed squared of scalar perturbations, $c_{\rm s}^2$;
 \item the sound speed squared of tensor perturbations, $c_{\rm T}^2$;
 \item $M^2$, which represents the effective Planck mass squared.
\end{itemize}
These stability conditions are implemented in our code \verb|EoS_class| as is the case in the \verb|hi_class| code.

\subsection{Specific classes of models used as examples in our analysis}\label{sect:Models}
In our numerical implementation, we assume an exact $\Lambda$CDM background expansion and specific phenomenological 
parametrisation of the $\alpha_{\rm x}$ functions, with ${\rm x}=\{{\rm K}, {\rm B}, {\rm M}, {\rm T}\}$. 
In particular, following the previous literature, we chose 
$\alpha_{\rm x}(a)=\alpha_{\rm x,0}\,\Omega_{\rm ds}(a)$,\footnote{We note that there is no absolute motivation for 
this, but it has been shown to fit a wide range of models \citep{Gleyzes2017}.} where $\Omega_{\rm ds}(a)$ is the dark 
energy density parameter and $\alpha_{\rm x,0}$ an arbitrary fixed number. We also introduce the \textit{effective 
Planck mass} $M^2$, via 
$$\alpha_{\rm M}\equiv\frac{\mathrm{d}\ln{M^2}}{\mathrm{d}\ln{a}}\,.$$ 
For $M^2$ to be uniquely determined, one needs to specify initial conditions. Here we choose $M^2/M_{\rm pl}^2=1$ at 
early times, i.e.\ $a\ll 1$. Following \cite{Zumalacarregui2017}, we will also fix $\alpha_{\rm K,0}=1$ for the 
kineticity term, as it has been shown that this function is very difficult to constrain with data, as observables 
are weakly dependent on $\alpha_{\rm K}$.

In contrast to \cite{Zumalacarregui2017}, we do not consider models with $\alpha_{\rm B}=\alpha_{\rm M}=0$ and 
$\alpha_{\rm T}\neq0$. Indeed, it is impossible to construct a Lagrangian with $\alpha_{\rm M}\equiv0$ and 
$\alpha_{\rm T}\neq 0$ because to achieve the first condition, excluding extreme fine-tuning between $G_4$ and $G_5$, 
one requires $M^2$ constant, and therefore $G_4$ and $G_5$ must be constant, which forces $\alpha_{\rm T}=0$, 
contradicting the original assumption. Further it is also impossible to have $a_{\rm B}=0$ and $a_{\rm T}\neq 0$ 
because, unless 
there is a fine-tuning (different from the previous one) between $G_3$, $G_4$ and $G_5$, $a_{\rm B}\equiv0$ also 
implies 
$G_4$ and $G_5$ constant and hence $a_{\rm T}=0$, again in disagreement with the original assumption. 
We note that, always under the assumption of $a_{\rm B}=0$, $G_3\equiv G_3(\phi)$, which due to an integration by 
parts, is equivalent to a $k$-essence model with $G_2\rightarrow G_2-G_{3,\phi}\nabla_{\mu}\phi\nabla^{\mu}\phi$.

Therefore, we consider the following classes of models:
\begin{enumerate}
 \item \label{itm:Q} \textit{$k$-essence-like models}: 
       $\alpha_{\rm K}\neq 0$, $\alpha_{\rm B}=\alpha_{\rm M}=\alpha_{\rm T}=0$. 
       This choice corresponds to Quintessence \citep{Ford1987,Peebles1988,Ratra1988a} and $k$-essence models 
       \citep{ArmendarizPicon1999,Mukhanov2006} and in terms of the Horndeski Lagrangian requires $G_2=G_2(\phi,X)$, 
       $G_{4}=M_{\rm pl}^2/2$ and $G_3=G_5=0$. 
       Expressions for the EoS for perturbations can be found in \cite{Battye2012,Battye2014}. 
       The numerical implementation of this class of models has been studied many times before and we will not 
       discuss these models further.
 
 \item \label{itm:fR} \textit{$f(R)$-like models}: 
       $\alpha_{\rm K}=\alpha_{\rm T}=0$, $\alpha_{\rm M}\neq0$, $\alpha_{\rm B}\neq 0$. 
       From the definition of the $\alpha_{\rm x}$s coefficients in Appendix~\ref{sect:perturbations} in terms of 
       the Horndeski functions $G_i$, again excluding fine-tuning between the Horndeski functions, one requires 
       $G_2=G_2(\phi)$, $G_4=G_4(\phi)$ and $G_3=G_5=0$. This is indeed the case for $f(R)$ gravity models
       \citep{Silvestri2009,Sotiriou2010,DeFelice2010}, which represent a particular subclass of these models where 
       $\alpha_{\rm M}=2\alpha_{\rm B}$ as $G_2=-\tfrac{1}{2}M_{\rm pl}^2(Rf_R-f)$ and 
       $G_4=\tfrac{1}{2}M_{\rm pl}^2(1+f_R)$, where $f_R=\mathrm{d}f/\mathrm{d}R$. Indeed, one can show 
       \citep[see for example,][]{DeFelice2010,Capozziello2007a} that $f(R)$ models are equivalent to Brans-Dicke 
       models \citep{Brans1961} with the parameter $\omega_{\rm BD}=0$ and with the identification 
       $\phi/M_{\rm pl}=1+f_R$ and scalar field potential $V(\phi)=-G_2$, where $R\equiv R(\phi)$. 
       In this paper we do not show numerical results for $f(R)$ gravity, as this was the subject of our previous 
       works \cite{Battye2016a,Battye2018a,Bellini2018}. 
       The EoS for $f(R)$ gravity can be found in \cite{Battye2016a,Battye2018a}. 
       We note that unlike \verb|hi_class|, our code is able to solve the equations of motion in this class of models.

 \item \label{itm:KGB} \textit{KGB-like models}: 
       $\alpha_{\rm K}\neq 0$, $\alpha_{\rm B}\neq 0$, $\alpha_{\rm M}=\alpha_{\rm T}=0$. These models have 
       $G_2=G_2(\phi,X)$, $G_3=G_3(\phi,X)$, $G_{4}=M_{\rm pl}^2/2$ and $G_5=0$ and describe KGB-like gravity 
       \citep{Deffayet2010,Pujolas2011,Kobayashi2010,Kimura2011}. The expressions for the EoS parameters have been 
       evaluated by \cite{Battye2013} and we verified that the expressions used here are equivalent to those found 
       there. As we will show later, it is possible to write the EoS in terms of the matter and dark sector fluid 
       variables, while \cite{Battye2013} used metric variables. To establish the equivalence of the two expressions, 
       it is useful to consider the gauge-invariant quantities $X$, $Y$ (introduced in Sect.~\ref{sect:attractor}) 
       defined as linear combinations of the metric perturbations and their time derivatives. By writing the entropy 
       as perturbations in both the Newtonian and synchronous gauge in terms of the gauge invariant quantities, it is 
       relatively easy to find what conditions the coefficients must obey in order to ensure gauge invariance. This 
       then leads to the equality of the expressions given here with those presented in \cite{Battye2013}.
 
 \item \label{itm:BDK0} $\alpha_{\rm K}\neq 0$, $\alpha_{\rm M}\neq 0$, $\alpha_{\rm B}=\alpha_{\rm T}=0$. A luminal 
       propagation of GW, $\alpha_{\rm T}=0$, requires $G_5=0$ and $G_4=G_4(\phi)$, while $\alpha_{\rm B}=0$ implies 
       $XG_{3,X}+G_{4,\phi}=0$. We are then free to write $G_4(\phi)=\tfrac{1}{2}M_{\rm pl}^2f(\phi/M_{\rm pl})$, with 
       $f(\phi/M_{\rm pl})$ a dimensionless function of the scalar field $\phi$, leads to 
       $G_3 = -\tfrac{1}{2}M_{\rm pl}\left[f^{\prime}\left(\phi/M_{\rm pl}\right)\ln{\left(X/m^4\right)}+
       g\left(\phi/M_{\rm pl}\right)\right]$, where the prime stands for the derivative with respect to 
       $\phi/M_{\rm pl}$, $m$ is an arbitrary mass scale and $g(\phi/M_{\rm pl})$ a dimensionless function of $\phi$. 
       These models also have $G_2=G_2(\phi,X)$.
 
 \item \label{itm:BDK1} $\alpha_{\rm K}\neq 0$, $\alpha_{\rm B}\neq 0$, $\alpha_{\rm M}\neq 0$, $\alpha_{\rm T}=0$. 
       This represents the most generic Horndeski model compatible with GW constraints. It is obtained by setting 
       $G_2=G_2(\phi,X)$, $G_3=G_3(\phi,X)$, $G_4=G_4(\phi)$ and $G_5=0$. In addition to models already mentioned 
       under classes 1--4, models falling in this category are non-minimally coupled $k$-essence 
       \citep{Kim2005,Sawicki2013}, MSG/Palatini $f(R)$ models \citep{Vollick2003,Carroll2006}, Galileon cosmology 
       \citep{Chow2009}. Note that often for these classes of models, there exists a relation between $\alpha_{\rm B}$ 
       and $\alpha_{\rm M}$. This is for example the case of the no slip gravity model proposed by \cite{Linder2018} 
       which has $\alpha_{\rm B}=\alpha_{\rm M}$ and of the minimally self-accelerating models of 
       \citep{Lombriser2016a,Lombriser2017} designed to break the dark degeneracy\footnote{The dark degeneracy 
       refers to Horndeski models whose cosmological background and linear scalar fluctuations are degenerate with the 
       $\Lambda$CDM cosmology, but differ from it in the tensor sector.}. 
       Model 4 is a subclass of model 5, provided that $G_3$ is suitably chosen to give $\alpha_{\rm B}=0$.
 
 \item \label{itm:GH} $\alpha_{\rm K}\neq 0$, $\alpha_{\rm B}\neq 0$, $\alpha_{\rm M}\neq 0$, $\alpha_{\rm T}\neq 0$. 
       This is the most general Horndeski model allowed by the theory. The free functions are all depending on 
       $\phi$ and $X$. Typical classes of models are given by Galileons \citep{Nicolis2009,Deffayet2009} and 
       Gauss-Bonnet models \citep{Carroll2005}. These models are in tension with gravitational waves measurements, but 
       we include them for completeness.
\end{enumerate}

\section{The Equation of State approach and its numerical implementation}\label{sect:EoS}
We have implemented the EoS approach for the scalar sector in a way similar to the implementation presented in our 
most recent analyses \cite{Battye2018a,Battye2019} with some minor differences. For clarity let us recall the main 
equations and definitions of our new implementation.

We use the density perturbation $\Delta$ and the rescaled velocity perturbation $\Theta$, defined as
\begin{equation}\label{eqn:DeltaTheta}
 \Delta \equiv \delta + \Theta\,, \quad \Theta \equiv 3H(1+w)\theta\,.
\end{equation}
In Eqn.~(\ref{eqn:DeltaTheta}), $H$ is the Hubble parameter, $w=\bar{P}/\bar{\rho}$ the background equation of state, 
$\delta=\delta\rho/\bar{\rho}$ the density perturbation and $\theta$ the divergence of the velocity perturbation. 
Overbarred variables refer to background quantities.

We introduce the gauge invariant rescaled velocity perturbation
\begin{equation}
 \hat{\Theta} \equiv \Theta + 3(1+w)T\,,
\end{equation}
where $T=0$ in the conformal Newtonian gauge (CNG) and $T=(h^{\prime}+6\eta^{\prime})/(2{\rm K}^2)$ in the synchronous 
gauge (SG), where $h$ and $\eta$ are the scalar metric perturbations, ${}^{\prime}$ denotes derivative with respect to 
$\ln{a}$ and ${\rm K}=k/(aH)$, with $k$ a wavenumber.

For the equations of motion we use the dimensionfull quantities 
$\tilde{\Delta}_{\rm x} \equiv \bar{\rho}_{\rm x}\Delta_{\rm x}$ and 
$\hat{\tilde{\Theta}}_{\rm x} \equiv \bar{\rho}_{\rm x}\hat{\Theta}_{\rm x}$. 
With these gauge-invariant variables, the equations of motion for the dark sector perturbations are
\begin{align}
 & \tilde{\Delta}_{\rm ds}^{\prime} + 3\tilde{\Delta}_{\rm ds} - 2\bar{\rho}_{\rm ds}w_{\rm ds}\Pi_{\rm ds} + 
   g_{\rm K}\epsilon_H\hat{\tilde{\Theta}}_{\rm ds} = 3(1+w_{\rm ds})\bar{\rho}_{\rm ds}X\,, \label{eqn:Delta} \\
 & \hat{\tilde{\Theta}}_{\rm ds}^{\prime} + \left[\epsilon_H + 3(1+c_{\rm a,ds}^2)\right]\hat{\tilde{\Theta}}_{\rm ds} 
   - 3c_{\rm a, ds}^2\tilde{\Delta}_{\rm ds} - 
   3\bar{\rho}_{\rm ds}w_{\rm ds}\zeta_{\rm ds} = 3(1+w_{\rm ds})\bar{\rho}_{\rm ds}Y\,, \label{eqn:Theta}
\end{align}
where $g_{\rm K}\equiv 1+{\rm K}^2/(3\epsilon_H)$, $\epsilon_H\equiv-H^{\prime}/H$ and 
$c_{\rm a,ds}^2\equiv \mathrm{d}\bar{P}_{\rm ds}/\mathrm{d}\bar{\rho}_{\rm ds}$ is the adiabatic sound speed. The 
quantity $w_{\rm ds}\zeta_{\rm ds}$ is a relevant linear combination of the anisotropic stress and entropy 
perturbations defined as
\begin{equation}\label{eqn:zeta_ds}
 w_{\rm ds}\zeta_{\rm ds} \equiv \frac{2}{3}w_{\rm ds}\Pi_{\rm ds} + w_{\rm ds}\Gamma_{\rm ds}\,.
\end{equation}
The two terms on the right hand side of Eqs.~(\ref{eqn:Delta}) and (\ref{eqn:Theta}), $X$ and $Y$, are also 
gauge-invariant quantities \cite{Battye2016a,Battye2017,Battye2018a} which read $X=\eta^\prime + \epsilon_HT$ and 
$Y=T^\prime+\epsilon_HT$ in the SG and $X=\phi^\prime + \psi$ and $Y=\psi$ in the CNG. These equations are equivalent 
to Eqs.~(122) and (123) in \cite{Gleyzes2014}.

The generic EoS for dark sector perturbations are the following expansions of the gauge invariant entropy perturbations 
$w_{\rm ds}\Gamma_{\rm ds}$ and anisotropic stress $w_{\rm ds}\Pi_{\rm ds}$:
\begin{align}
 \bar{\rho}_{\rm ds}w_{\rm ds}\Gamma_{\rm ds} & = C_{\Gamma\Delta_{\rm ds}}\tilde{\Delta}_{\rm ds} + 
                                  C_{\Gamma\Theta_{\rm ds}}\hat{\tilde{\Theta}}_{\rm ds} + 
                                  C_{\Gamma\Delta_{\rm m}}\tilde{\Delta}_{\rm m} + 
                                  C_{\Gamma\Theta_{\rm m}}\hat{\tilde{\Theta}}_{\rm m} + 
                                  C_{\Gamma\Gamma_{\rm m}}\bar{\rho}_{\rm m}w_{\rm m}\Gamma_{\rm m}\,, 
                                  \label{eqn:wGamma}\\
 \bar{\rho}_{\rm ds}w_{\rm ds}\Pi_{\rm ds} & = C_{\Pi\Delta_{\rm ds}}\tilde{\Delta}_{\rm ds} + 
                               C_{\Pi\Theta_{\rm ds}}\hat{\tilde{\Theta}}_{\rm ds} + 
                               C_{\Pi\Delta_{\rm m}}\tilde{\Delta}_{\rm m} + 
                               C_{\Pi\Theta_{\rm m}}\hat{\tilde{\Theta}}_{\rm m} + 
                               C_{\Pi\Pi_{\rm m}}\bar{\rho}_{\rm m}w_{\rm m}\Pi_{\rm m}\,, \label{eqn:wPi}
\end{align}
where the coefficients $C_{XY}$ are, in general, functions of $a$ and the rescaled wavenumber ${\rm K}$ and the 
matter fluid quantities are directly evaluated in \verb|CLASS| via
\begin{align}
 aH{\rm K}^2\hat{\Theta}_{\rm m} = &\,
 3\left\langle(\bar{\rho}_{\rm m}+\bar{P}_{\rm m}){\theta_{{{\rm m}}}^{{{\rm class}}}}\right\rangle
 /(\bar{\rho}_{\rm m} + \bar{P}_{\rm m})\,,\\
 \Delta_{\rm m} = &\, \left\langle\delta\rho_{\rm m}^{{{\rm class}}}\right\rangle/\bar{\rho}_{\rm m} + 
                      \Theta_{\rm m}\,,\\
 \bar{\rho}_{\rm m}w_{\rm m}\Gamma_{\rm m}  = &\, \left\langle\delta P_{\rm m}^{{{\rm class}}}\right\rangle - 
                                          c_{a,{\rm m}}^2\bar{\rho}_{\rm m}(\Delta_{\rm m} - \hat{\Theta}_{\rm m})\,,\\
 \bar{\rho}_{\rm m}w_{\rm m}\Pi_{\rm m} = &\, -\frac{3}{2}\left\langle(\bar{\rho}_{\rm m}+\bar{P}_{\rm m})
                                               \sigma_{\rm m}^{\rm class}\right\rangle\,.
\end{align}
The matter adiabatic sound speed is defined as
\begin{equation}
 c_{a,{\rm m}}^2 = \frac{w_{\rm m}\Omega_{\rm m} + \left\langle w_{\rm m}^2\Omega_{\rm m}\right\rangle}
                        {\left(1+w_{\rm m}\right)\Omega_{\rm m}}\;,
\end{equation}
with $\Omega_{\rm m}$ and $w_{\rm m} \equiv \left\langle w_{\rm m}\Omega_{\rm m} \right\rangle/\Omega_{\rm m}$ the 
matter density parameter and background equation of state, respectively, and where the brackets $\langle\cdots\rangle$ 
indicate the sum over all matter components. Our definitions of peculiar velocity $\theta$ and anisotropic stress 
$w\Pi$ are different from those used in \verb|CLASS|, which are defined as in \cite{Ma1995}. The following 
correspondence holds: $\theta^{\rm class}=\frac{k^2}{a}\theta$ and $(\rho+P)\sigma^{\rm class}=-\tfrac{2}{3}\rho w\Pi$.

A specific dark energy or modified gravity model is characterised by its EoS for perturbations and therefore by the 
functional form of the coefficients $C_{XY}$ in Eqs.~(\ref{eqn:wGamma}) and (\ref{eqn:wPi}). We have computed these 
coefficients for the generic Horndeski Lagrangian and for each of the specific models listed in the previous section. 
Their expressions are reported in Appendix~\ref{sect:coefficientsEoS}.

Integrating the equations of motion for the dark sector perturbations requires the update of the total stress-energy 
tensor which is done as follows in \verb|EoS_class|:
\begin{align}
 \delta\rho_{\rm tot} = &\, \left\langle \delta\rho_{\rm m}\right\rangle + 
                            (\tilde{\Delta}_{\rm ds} - \hat{\tilde{\Theta}}_{\rm ds})\,, \nonumber\\
 (\bar{\rho}_{\rm tot} + \bar{P}_{\rm tot})\theta_{\rm tot}^{\rm class} = &\, 
       \left\langle(\bar{\rho}_{\rm m}+\bar{P}_{\rm m}){\theta_{{{\rm m}}}^{{{\rm class}}}}\right\rangle + 
       \frac{1}{3}aH{\rm K}^2\hat{\tilde{\Theta}}_{\rm ds}\,,\nonumber\\
 (\bar{\rho}_{\rm tot}+\bar{P}_{\rm tot})\sigma_{\rm tot}^{\rm class} = &\;
       \left\langle(\rho_{\rm m}+P_{\rm m}){\sigma_{{{\textrm{m}}}}^{{{\textrm{class}}}}}\right\rangle - 
       \frac{2}{3}\bar{\rho}_{\rm ds}w_{\rm ds}\Pi_{\rm ds}\,,\nonumber\\
 \delta P_{\rm tot} = &\; \left\langle \delta P_{\rm m}\right\rangle + \bar{\rho}_{\rm ds}w_{\rm ds}\Gamma_{\rm ds} + 
        c_{a,{\rm ds}}^2(\tilde{\Delta}_{\rm ds} - \hat{\tilde{\Theta}}_{\rm ds})\,.\nonumber
\end{align}

We set the initial conditions for dark sector perturbations at early time, $a_{\rm ini}$, when the dark sector density 
is subdominant compared to matter density. Note that $a_{\rm ini}$ represents the time when dark sector perturbations 
are switched on, and it does not affect the standard implementation in \verb|CLASS|, where matter perturbations start 
at a much earlier time. Our choice is on the same line of \verb|EFTCAMB| and it is done for practical reasons, as 
numerical instabilities may arise as at early time all the models considered reduce to General Relativity. 
The default setting in our numerical implementation is $a_{\rm ini}=10^{-4}$. 
We ran several tests and checked that our spectra are unaffected by the exact choice of $a_{\rm ini}$, as long as 
$\rho_{\rm ds}(a_{\rm ini})/\rho_{\rm m}(a_{\rm ini})\ll 1$.

Although to our knowledge this has not been rigorously proven in the generic case, it appears that dark sector 
perturbations follow attractor solutions. In other words there is no significant sensitivity to initial conditions. 
So, in principle, one could set $\Delta_{\rm ds}(a_{\rm ini})=\Theta_{\rm ds}(a_{\rm ini})=0$ at early times because 
$\Delta_{\rm ds}$ and $\Theta_{\rm ds}$ would anyway rapidly converge to the attractor solution. However, this implies 
that if we are able to derive an analytical approximation for the attractor solution, setting the initial conditions 
to the attractor is numerically more stable and efficient.

In \cite{Battye2018b} we derived the generic form of the attractor solution for $\Delta_{\rm ds}$ using the EoS 
formalism, under the assumption that $w_{\rm ds}$ is constant and velocity perturbations are subdominant with respect 
to density perturbations, i.e.\ $\hat{\Theta}\ll\Delta$.

Here, we improve upon this result and derive the generic form of the attractor solution for $\Delta_{\rm ds}$ just 
under the assumption of subdominant velocity perturbations. We present this result in Section~\ref{sect:ds} where 
we also show that the assumption of subdominant velocity has a wide range of validity, which only ends at the largest 
cosmological scales (${\rm K}\simeq 1$). Hence, we always set $\Delta_{\rm ds}(a_{\rm ini})$ to the attractor 
solution given by Eqn.~(\ref{eqn:dAIC}).

Regarding the attractor solution of dark sector velocity perturbations, we have been able to obtain an analytical 
formula that is a good approximation at early times for some but not all the models we studied. We used this formula to 
set the initial conditions for $\Theta_{\rm ds}$.

Note that attractor solutions to determine the initial conditions for modified gravity models are also used in 
\verb|EFTCAMB| for the St\"uckelberg field representing the perturbations in the dark sector \citep{Hu2014b}. 
The main difference between the two approaches is that \verb|EFTCAMB| assumes that the theory is close to General 
Relativity at sufficiently early times so that the attractor is only valid for a limited time range, while in our case 
the attractor is valid even at late time, provided ${\rm K}\gg 1$.

\begin{figure}[!ht]
 \centering
 \includegraphics[width=6.5cm,angle=0]{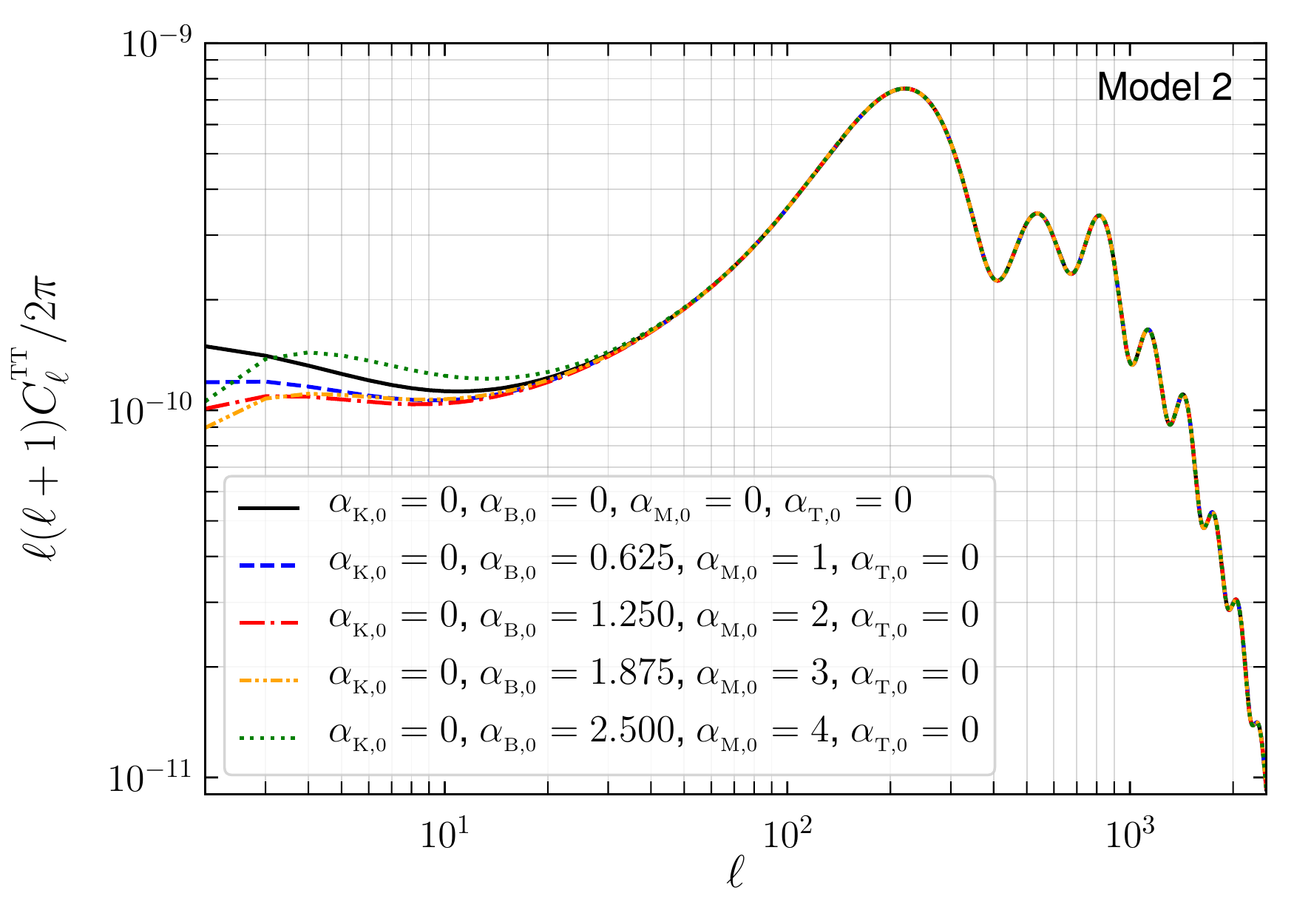}
 \includegraphics[width=6.5cm,angle=0]{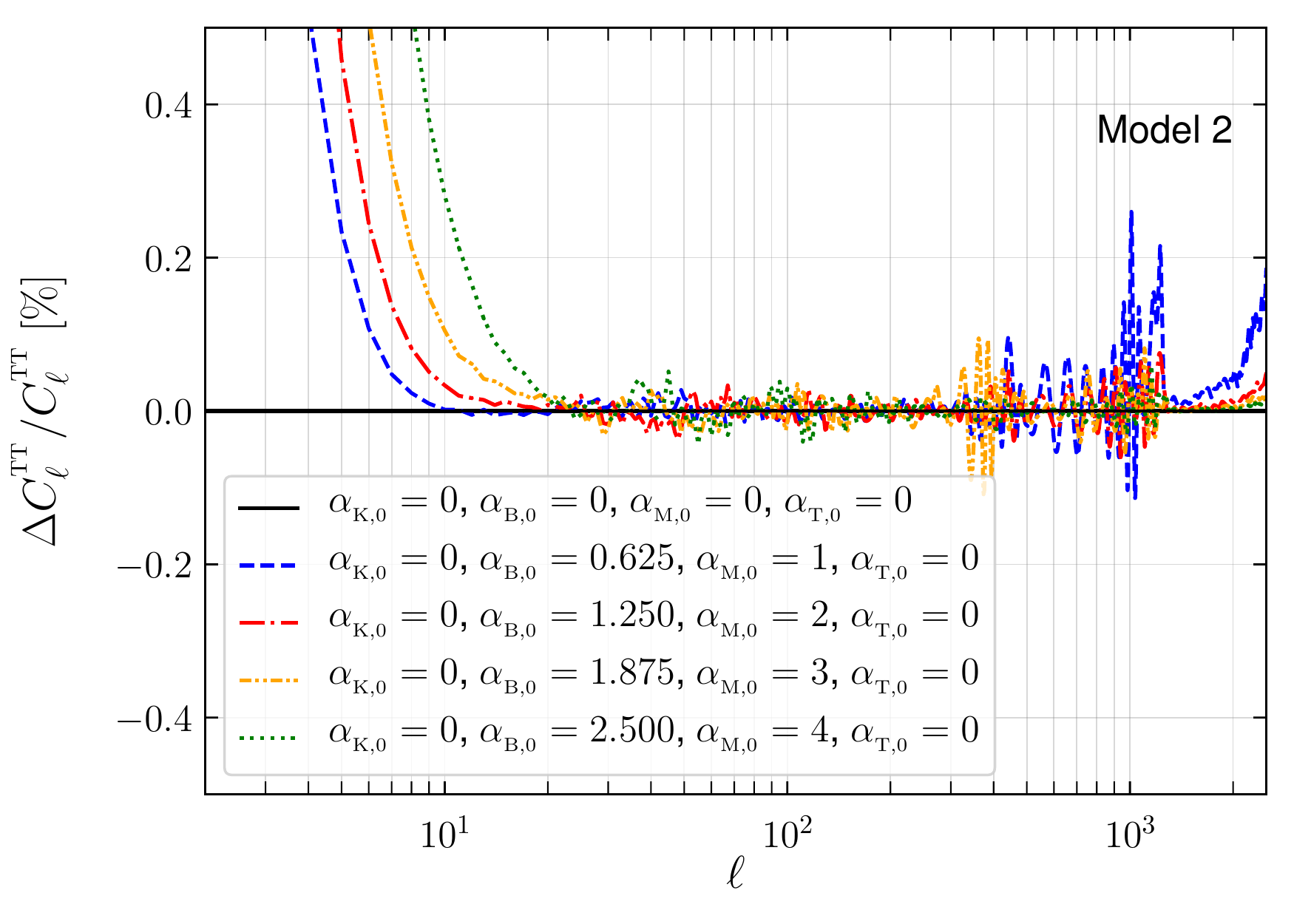}
 \includegraphics[width=6.5cm,angle=0]{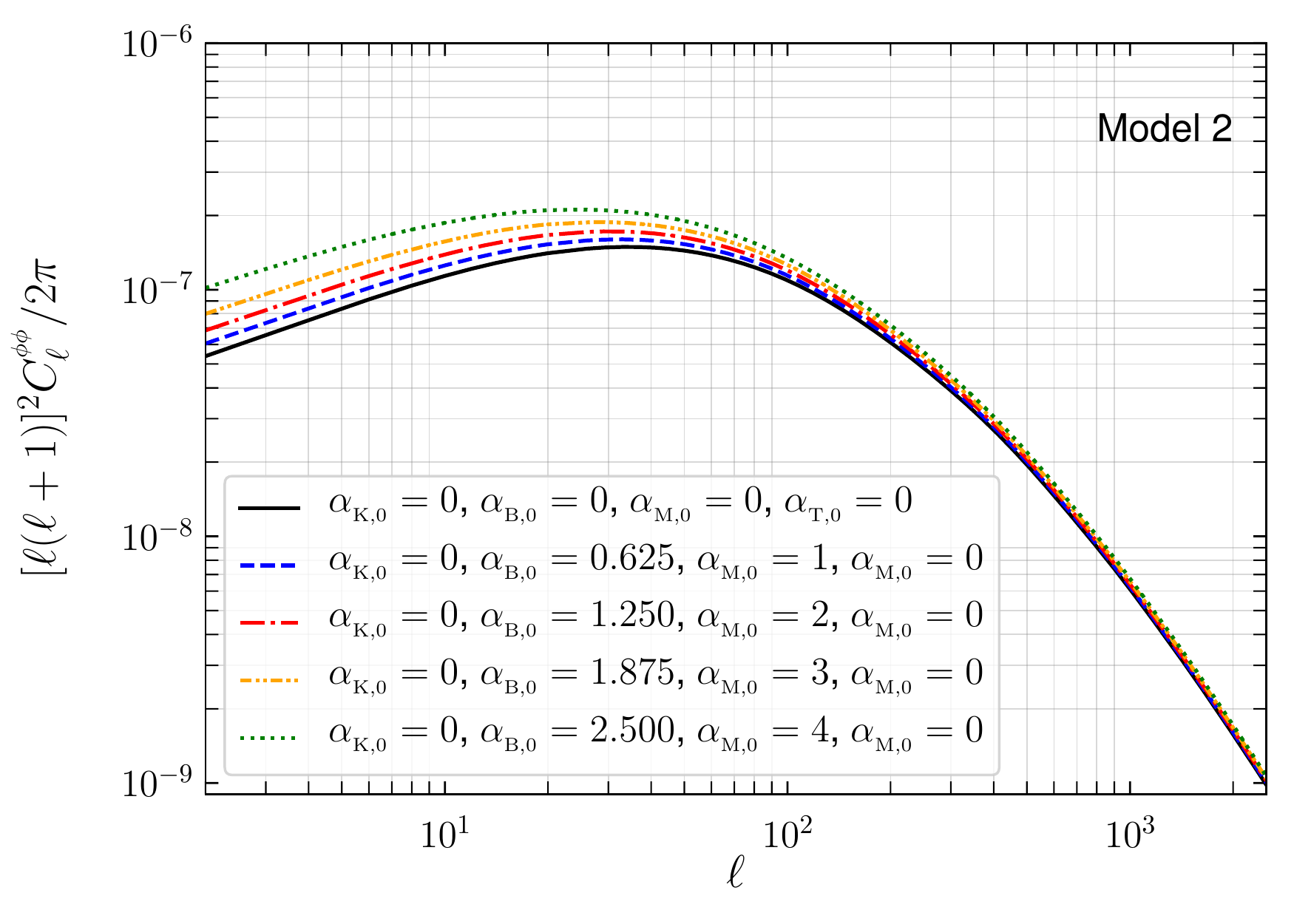}
 \includegraphics[width=6.5cm,angle=0]{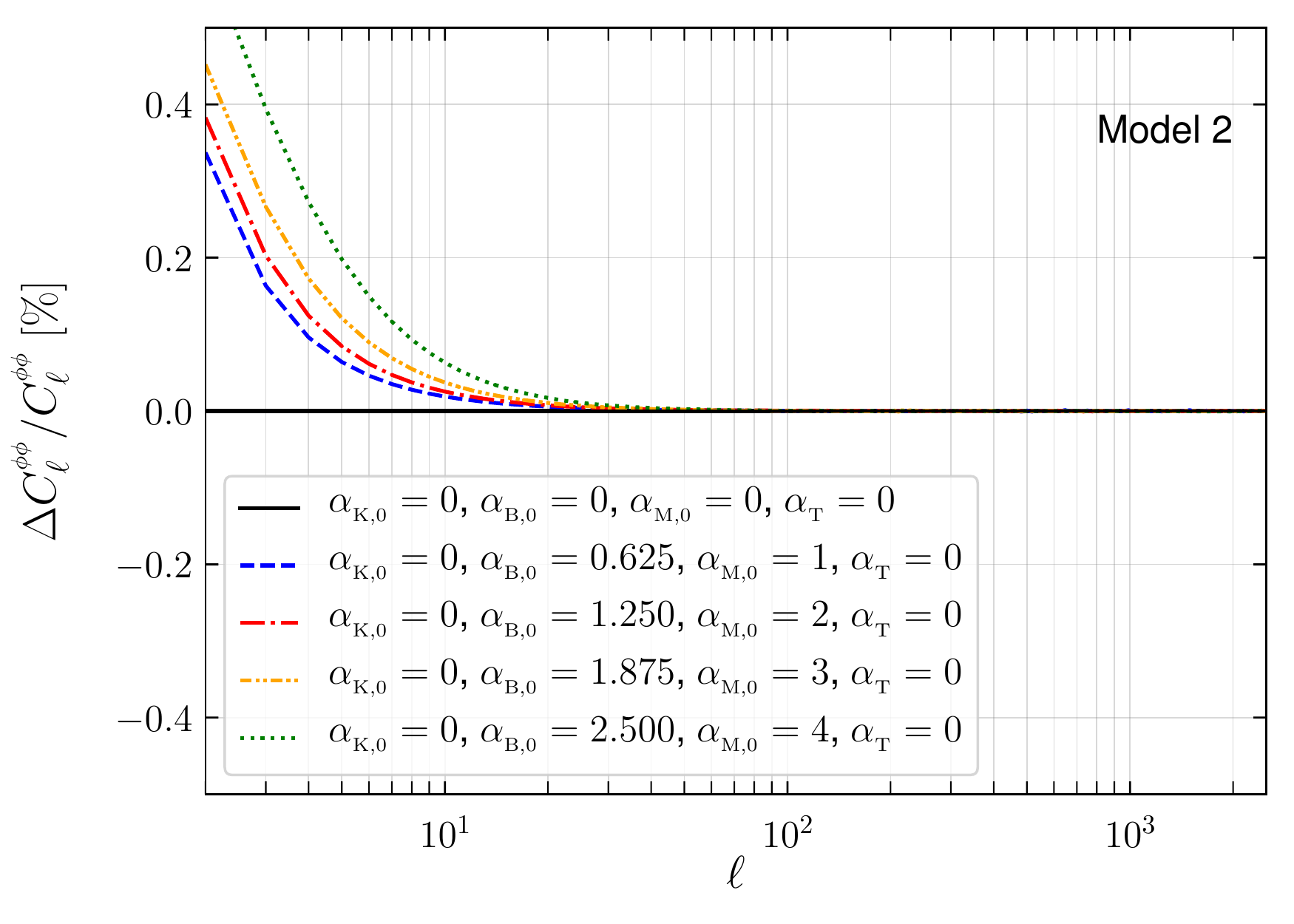}
 \includegraphics[width=6.5cm,angle=0]{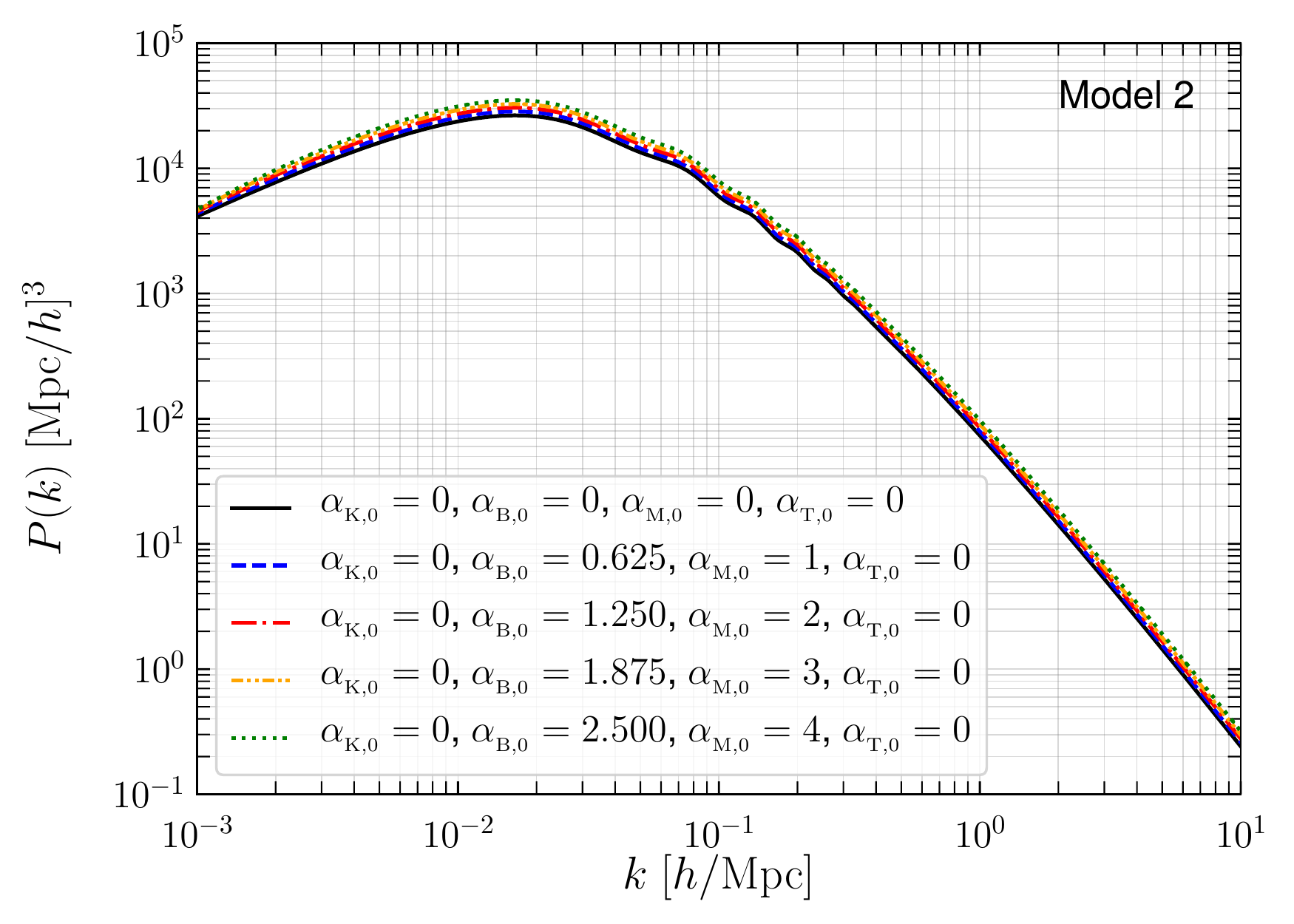}
 \includegraphics[width=6.5cm,angle=0]{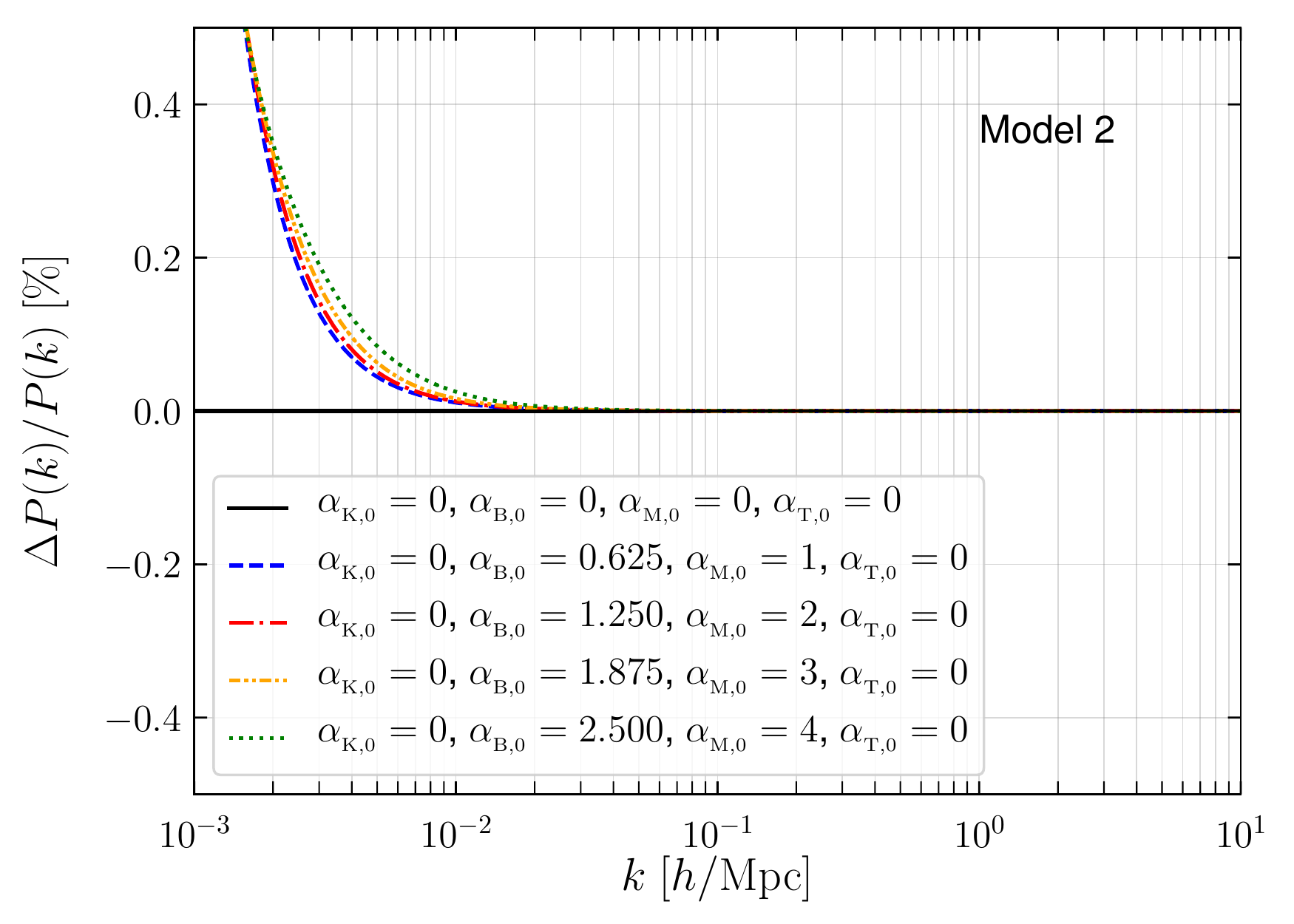}
 \cprotect\caption[justified]{Comparison for the spectra for model 2 ($f(R)$-like). In the left hand panels we present 
 the spectra obtained with \verb|EoS_class| and on the right the relative difference with the spectra of model 5 with 
 $\alpha_{\rm K,0}=1$, as explained in the text. Top panels show the angular temperature anisotropy power spectrum, 
 middle panels the angular power spectrum of the lensing potential, while bottom panels present the linear matter 
 power spectrum. 
 The black solid line represents the model with $\alpha_{\rm B,0}=\alpha_{\rm M,0}=0$, the dashed blue 
 (red dot-dashed) stands for $\alpha_{\rm B,0}=0.625$ and $\alpha_{\rm M,0}=1$ ($\alpha_{\rm B,0}=1.25$ and 
 $\alpha_{\rm M,0}=2$), the orange dot-dotted-dashed (dotted green) curve is for $\alpha_{\rm B,0}=1.875$ and 
 $\alpha_{\rm M,0}=3$ ($\alpha_{\rm B,0}=2.5$ and $\alpha_{\rm M,0}=4$), respectively. For all the models, 
 $\alpha_{\rm K,0}=0$.}
 \label{fig:aBaM}
\end{figure}

\section{Code validation}\label{sect:validation}
In this section we compare spectra obtained with our code \verb|EoS_class| with results from \verb|hi_class|. We 
compute the dimensionless CMB angular temperature anisotropy power spectra $C_{\ell}^{\rm TT}$, the dimensionless 
angular power spectrum of the lensing potential $C_{\ell}^{\phi\phi}$ and the total linear matter power spectrum 
$P(k)$ in units (Mpc/$h$)$^3$.

We do not report here a similar comparison with \verb|EFTCAMB|, as a detailed analysis for Horndeski models between 
these two codes has already been done in \cite{Bellini2018}. Note that in \cite{Bellini2018} we had also demonstrated 
the accuracy of our implementation of the EoS approach for $f(R)$ models compared to \verb|EFTCAMB|.

\begin{figure*}[!ht]
 \centering
 \includegraphics[width=6.5cm,angle=0]{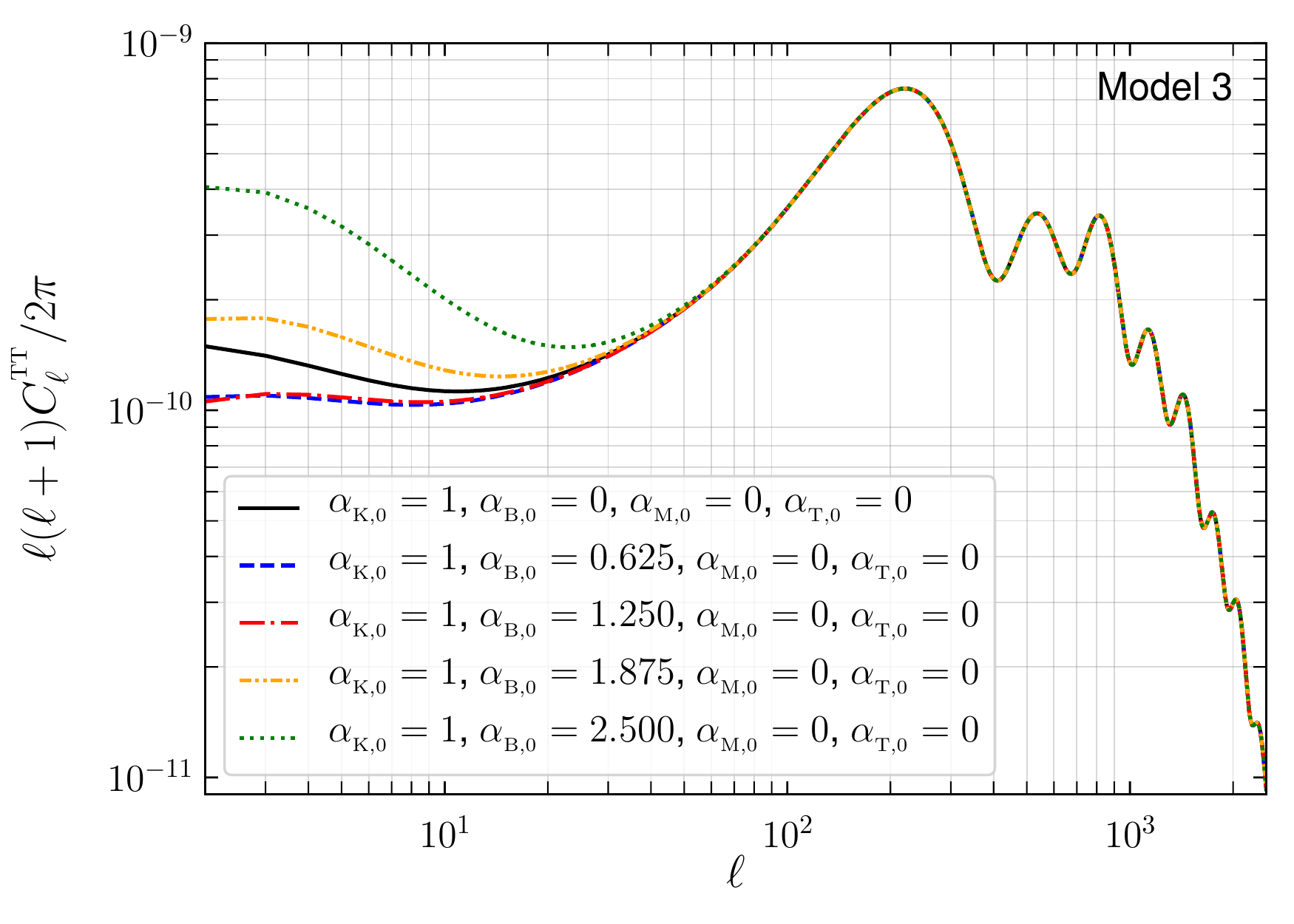}
 \includegraphics[width=6.5cm,angle=0]{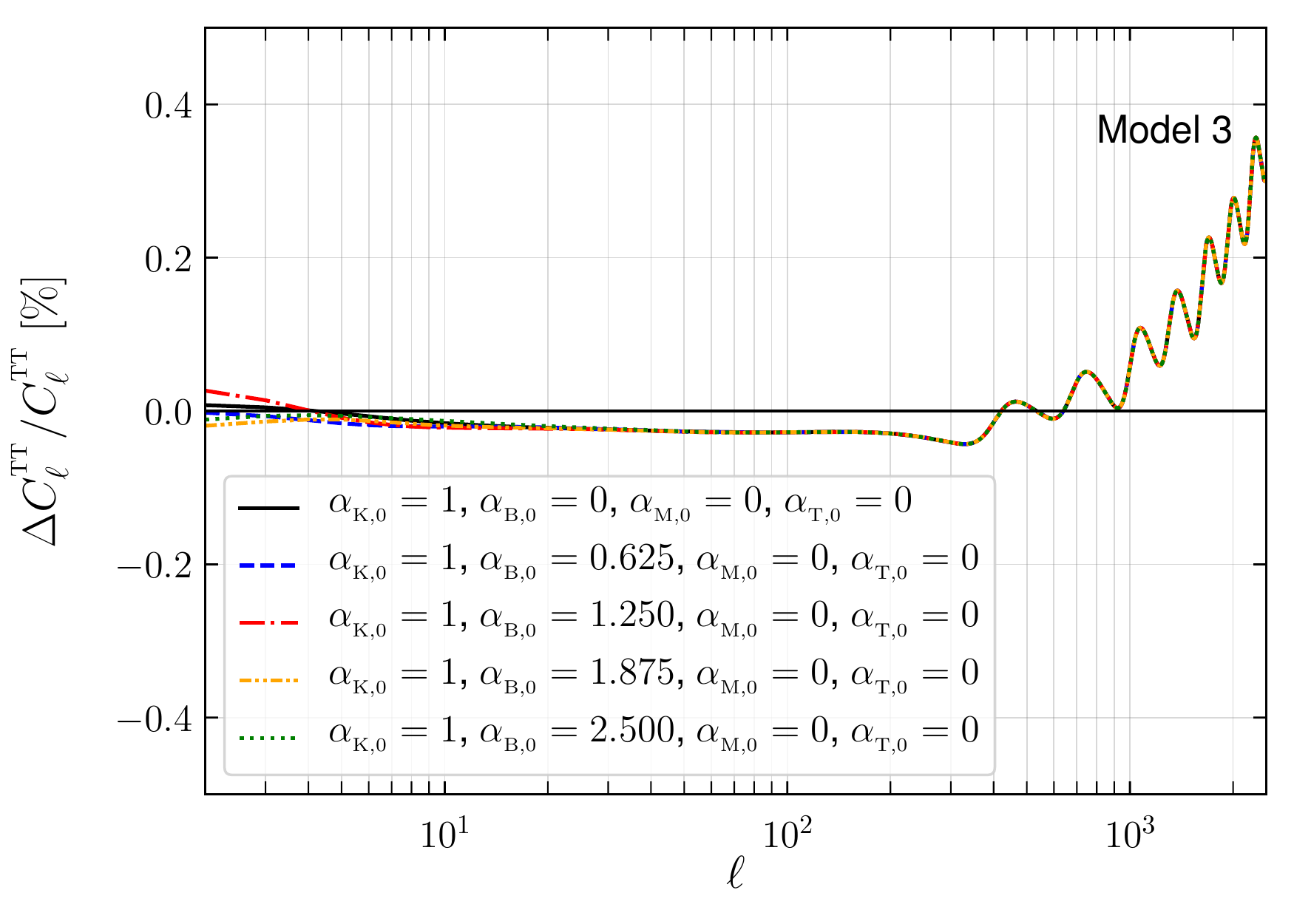}
 \includegraphics[width=6.5cm,angle=0]{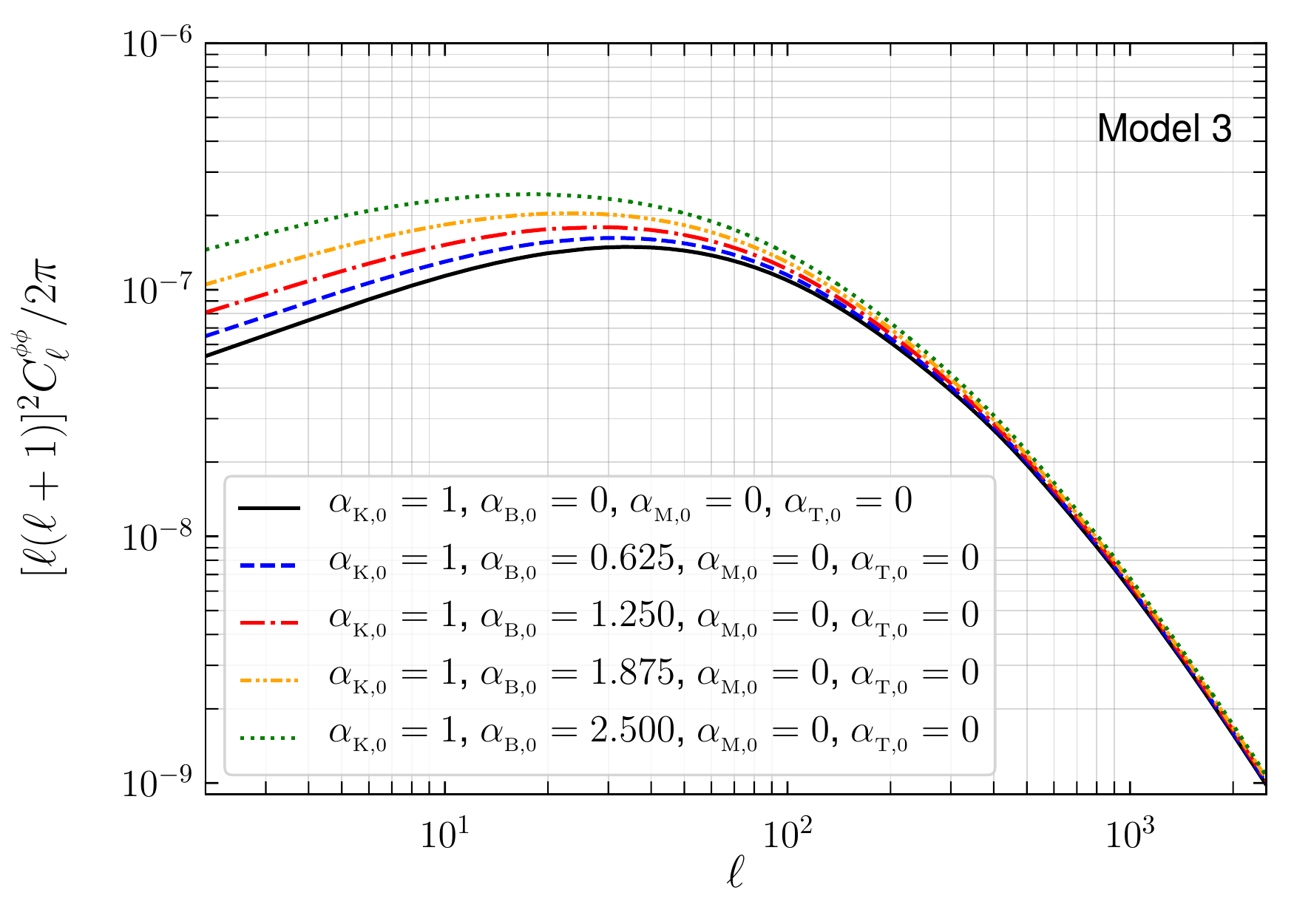}
 \includegraphics[width=6.5cm,angle=0]{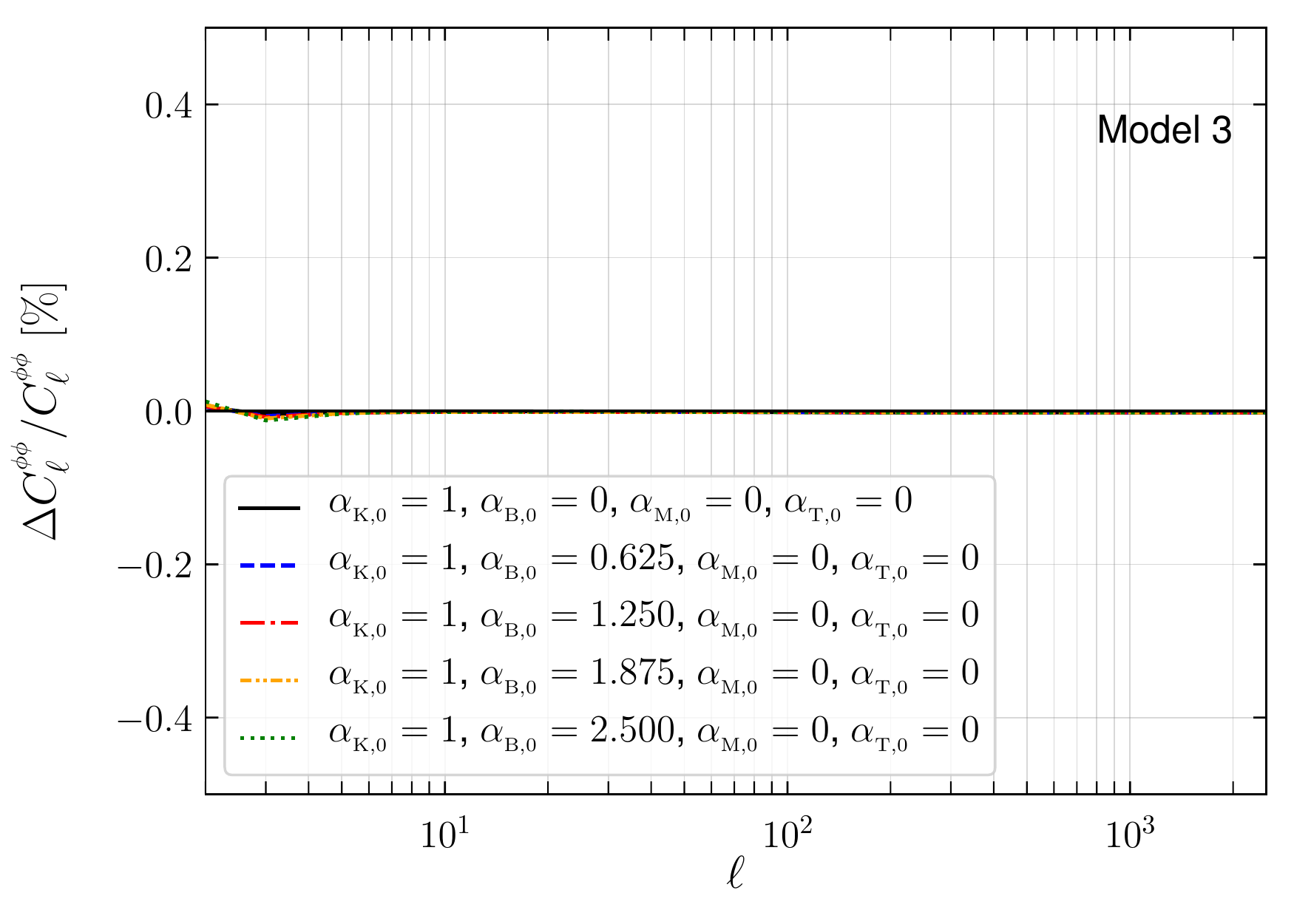}
 \includegraphics[width=6.5cm,angle=0]{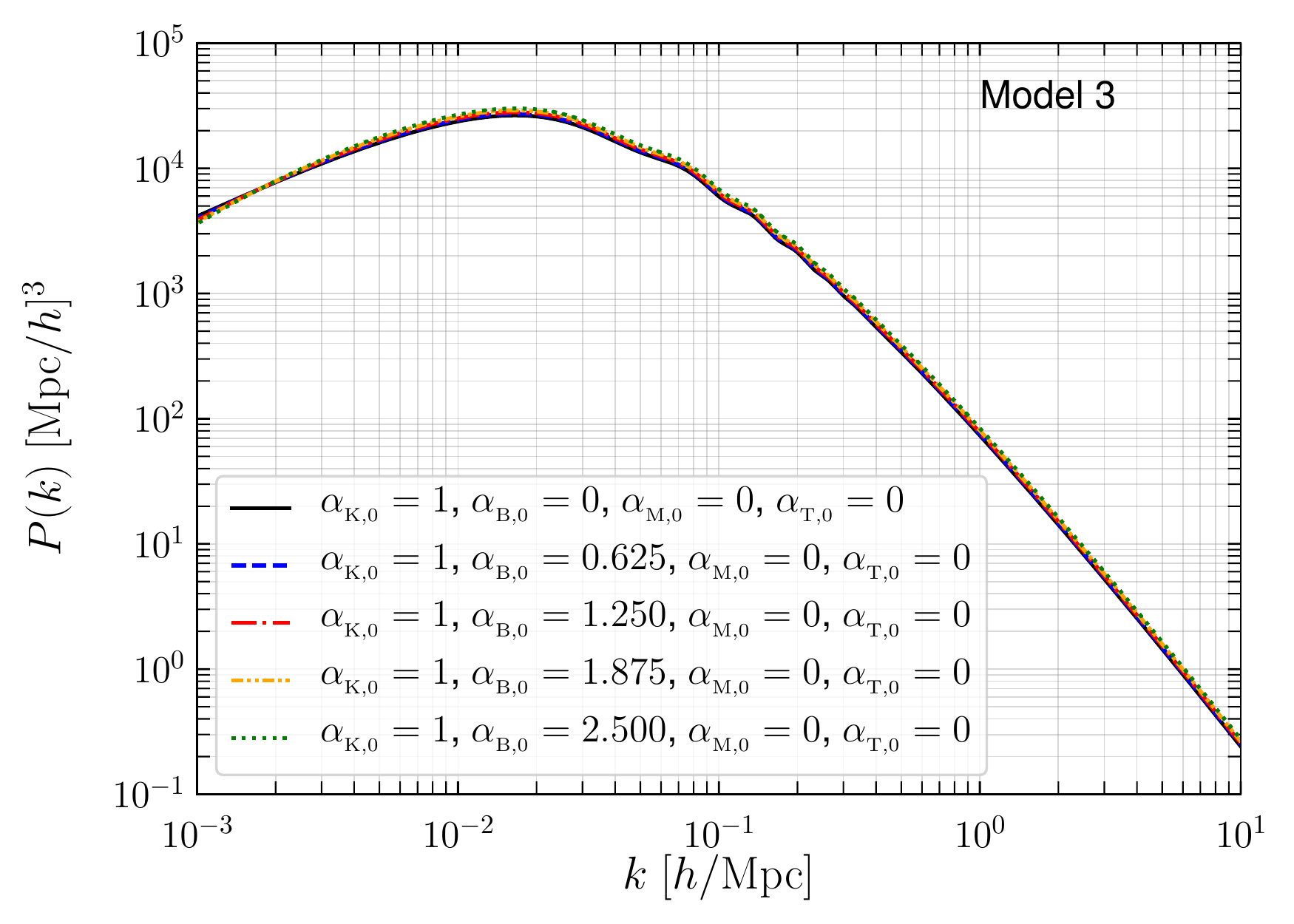}
 \includegraphics[width=6.5cm,angle=0]{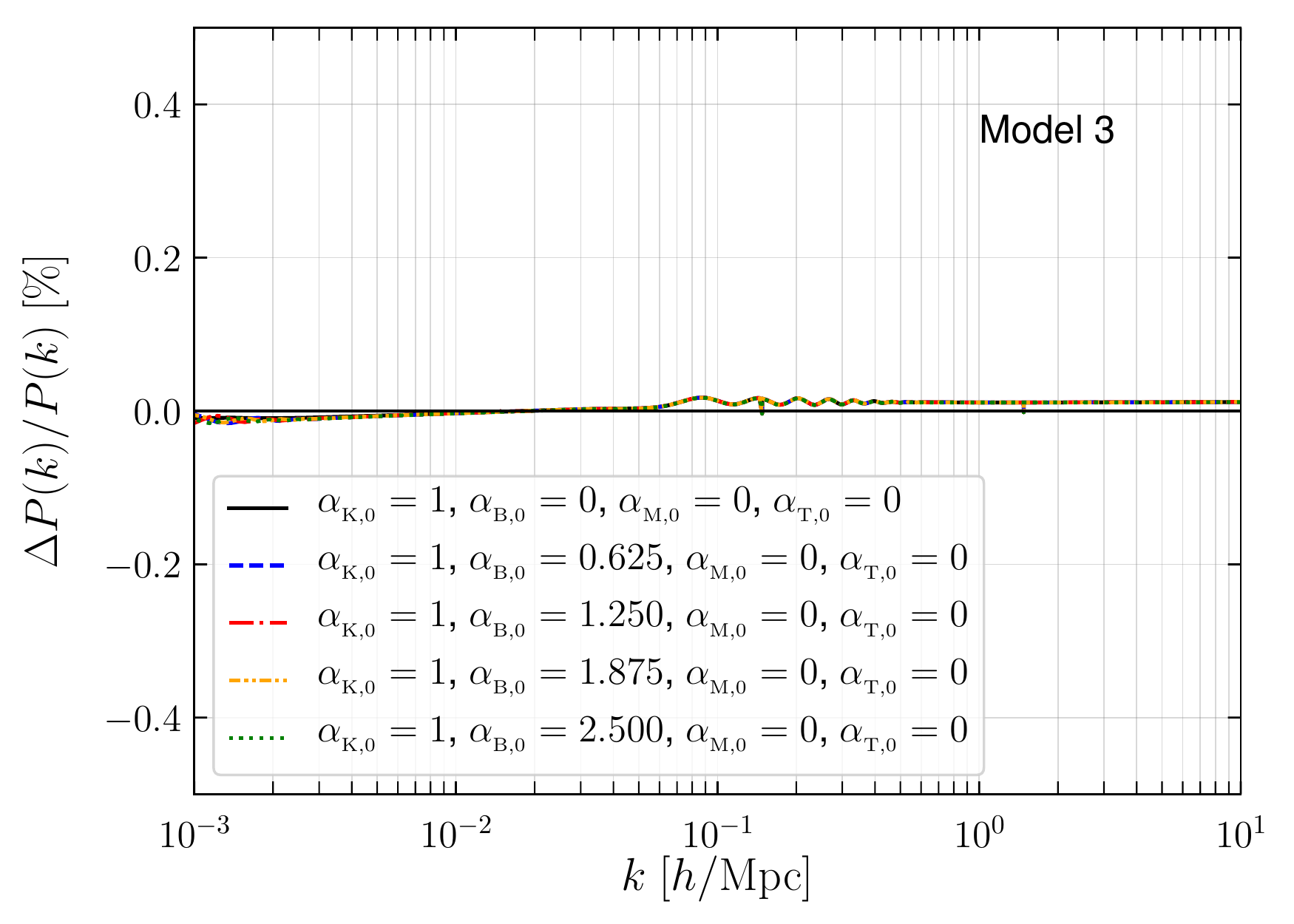}
 \cprotect\caption[justified]{Comparison for the spectra for model 3 (KGB-like). In the left hand panels we present the 
 spectra obtained with \verb|EoS_class| and on the right the relative difference with the corresponding spectra 
 obtained with \verb|hi_class|. Top panels show the angular temperature anisotropy power spectrum, middle panels the 
 angular power spectrum of the lensing potential, while bottom panels present the linear matter power spectrum. 
 The black solid line represents the model with $\alpha_{\rm B,0}=0$, the dashed blue (red dot-dashed) stands for 
 $\alpha_{\rm B,0}=0.625$ ($\alpha_{\rm B,0}=1.25$), the orange dot-dotted-dashed (dotted green) curve is for 
 $\alpha_{\rm B,0}=1.875$ ($\alpha_{\rm B,0}=2.5$), respectively. For all the models, $\alpha_{\rm K,0}=1$.}
 \label{fig:aKaB}
\end{figure*}

\begin{figure*}[!ht]
 \centering
 \includegraphics[width=6.5cm,angle=0]{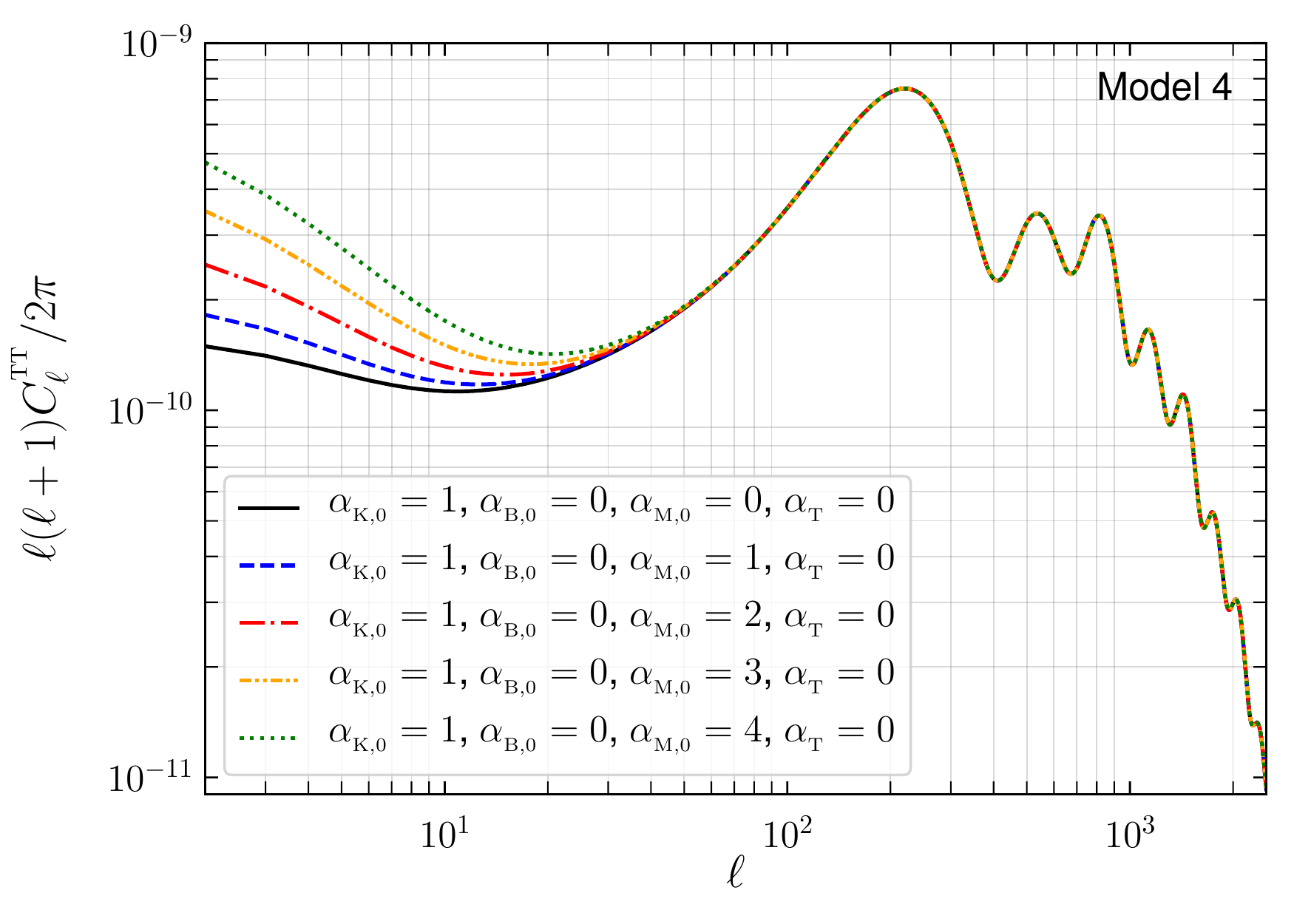}
 \includegraphics[width=6.5cm,angle=0]{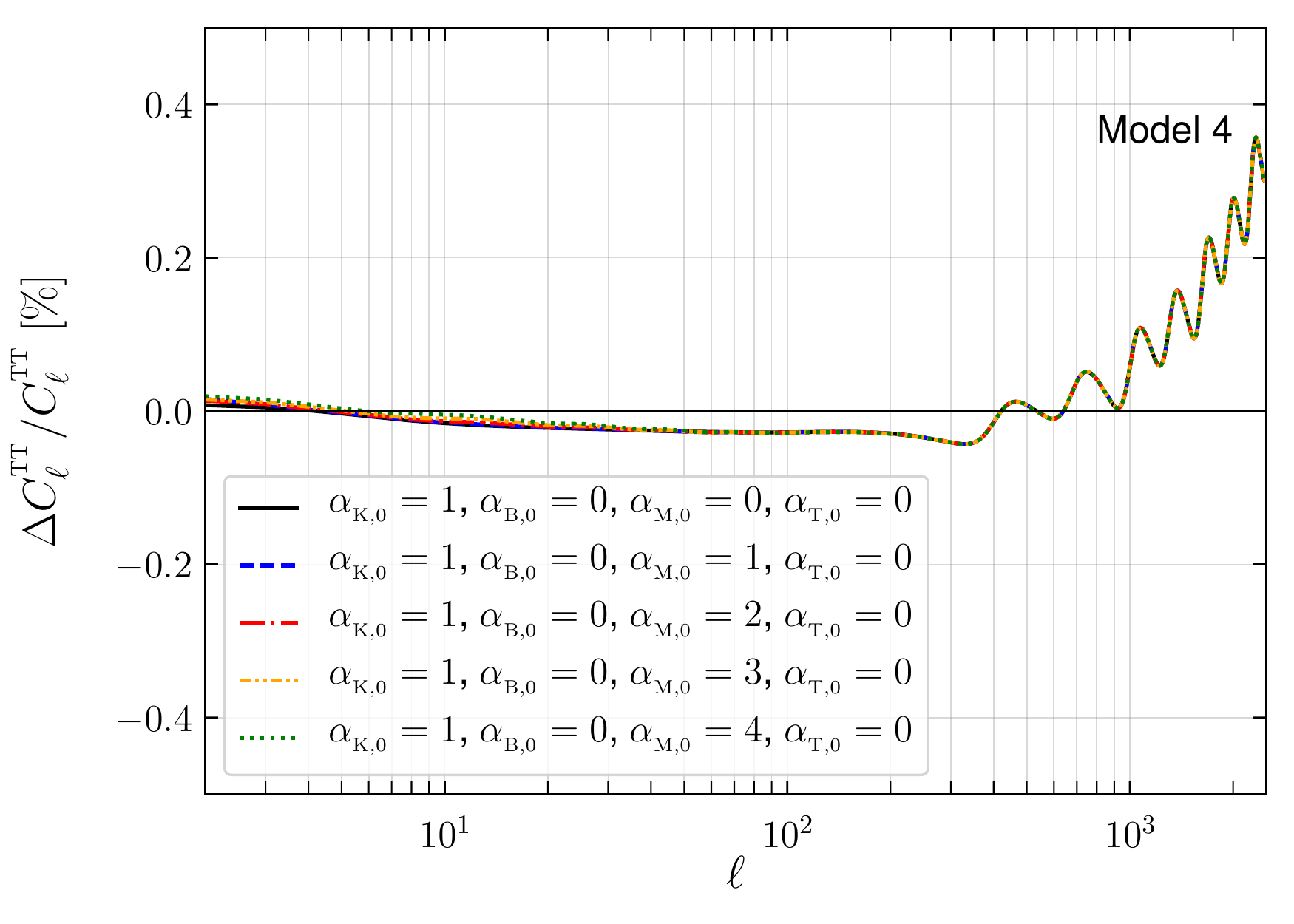}
 \includegraphics[width=6.5cm,angle=0]{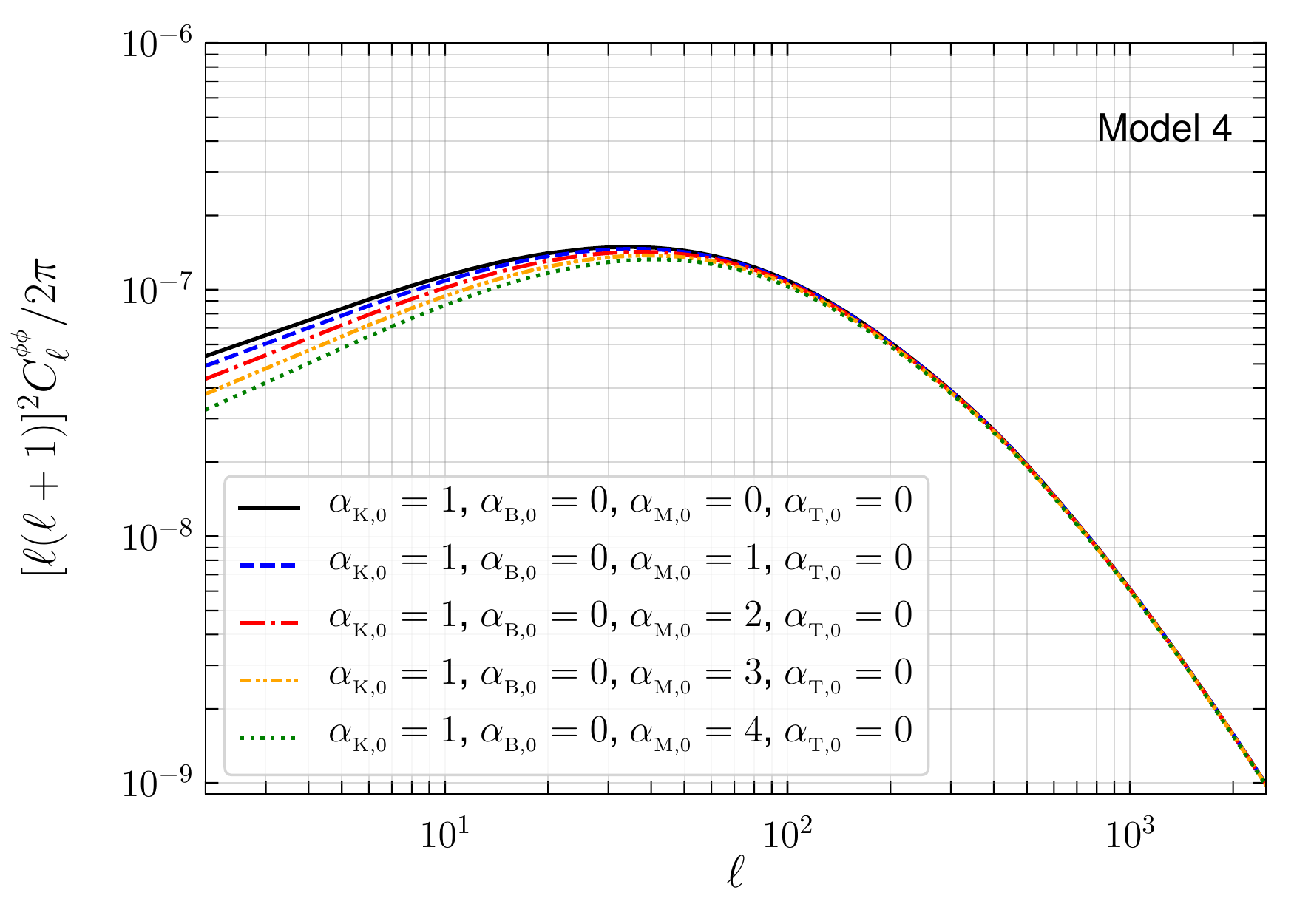}
 \includegraphics[width=6.5cm,angle=0]{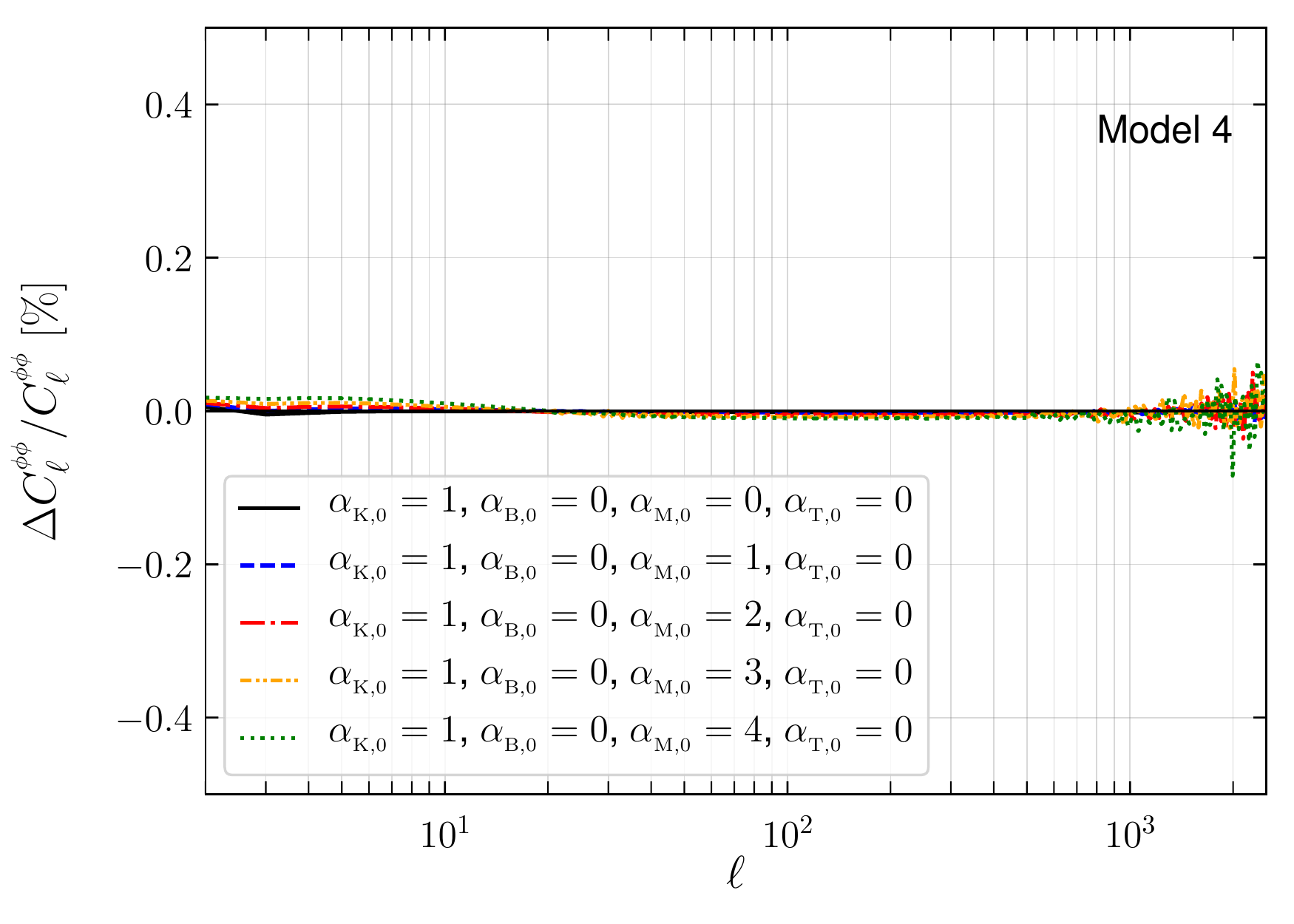}
 \includegraphics[width=6.5cm,angle=0]{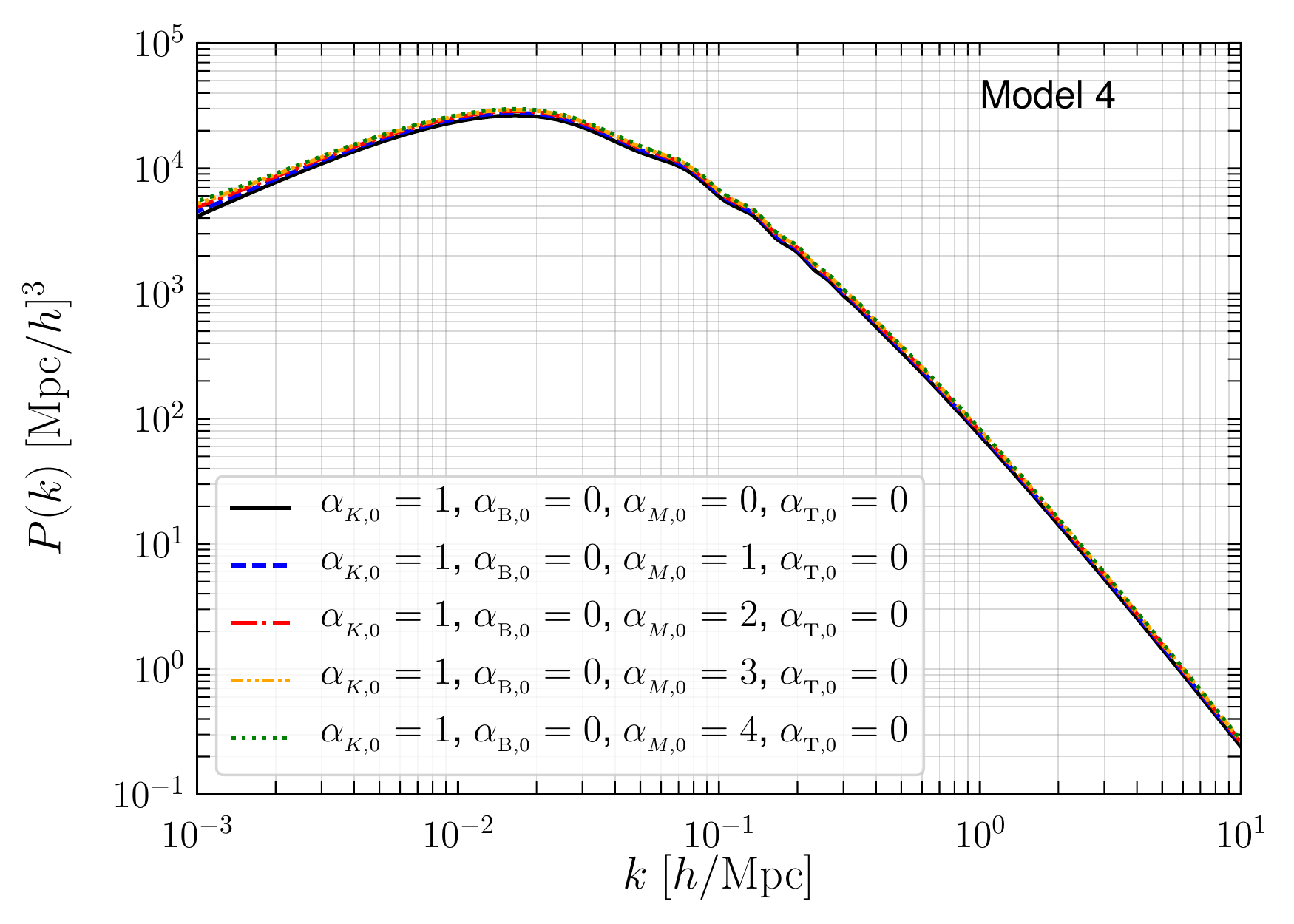}
 \includegraphics[width=6.5cm,angle=0]{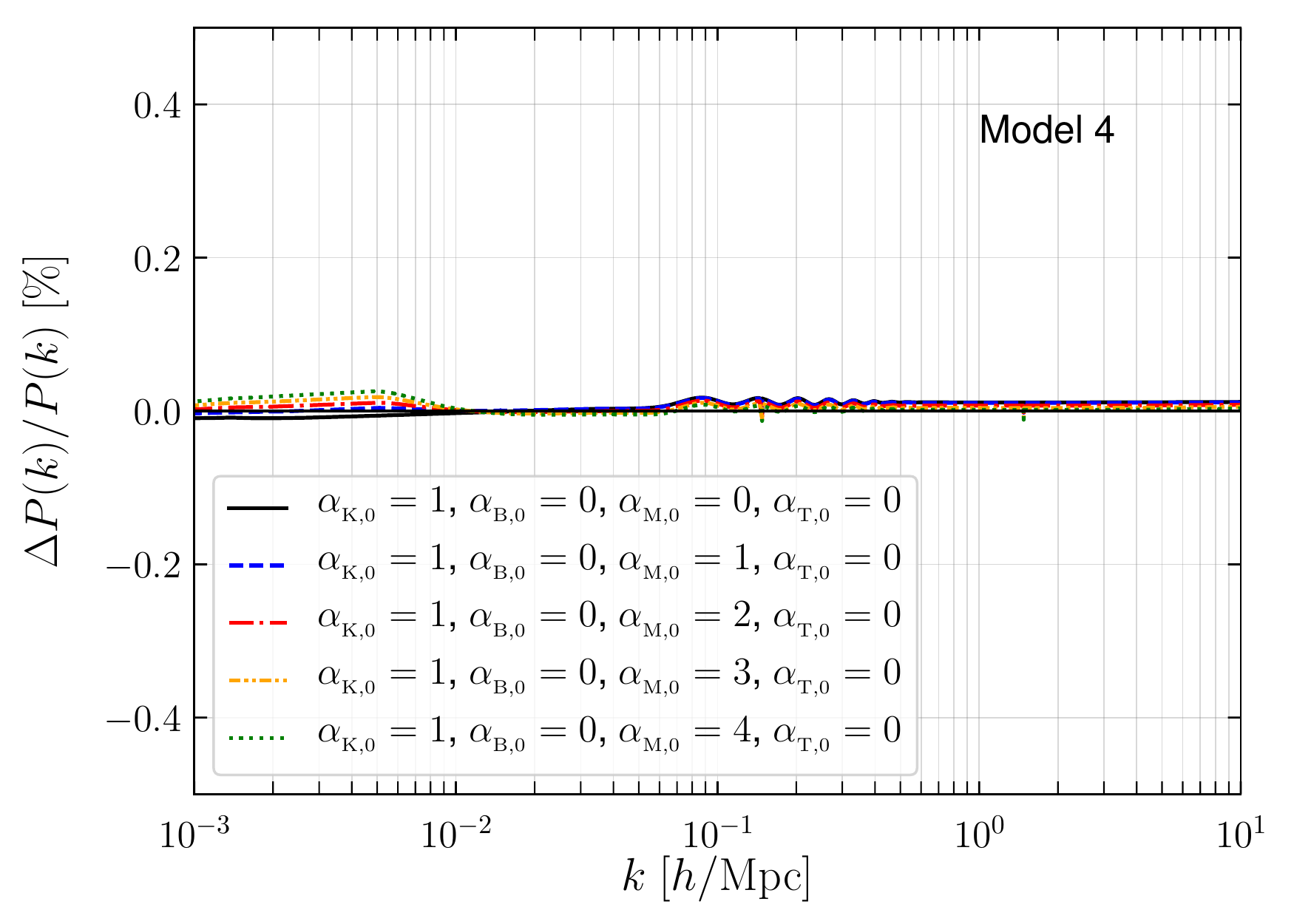}
 \cprotect\caption[justified]{Comparison for the spectra for model 4. In the left hand panels we present the spectra 
 obtained with \verb|EoS_class| and on the right the relative difference with the corresponding spectra obtained with 
 \verb|hi_class|. Top panels show the angular temperature anisotropy power spectrum, middle panels the angular power 
 spectrum of the lensing potential, while bottom panels present the linear matter power spectrum. 
 The black solid line represents the model with $\alpha_{\rm M,0}=0$, the dashed blue (red dot-dashed) stands for 
 $\alpha_{\rm M,0}=1$ ($\alpha_{\rm M,0}=2$), the orange dot-dotted-dashed (dotted green) curve is for 
 $\alpha_{\rm M,0}=3$ ($\alpha_{\rm M,0}=4$), respectively. For all the models, $\alpha_{\rm K,0}=1$.}
 \label{fig:aKaM}
\end{figure*}

\begin{figure*}[!ht]
 \centering
 \includegraphics[width=6.5cm,angle=0]{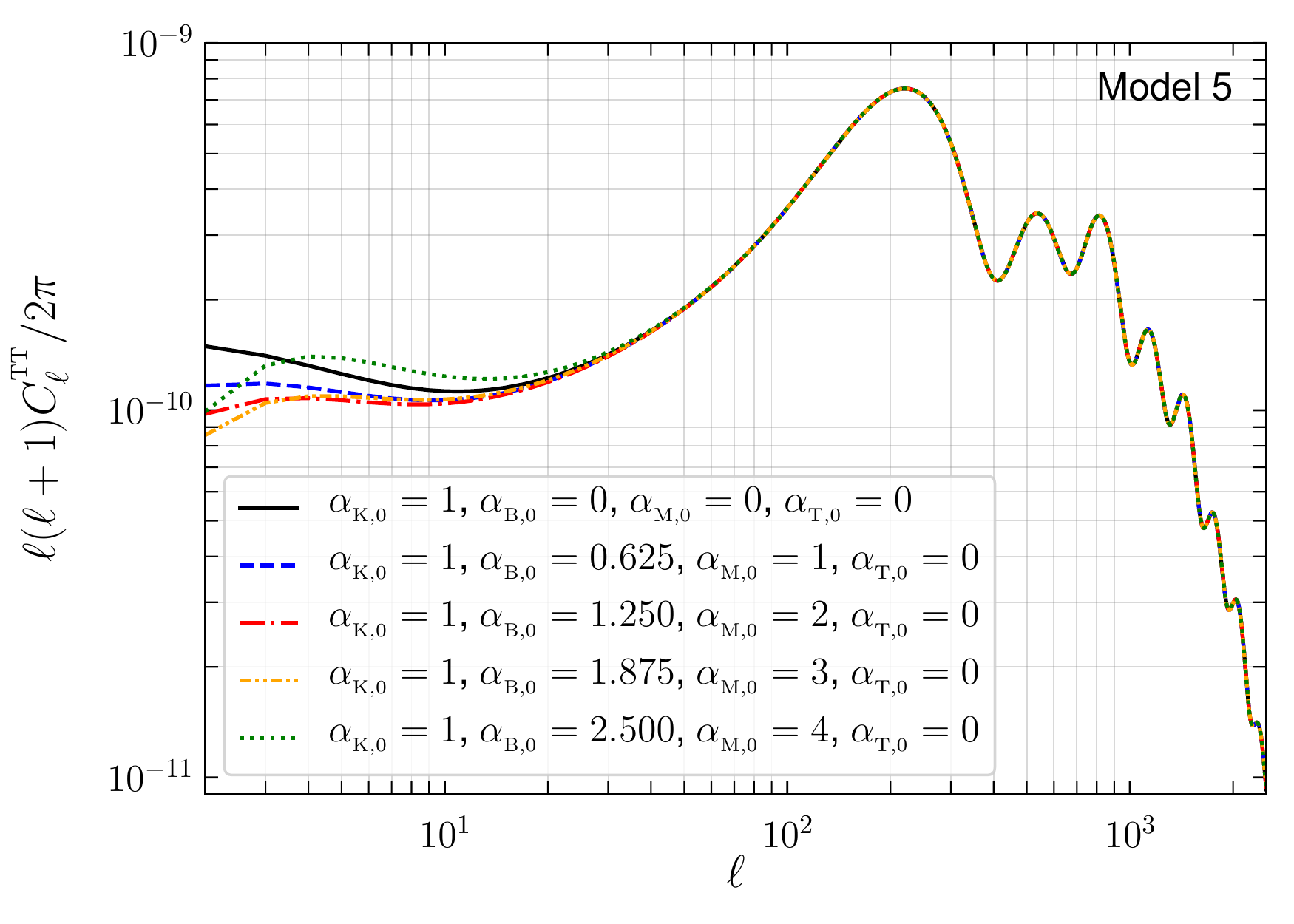}
 \includegraphics[width=6.5cm,angle=0]{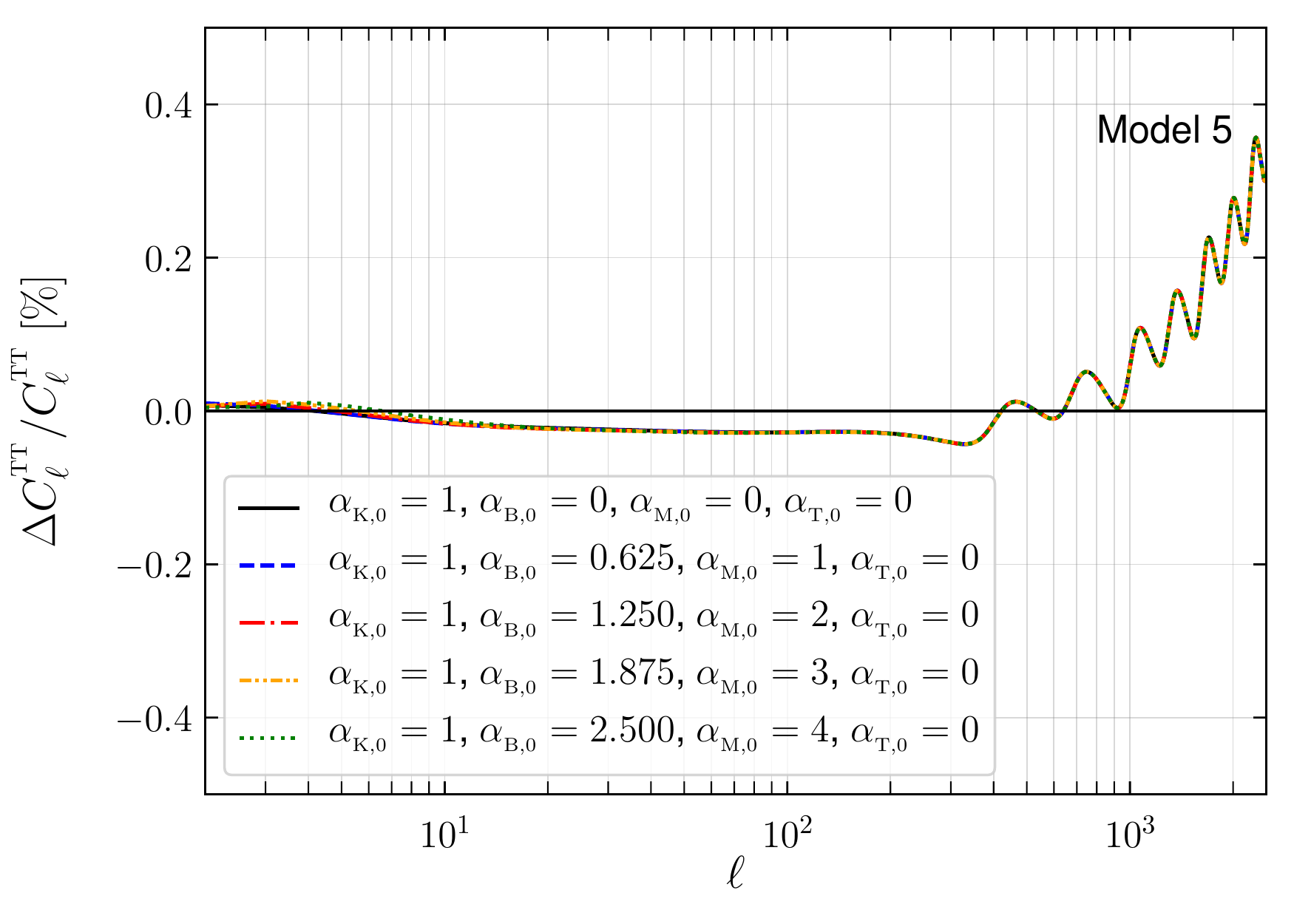}
 \includegraphics[width=6.5cm,angle=0]{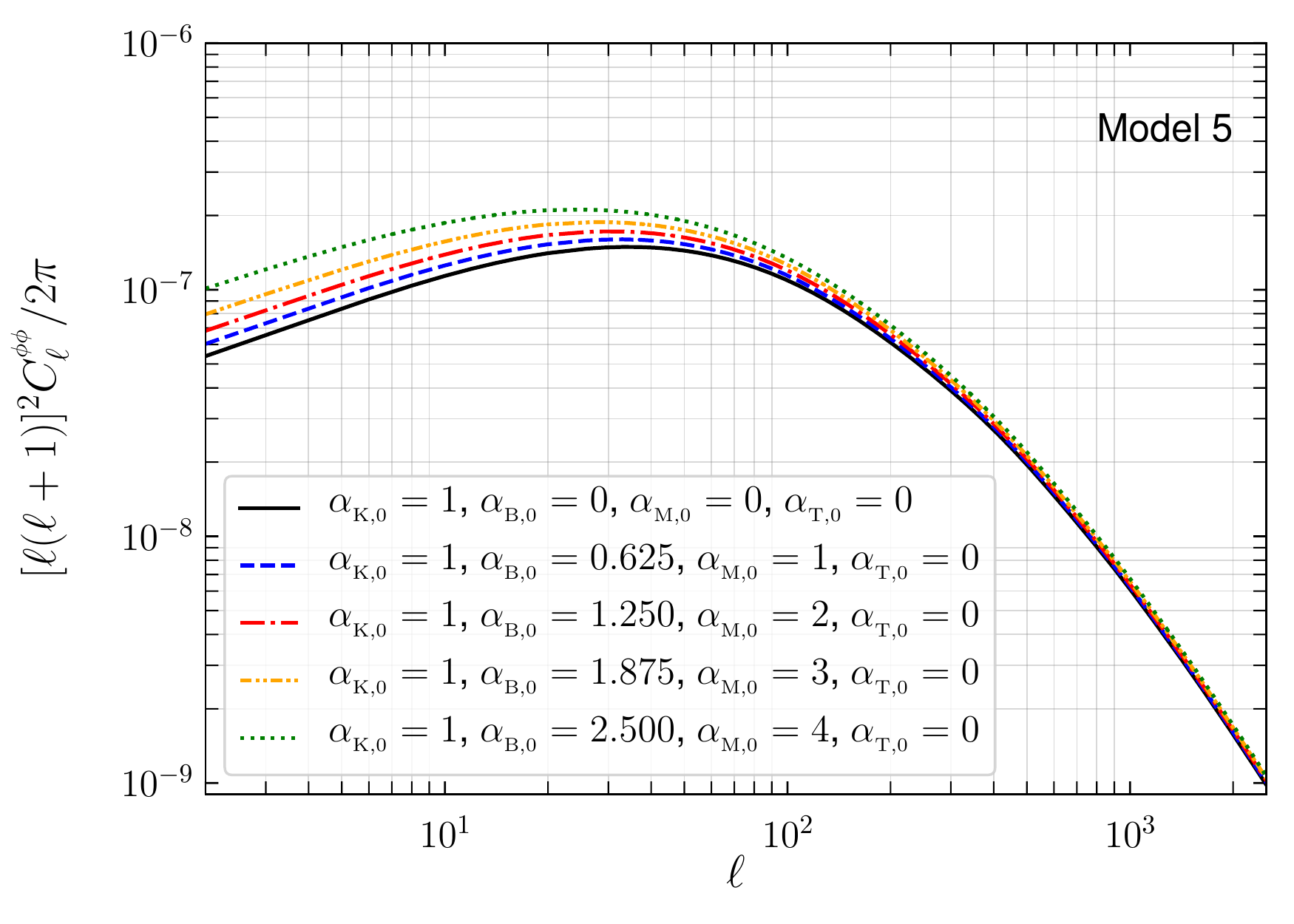}
 \includegraphics[width=6.5cm,angle=0]{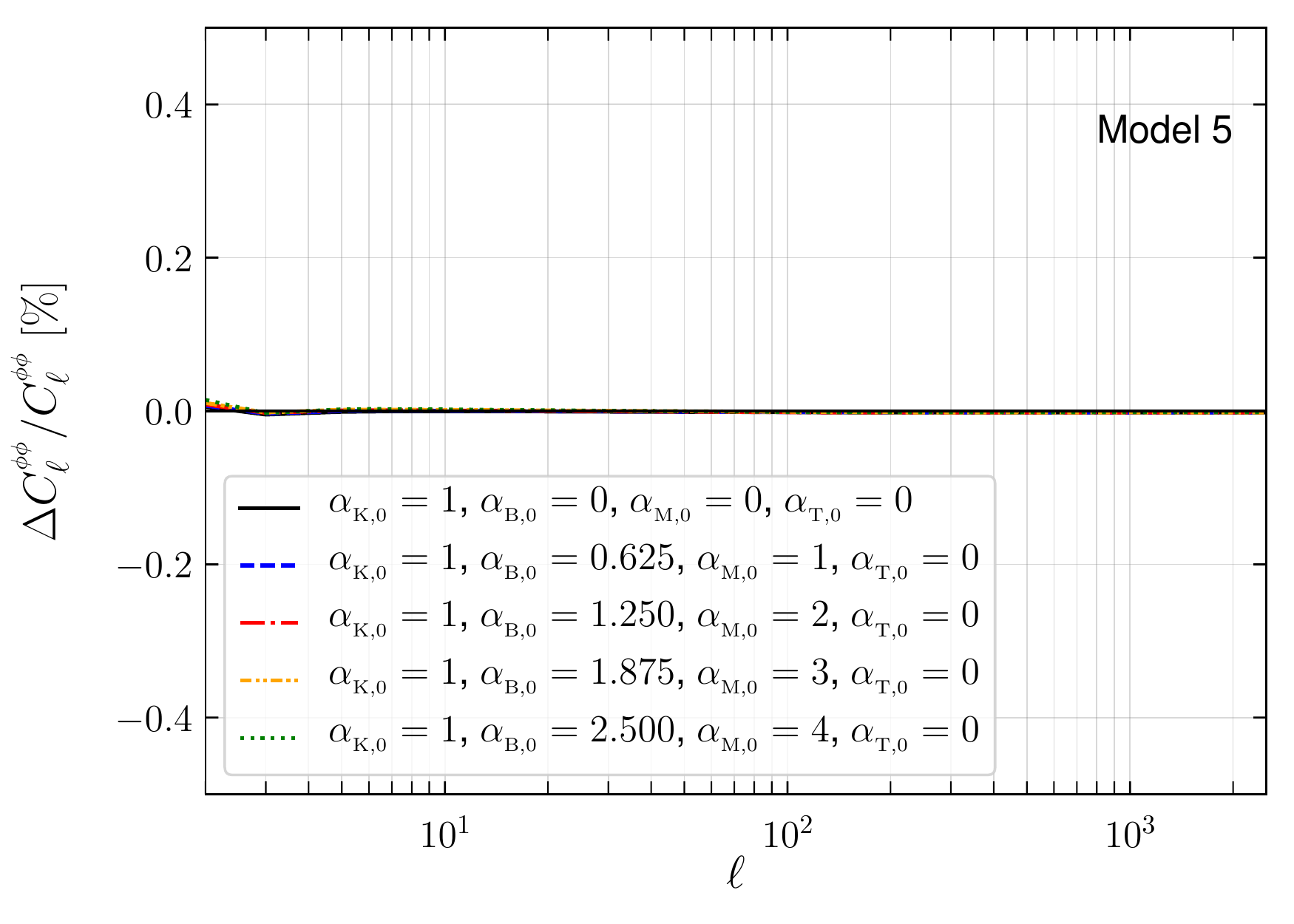}
 \includegraphics[width=6.5cm,angle=0]{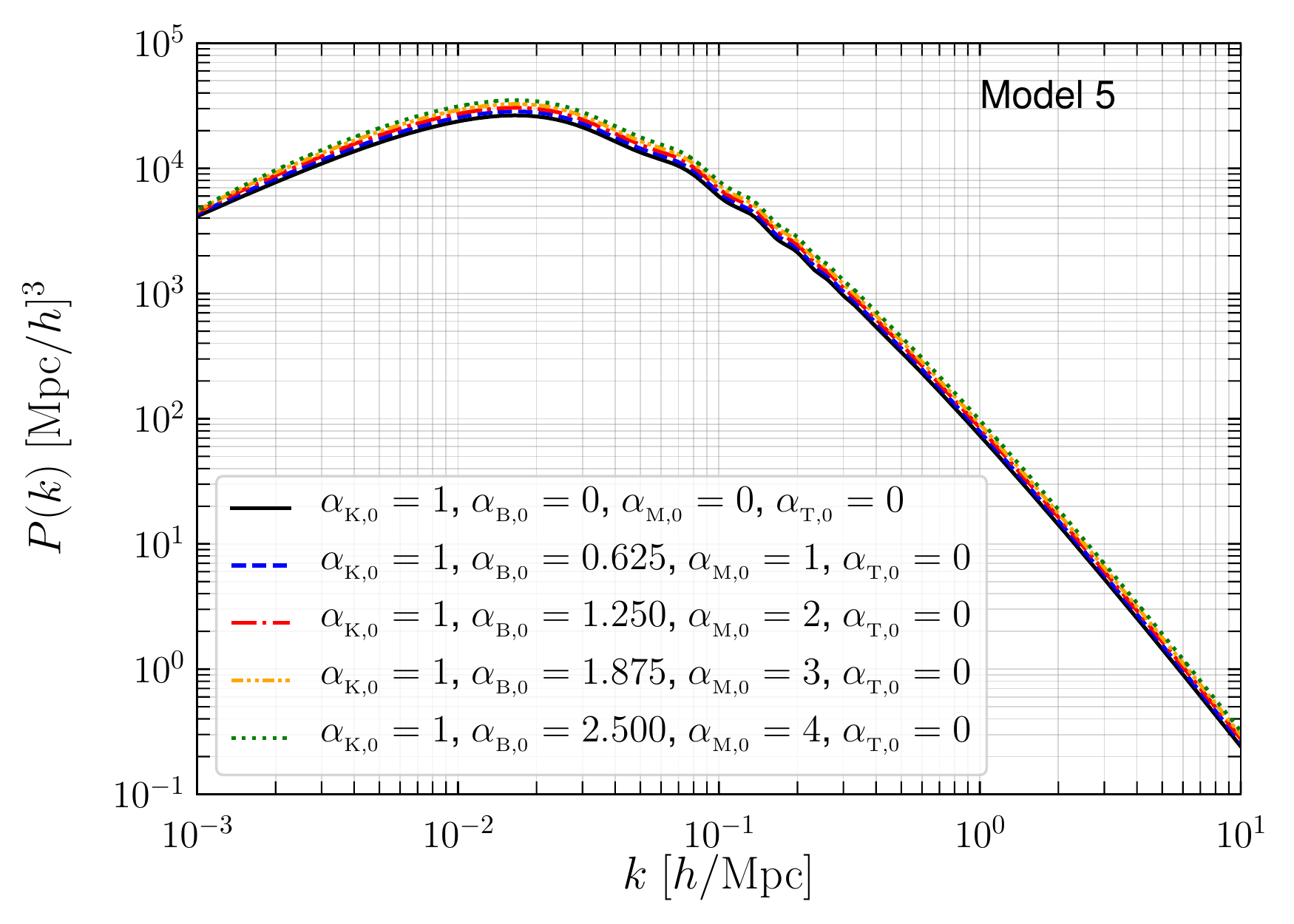}
 \includegraphics[width=6.5cm,angle=0]{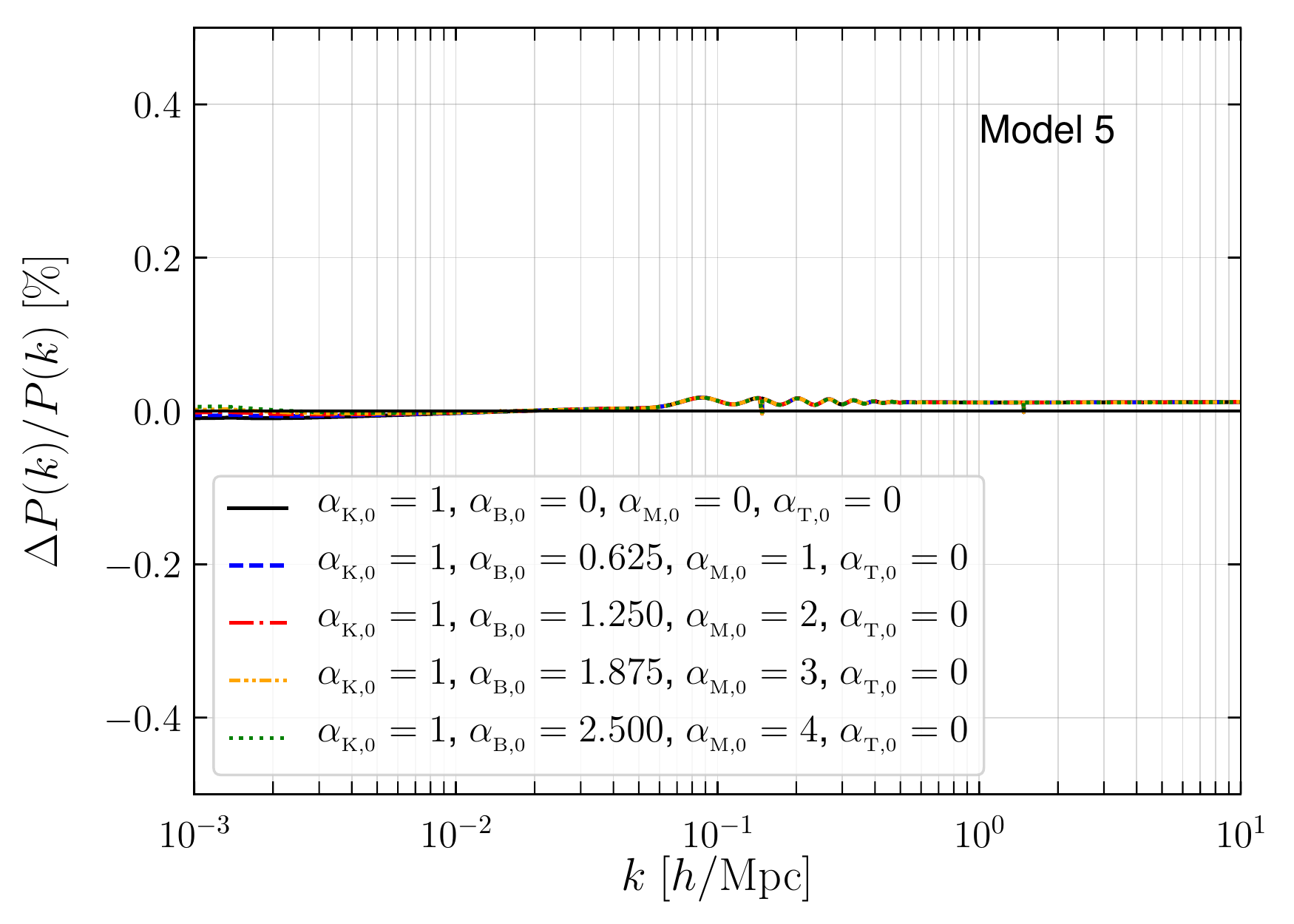}
 \cprotect\caption[justified]{Comparison for the spectra for model 5. In the left hand panels we present the spectra 
 obtained with \verb|EoS_class| and on the right the relative difference with the corresponding spectra obtained 
 with \verb|hi_class|. Top panels show the angular temperature anisotropy power spectrum, middle panels the angular 
 power spectrum of the lensing potential, while bottom panels present the linear matter power spectrum. The black 
 solid line represents the model with $\alpha_{\rm B,0}=\alpha_{\rm M,0}=0$, the dashed blue (red dot-dashed) stands 
 for $\alpha_{\rm B,0}=0.625$ and $\alpha_{\rm M,0}=1$ ($\alpha_{\rm B,0}=1.25$ and $\alpha_{\rm M,0}=2$), the orange 
 dot-dotted-dashed (dotted green) curve is for $\alpha_{\rm B,0}=1.875$ and $\alpha_{\rm M,0}=3$ 
 ($\alpha_{\rm B,0}=2.5$ and $\alpha_{\rm M,0}=4$), respectively. For all the models, $\alpha_{\rm K,0}=1$.}
 \label{fig:aKaBaM}
\end{figure*}

\begin{figure*}[!ht]
 \centering
 \includegraphics[width=6.5cm,angle=0]{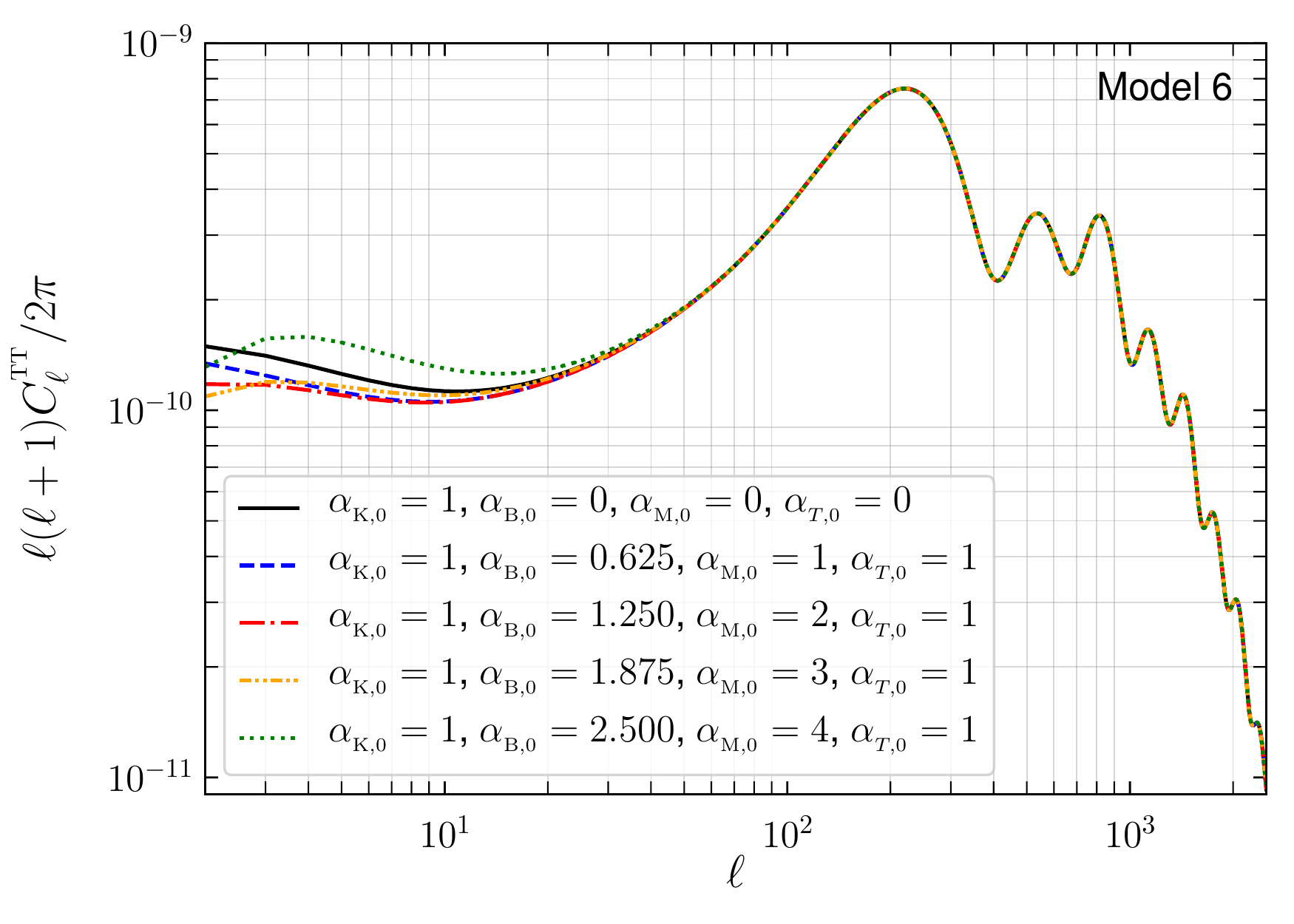}
 \includegraphics[width=6.5cm,angle=0]{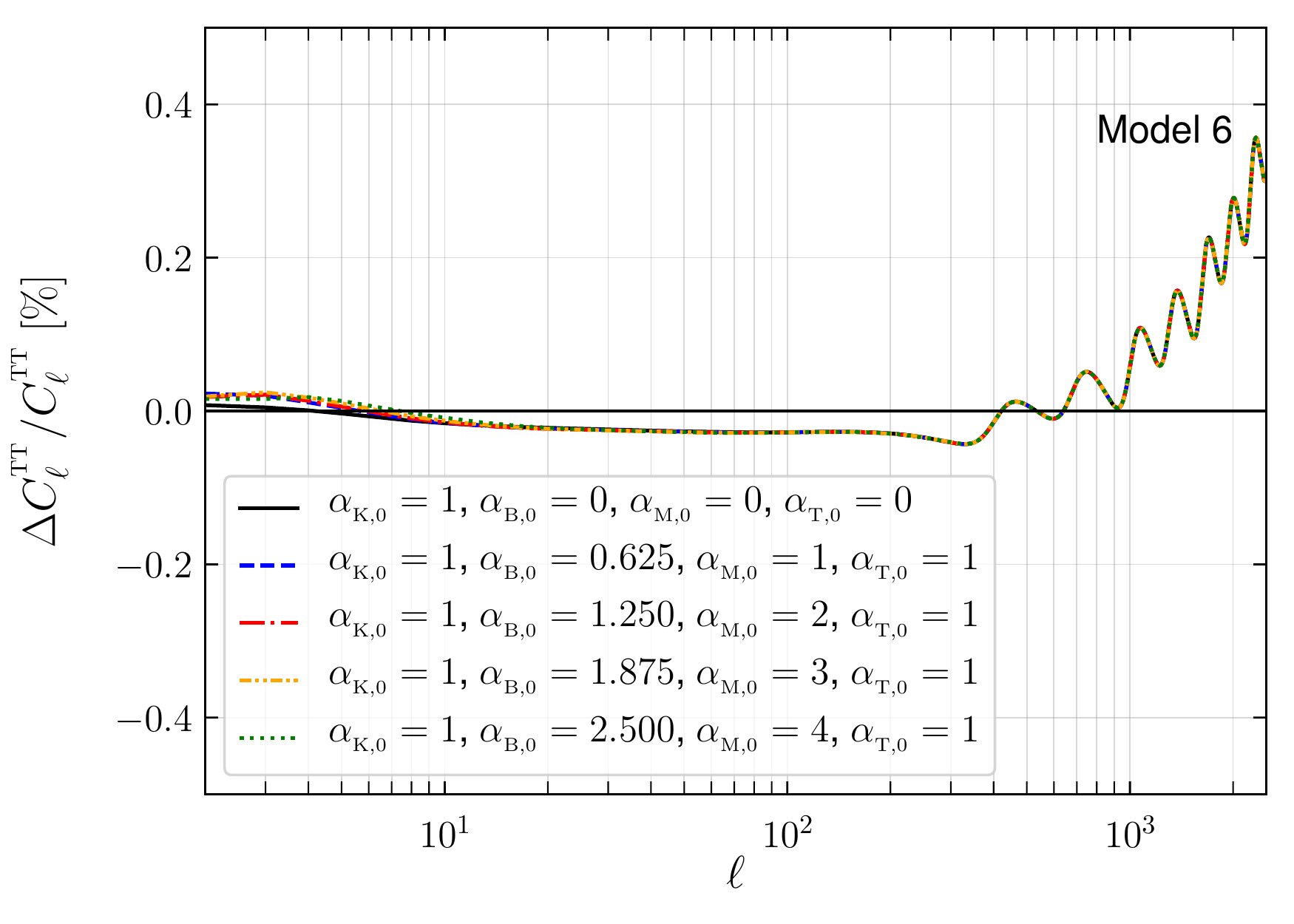}
 \includegraphics[width=6.5cm,angle=0]{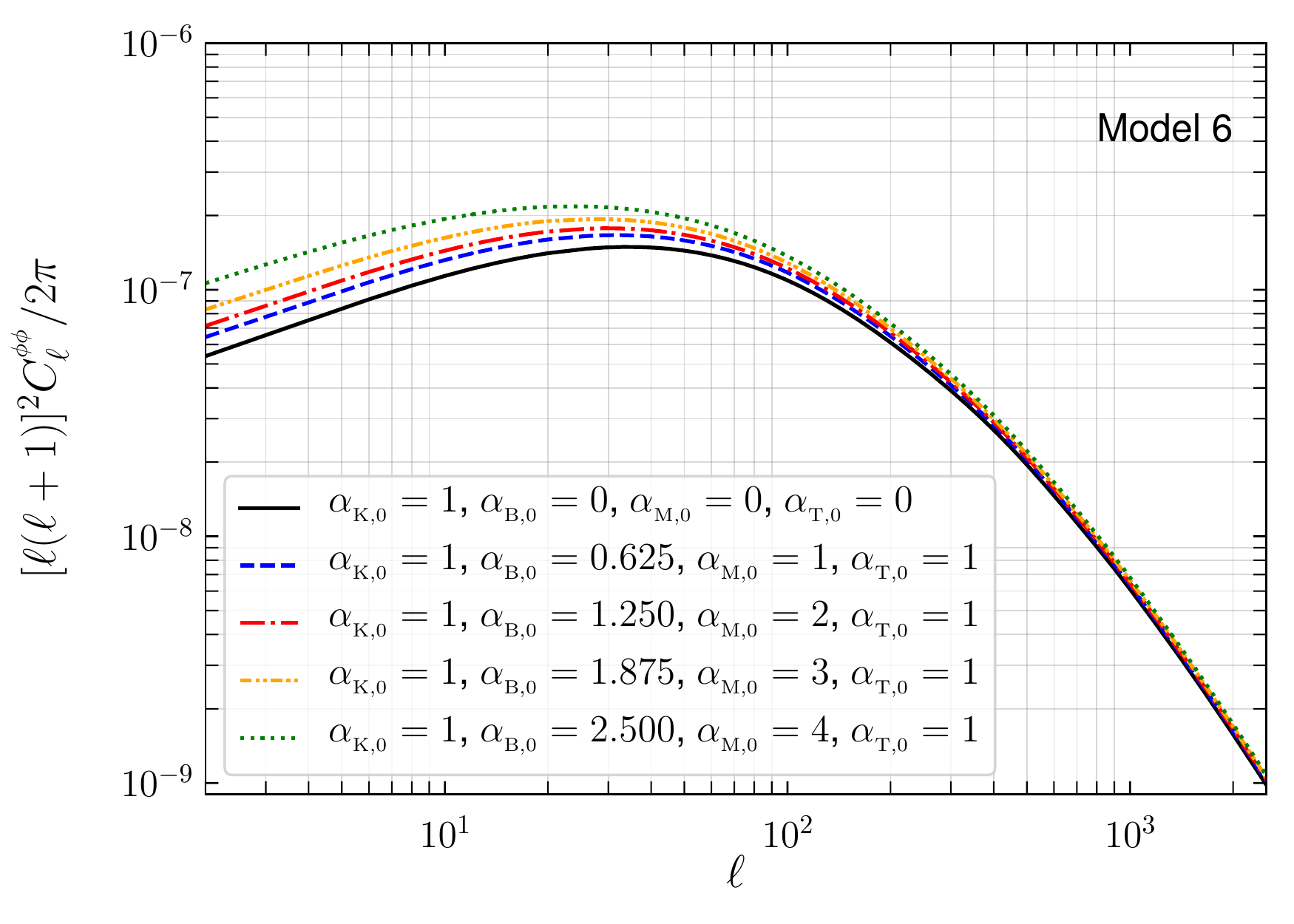}
 \includegraphics[width=6.5cm,angle=0]{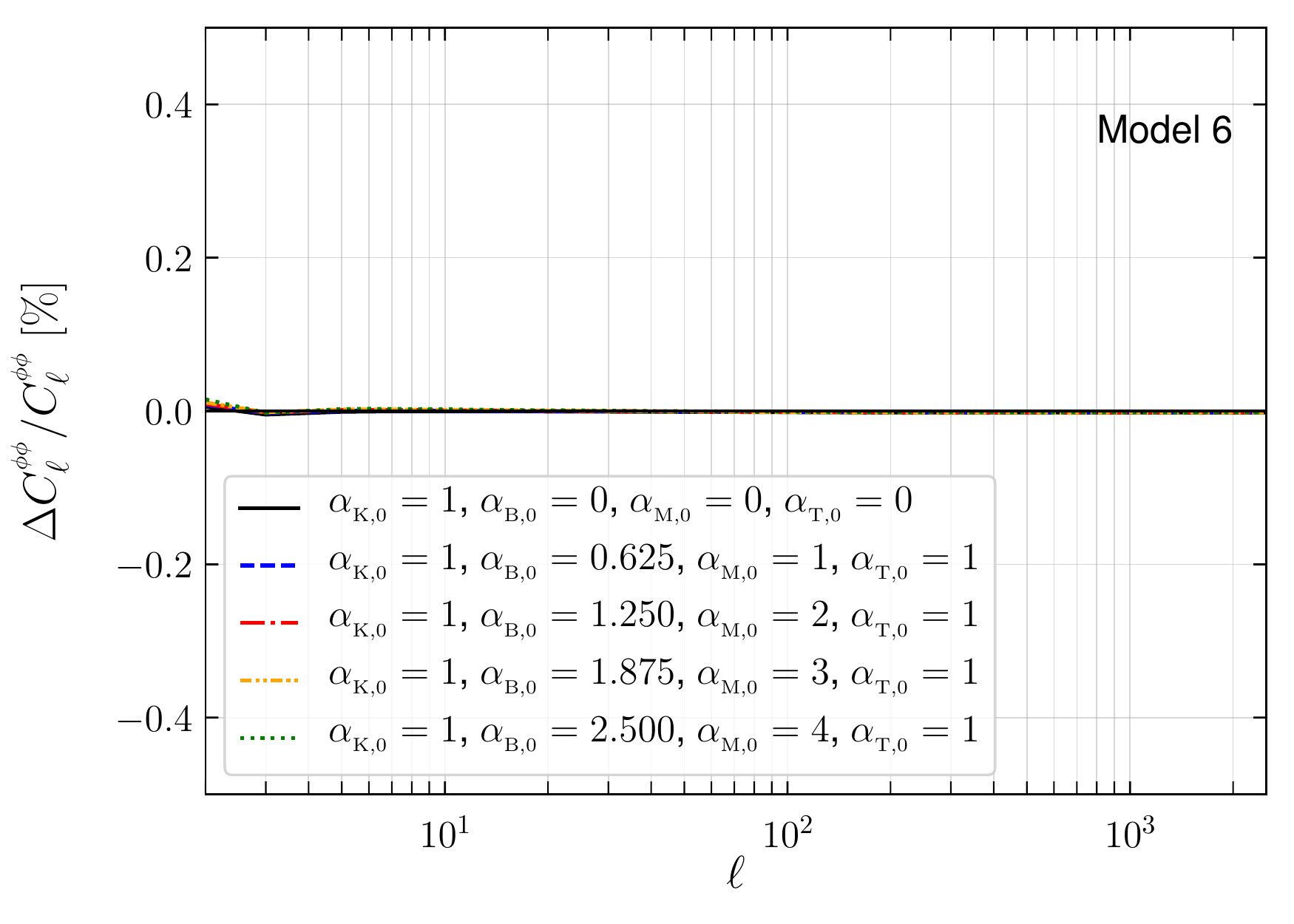}
 \includegraphics[width=6.5cm,angle=0]{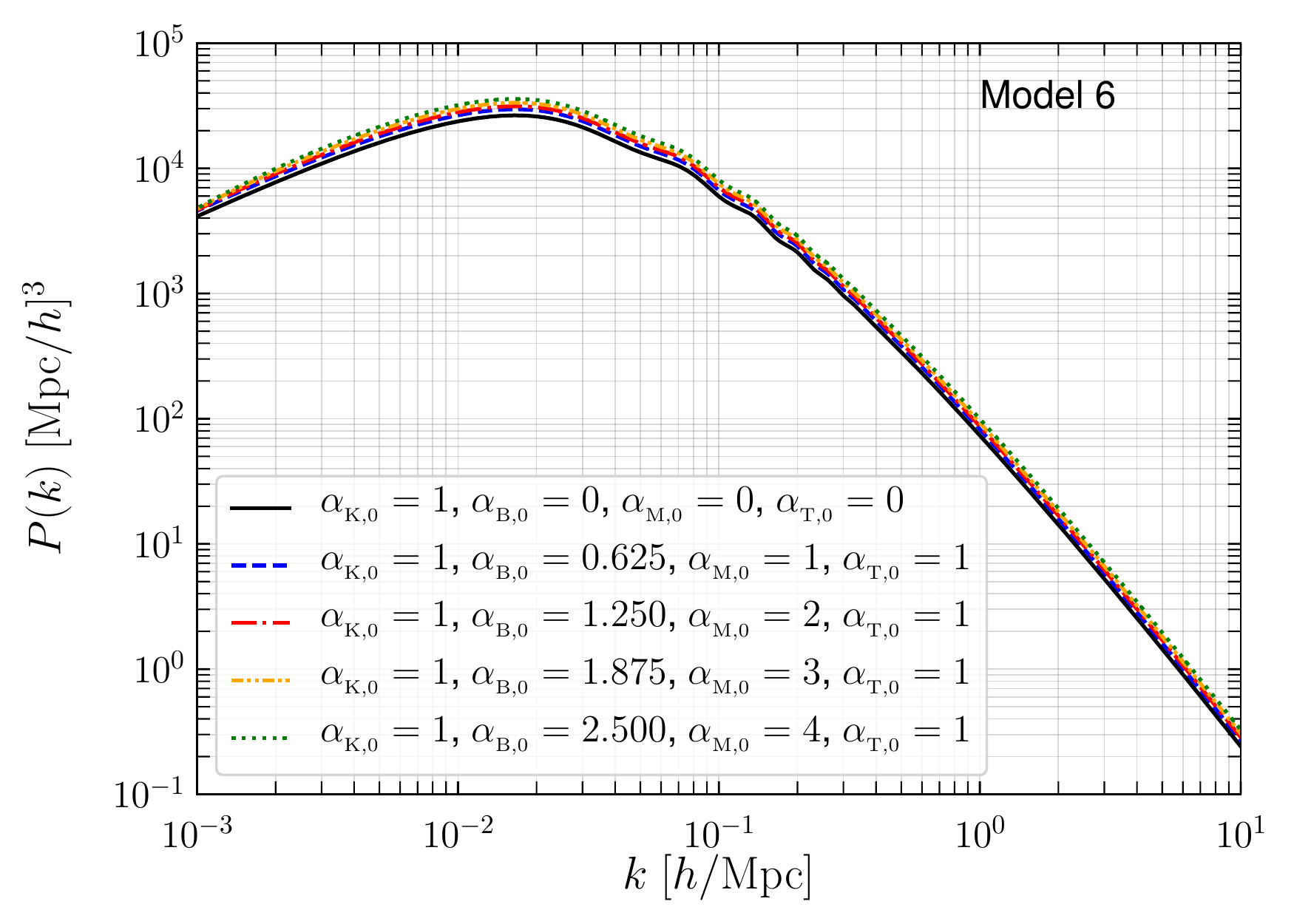}
 \includegraphics[width=6.5cm,angle=0]{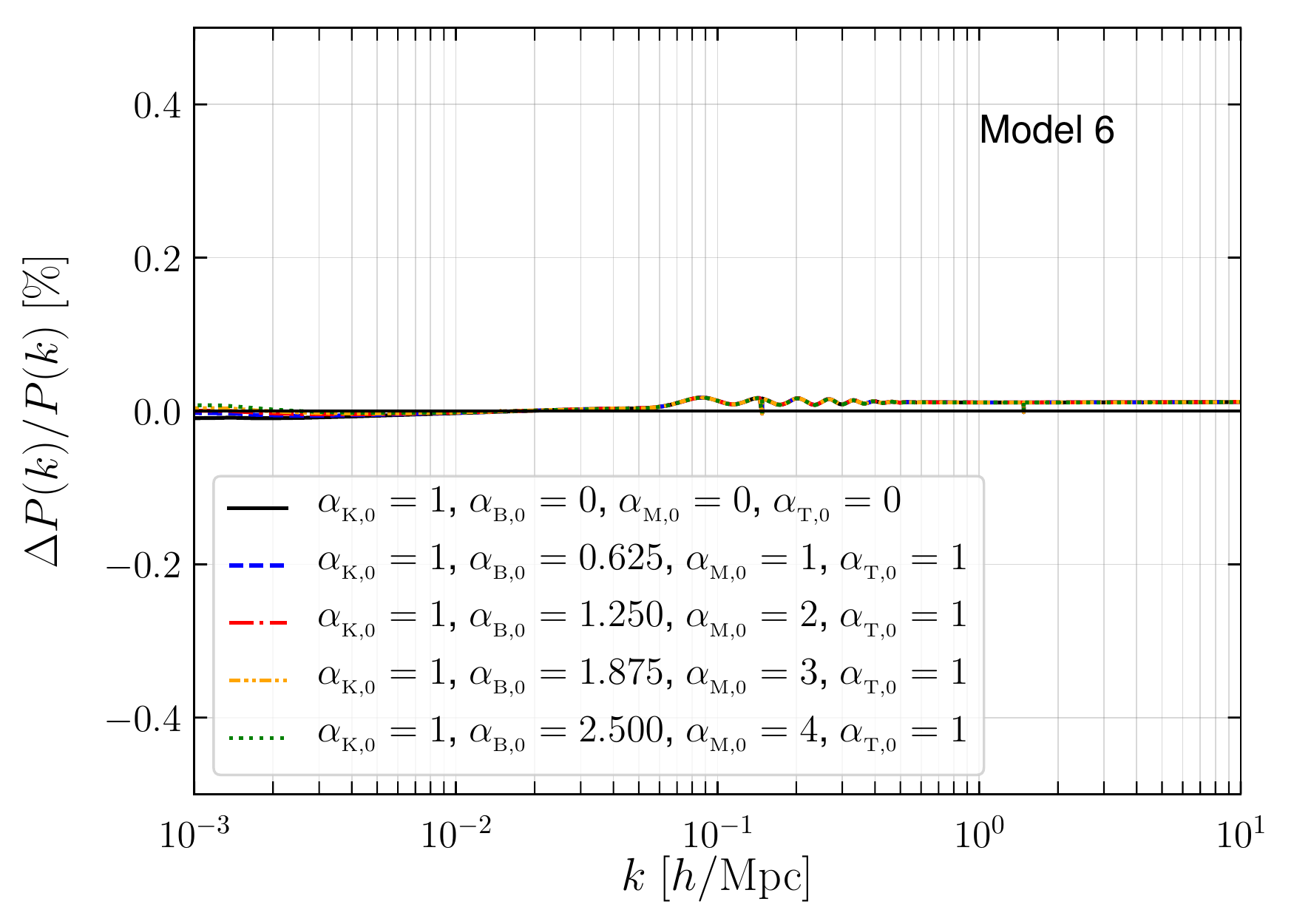}
 \cprotect\caption[justified]{Comparison for the spectra for model 6. In the left hand panels we present the spectra 
 obtained with \verb|EoS_class| and on the right the relative difference with the corresponding spectra obtained with 
 \verb|hi_class|. Top panels show the angular temperature anisotropy power spectrum, middle panels the angular power 
 spectrum of the lensing potential, while bottom panels present the linear matter power spectrum. 
 The black solid line represents the model with $\alpha_{\rm B,0}=\alpha_{\rm M,0}=0$, the dashed blue 
 (red dot-dashed) stands for $\alpha_{\rm B,0}=0.625$ and $\alpha_{\rm M,0}=1$ ($\alpha_{\rm B,0}=1.25$ and 
 $\alpha_{\rm M,0}=2$), the orange dot-dotted-dashed (dotted green) curve is for $\alpha_{\rm B,0}=1.875$ and 
 $\alpha_{\rm M,0}=3$ ($\alpha_{\rm B,0}=2.5$ and $\alpha_{\rm M,0}=4$), respectively. All the models have 
 $\alpha_{\rm T,0}=1$. For all the models, $\alpha_{\rm K,0}=1$.}
 \label{fig:aKaBaMaT}
\end{figure*}

In Figs.~\ref{fig:aKaB}--\ref{fig:aKaBaMaT} we present the comparison for models 3--6, respectively. Model 1, that 
corresponds to setting $\alpha_{\rm B}=\alpha_{\rm M}=\alpha_{\rm T}=0$ and $\alpha_{\rm K}\neq 0$, is shown as the 
black lines on Figs.~\ref{fig:aKaB}--\ref{fig:aKaBaMaT}. 
In the right hand panels we show the relative difference $\Delta C/C$, with 
$C=\{C_{\ell}^{\rm TT}, C_{\ell}^{\phi\phi}, P(k)\}$ and 
$\Delta C\equiv C^{\verb|EoS_class|}-C^{\verb|hi_class|}$ and in the denominator we use $C=C^{\verb|EoS_class|}$. 
In each figure we have chosen a set of five specific values for the $\alpha_{\rm x}$s, the same values that were used 
for the \verb|hi_class| paper in \cite{Zumalacarregui2017}.

For all the models we considered, we achieve sub-percent agreement with the \verb|hi_class| results. Since 
\verb|hi_class| does not work for model 2 ($f(R)$-like models), we could not compare our results with \verb|hi_class| 
for this class of models. Instead, in Fig.~\ref{fig:aBaM}, we compare model 2 with model 5 with the same parameters, 
except $\alpha_{\rm K,0}=1$. Indeed, the only difference between both classes of models is that in model 5 
$\alpha_{\rm K}\neq 0$. In this figure, $\Delta C = C^{\rm Model~2}-C^{\rm Model~5}$ and in the denominator 
$C=C^{\rm Model~2}$. From this figure, we see that $\alpha_{\rm K}$ plays a role only for the largest cosmological 
scales with $k\lesssim 10^{-2}~h/{\rm Mpc}$, or $\ell\lesssim 10$. This result has also been found in 
\cite{Frusciante2019}, where the authors showed that $\alpha_{\rm K}$ only affects low multipoles in the ISW tail and 
that the effect is not measurable since it is smaller than cosmic variance.

While we get sub-percent agreement with \verb|hi_class| ($<0.1\%$ in almost all the cases), we note that the largest 
differences between \verb|hi_class| and \verb|EoS_class| arise for the temperature anisotropy power spectrum at 
$\ell\gtrsim 1000$. 
In fact we identified the source of this difference to be associated with the different versions of the \verb|CLASS| 
code: \verb|hi_class| is based on \verb|CLASS| version 2.4.5, while for \verb|EoS_class| we used a more recent version 
(2.6.3)\footnote{We thank Emilio Bellini for pointing this out to us.}. Hence, we conclude that our implementation 
reproduces the results of \verb|hi_class| for models 1 and 3--6 to high precision and appears to also be working for 
model 2.

\section{Dark sector evolution in Horndeski theories: analytical results}\label{sect:ds}
In this section we carry out an analytical analysis of Horndeski theories. 
In subsection~\ref{sect:attractor} we present our analytical approximation for the attractor solution of the dark 
sector fluid variables. In subsection~\ref{sect:MGparams} we present new expressions for the standard modified gravity 
parameters for cosmological perturbations: $\mu$ (or $G_{\rm eff}$), $\eta$ and $\Sigma$. 
Finally, in subsection~\ref{sect:simplified} we present simplified forms of the EoS for perturbations in Horndeski 
models.

\subsection{Analytical approximations for the attractor solution}\label{sect:attractor}
The existence of an attractor solution for the dark sector variables in modified gravity or dark energy models has been 
recognised and used in several previous analysis, for instance \cite{Ballesteros2010,Hu2014b,Battye2018a}. Here we 
derive an analytical approximation for the attractor solution of Horndeski theories. Our derivation relies on two 
single assumptions: that the mode as comoving wavelength is well inside the Hubble horizon, ${\rm K}^2\gg1$, and 
that time derivatives can be neglected.

This assumption on the modes wavenumber determines the range of validity of our approximation in terms of scale. In 
fact, as can be seen in the top panel of Fig.~2 of \cite{Battye2018a}, the condition ${\rm K}^2\gg 1$ translates into 
$k\gg 10^{-4}~{\rm Mpc}^{-1}$ today or $k\gg 10^{-3}~{\rm Mpc}^{-1}$ at $z\approx 100$. This includes the range of 
wavenumbers of observational interest, so this condition is not restrictive for our purposes.

This assumption also naturally implies that the gauge-invariant velocity perturbation $\hat{\Theta}$ is small compared 
to the gauge invariant density perturbation $\Delta$ in both the matter and dark sector. To see why this is the case, 
we write Einstein field equations in terms of the gauge-invariant quantities introduced in \citep{Battye2016a}
\begin{subequations}\label{eqn:EFE}
 \begin{align}
  -\frac{2}{3}{\rm K}^2Z = &\, \sum_i\Omega_i\Delta_i\,, \label{eqn:00} \\
   2X = &\, \sum_i\Omega_i\hat{\Theta}_i\,, \label{eqn:0i}\\
   \frac{1}{3}{\rm K}^2(Y-Z) = &\, \sum_i\Omega_iw_i\Pi_i\,, \label{eqn:ij}
 \end{align}
\end{subequations}
where $Z$, $Y$ and $X$ on the left hand side are linear combinations of the metric perturbations and their derivatives 
with respect to $\ln{a}$, and the sum over $i$ on the right hand side means matter plus dark sector quantities. 
The different metric perturbations and their derivatives are generally all of the same order. Thus since 
${\rm K}^2\gg 1$, we understand from Eqs.~(\ref{eqn:EFE}) that the gauge-invariant velocity perturbations are 
automatically smaller than the gauge-invariant density perturbations by a factor $\sim 1/{\rm K}^2$.

When we neglect velocity perturbations and take the derivative of Eqn.~(\ref{eqn:Delta}) we obtain
\begin{equation}\label{eqn:gf_ds}
 \Delta_{\rm ds}^{\prime\prime} + \left(2+3c_{\rm a,ds}^2-6w_{\rm ds}-\epsilon_H-2C_{\Pi\Delta_{\rm ds}}\right)
  \Delta_{\rm ds}^{\prime} +\left(c_{\rm a,ds}^2+C_{\zeta\Delta_{\rm ds}}\right){\rm K}^2 \Delta_{\rm ds} 
  = -\frac{\Omega_{\rm m}}{\Omega_{\rm ds}}C_{\zeta\Delta_{\rm m}}{\rm K}^2\Delta_{\rm m}\,,
\end{equation}
where we replaced $\hat{\Theta}_{\rm ds}^{\prime}$ with Eqn.~(\ref{eqn:Theta}). Note that in this procedure, one has 
to initially keep terms proportional to ${\rm K}^2\hat{\Theta}_{\rm ds}$ since they are of the same order of 
magnitude as $\Delta_{\rm ds}$, when taking the derivative of Eqn.~(\ref{eqn:Delta}) and before additionally 
neglecting the contribution of velocity perturbations.

The differential equation for the density perturbation $\Delta_{\rm ds}$ (\ref{eqn:gf_ds}) is similar to a damped 
harmonic oscillator sourced by matter perturbations. The time dependent frequency is 
$\omega^2=\left(c_{\rm a,ds}^2+C_{\zeta\Delta_{\rm ds}}\right){\rm K}^2\gg1$ and is, in general, much smaller than the 
damping time scale represented by the Hubble expansion rate. This implies that the homogeneous solution becomes 
subdominant very quickly with respect to the particular solution, which, therefore, becomes the attractor of the 
evolution of the dark sector perturbations. In other words, similarly to what is normally done for the 
quasi-static approximation, we neglect the time derivatives of the density perturbation $\Delta_{\rm ds}$.

The analytical approximation for the attractor can be obtained by equating the last term of the left hand side 
with the term on the right hand side. This gives
\begin{equation}\label{eqn:dAIC}
 \Delta_{\rm ds} = -\left(\frac{C_{\zeta\Delta_{\rm m}}}{c_{\rm a,ds}^2+C_{\zeta\Delta_{\rm ds}}}\right)
                    \frac{\Omega_{\rm m}}{\Omega_{\rm ds}}\Delta_{\rm m}\;,
\end{equation}
where in accordance to our definition of $w_{\rm ds}\zeta_{\rm ds}$ in Eqn.~(\ref{eqn:zeta_ds}), 
$C_{\zeta\Delta_{i}}=\tfrac{2}{3}C_{\Pi\Delta_{i}}+C_{\Gamma\Delta_{i}}$, with $i=\{{\rm m}, {\rm ds}\}$. Note that 
here we used the dimensionless perturbed fluid variables, not the tilde quantities. This expression generalises the 
one found in our previous paper \citep{Battye2018b} to non-constant $w_{\rm ds}$.\footnote{Note that with respect to 
\cite{Battye2018b}, the additional $\Omega_{\rm m}/\Omega_{\rm ds}$ originates from a different definition of the 
coefficients in the entropy perturbation and anisotropic stress.}

\begin{figure*}[!ht]
 \centering
 \includegraphics[width=6.5cm,angle=0]{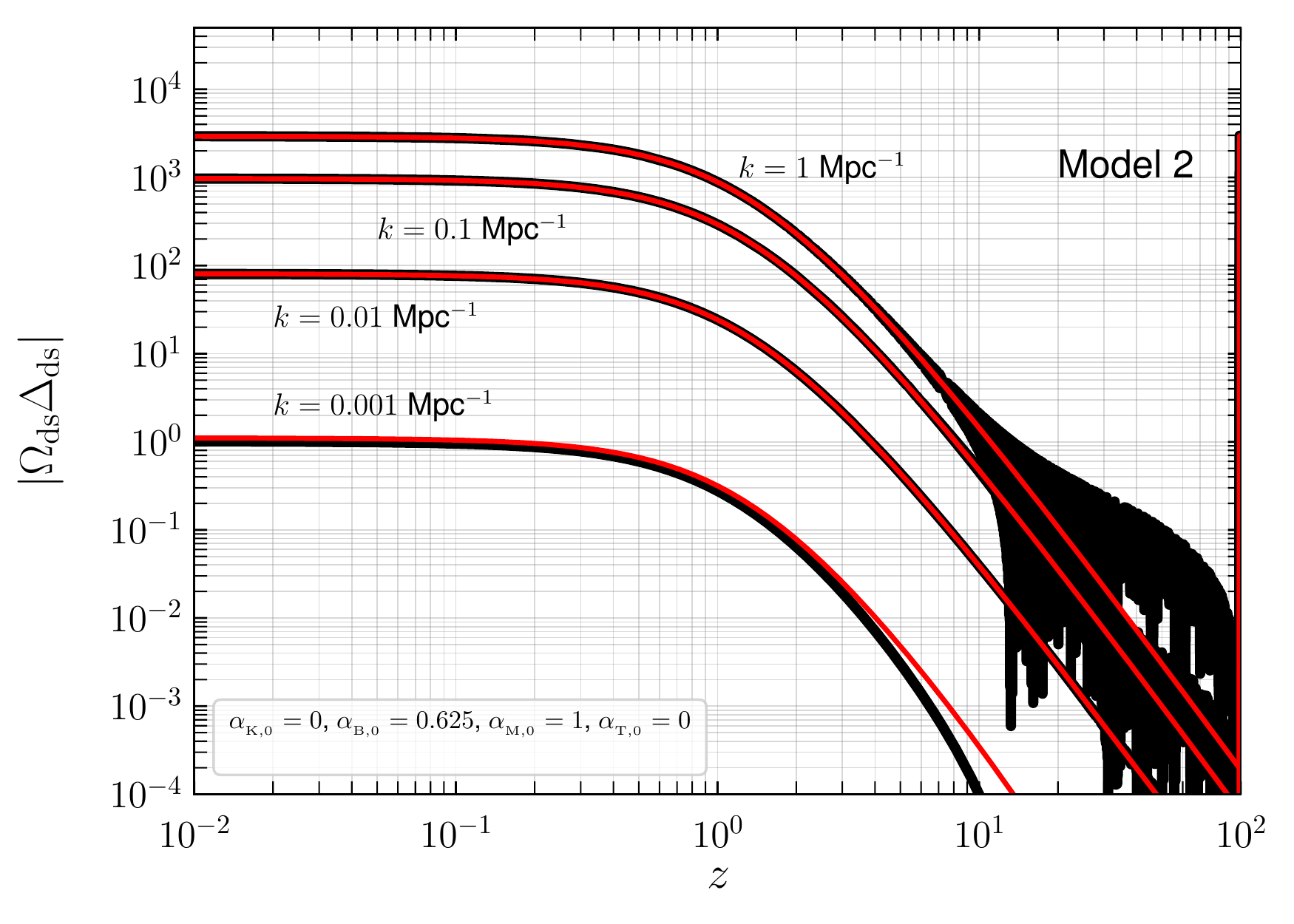}
 \includegraphics[width=6.5cm,angle=0]{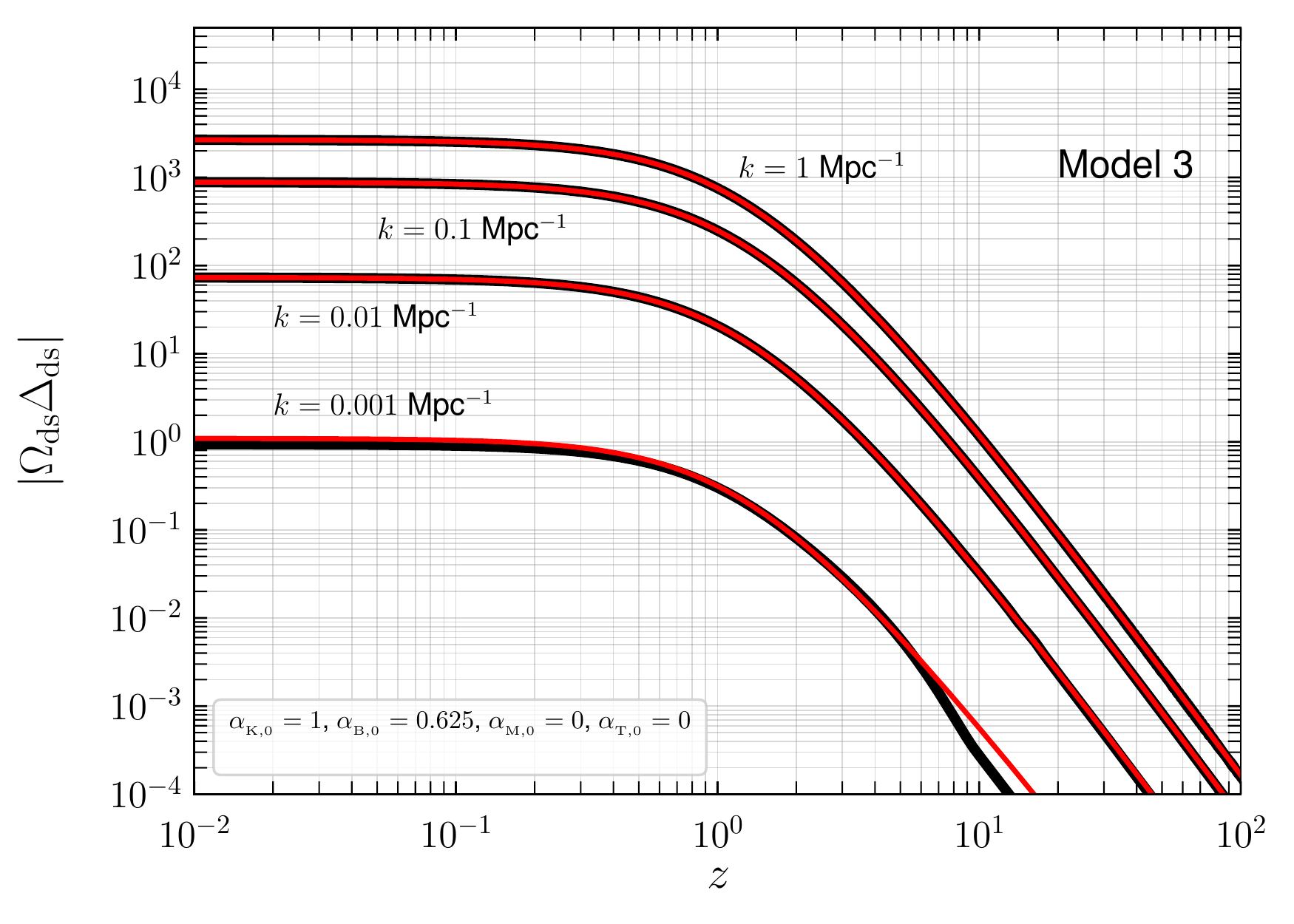}
 \includegraphics[width=6.5cm,angle=0]{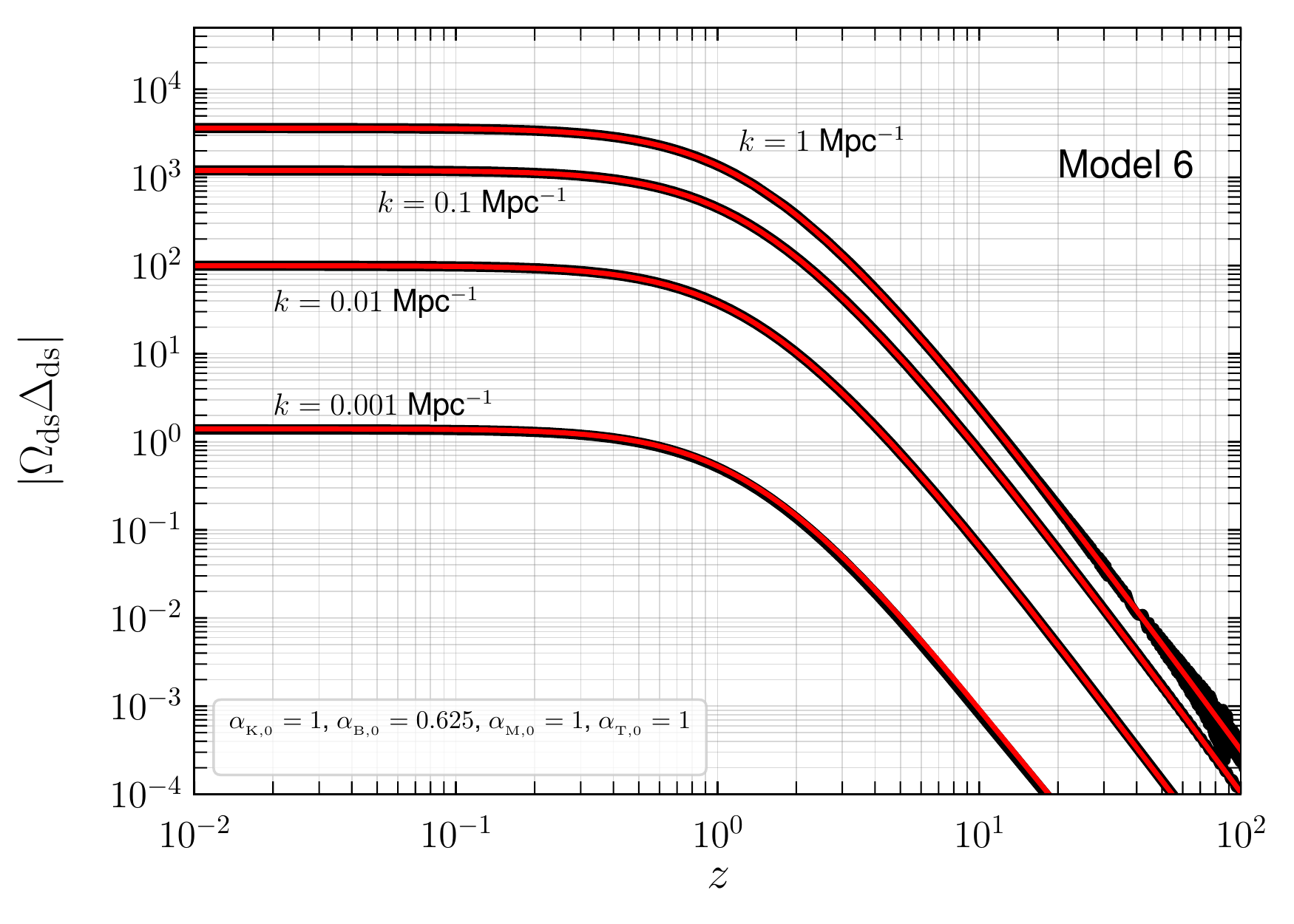}
 \includegraphics[width=6.5cm,angle=0]{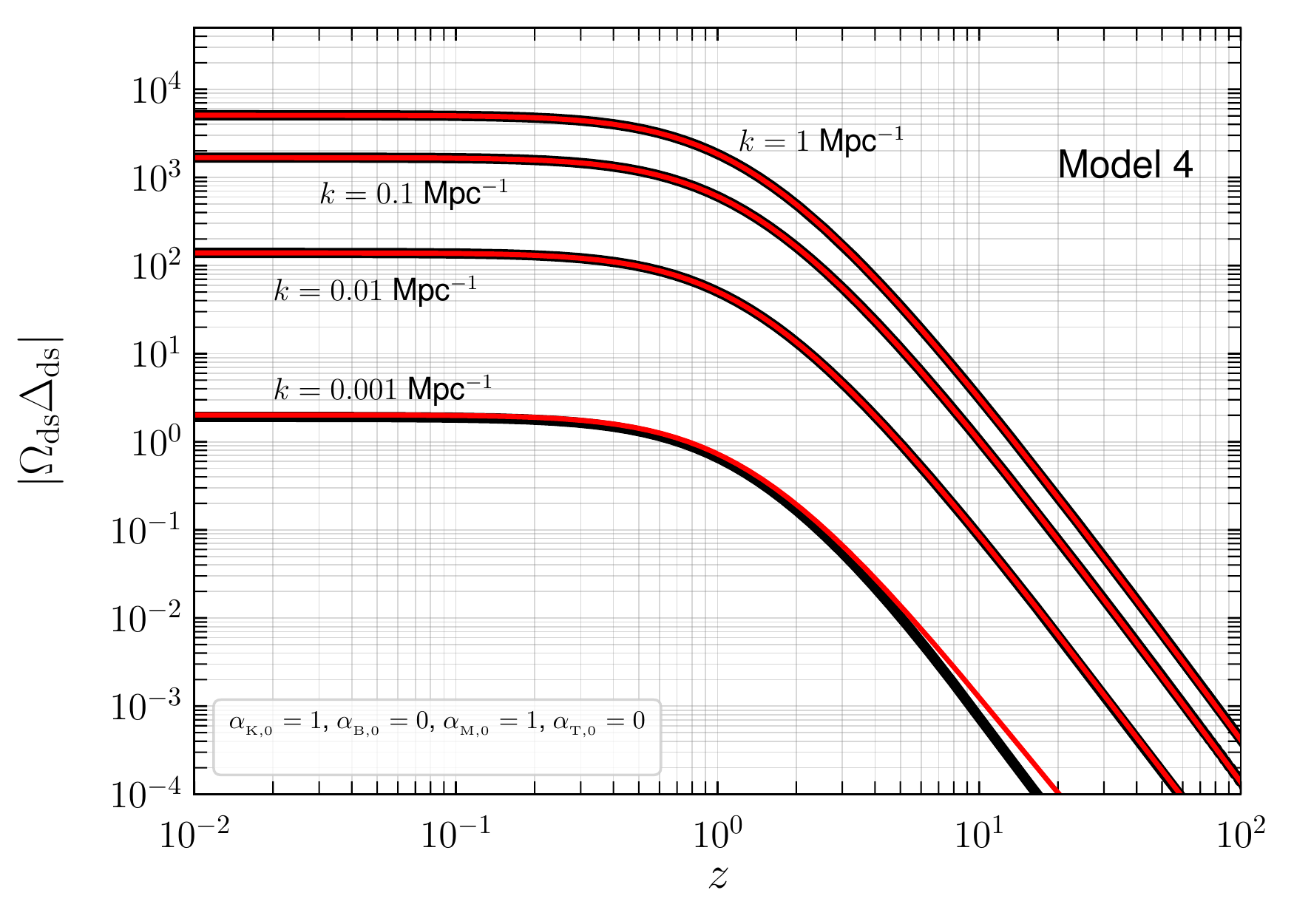}
 \includegraphics[width=6.5cm,angle=0]{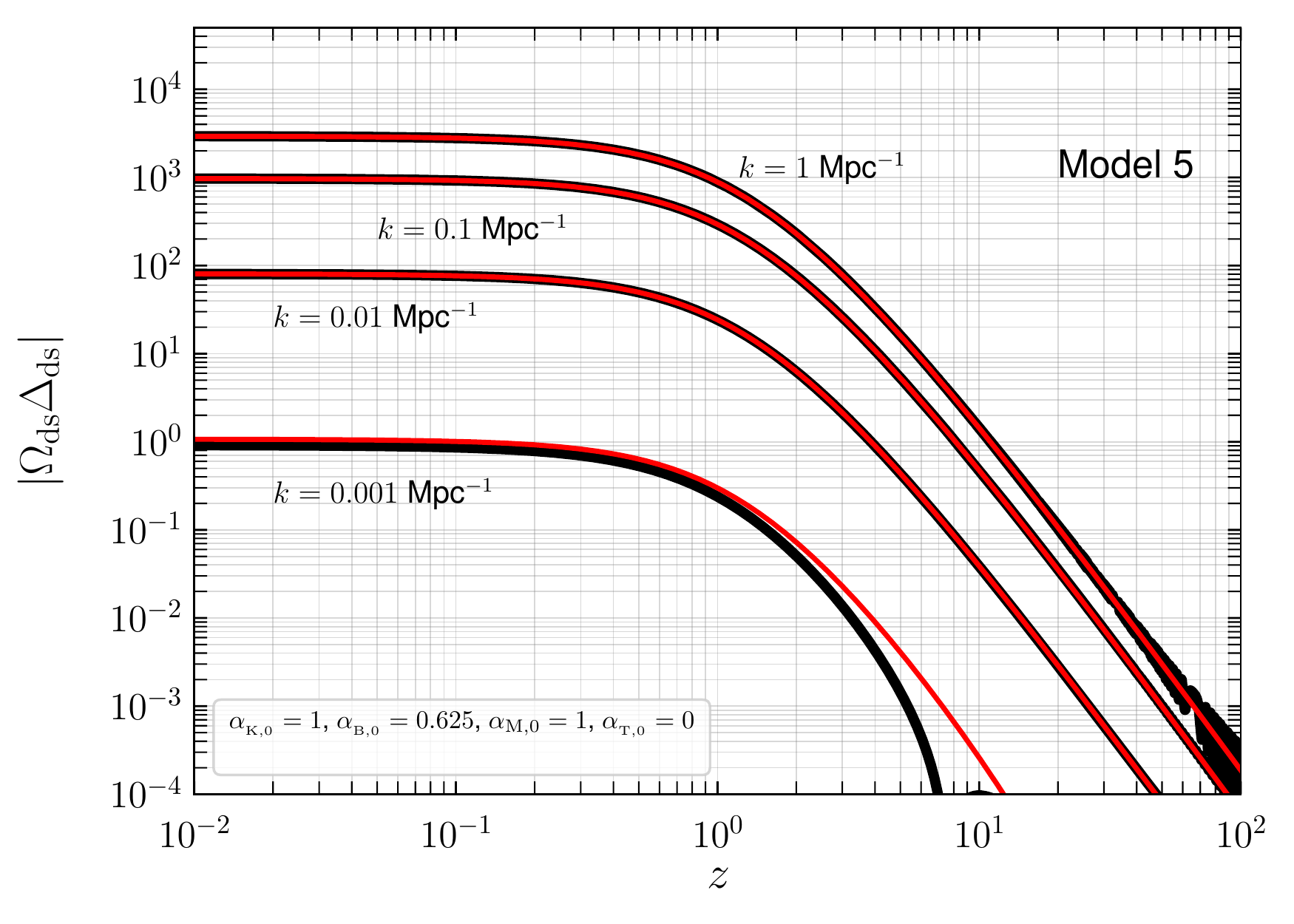}
 \caption[justified]{Evolution of the dark sector gauge invariant density perturbation $\Delta_{\rm ds}$ as a 
 function of redshift $z$. The normalisation of the $\alpha$ functions is as follows: $\alpha_{\rm B,0}=0.625$, 
 $\alpha_{\rm M,0}=1$ and $\alpha_{\rm T,0}=1$. From top left to bottom left, in clockwise order, we show the results 
 for models 2--6, respectively. We consider four different wavelengths, $k=10^{-3}$~Mpc$^{-1}$, $k=10^{-2}$~Mpc$^{-1}$, 
 $k=10^{-1}$~Mpc$^{-1}$ and $k=1$~Mpc$^{-1}$. 
 The upper black solid lines represent the absolute value of $\Omega_{\rm ds}\Delta_{\rm ds}$ while the red lines on 
 top of the black ones represent the attractor solution [Eqs.~(\ref{eqn:dAIC})]. For a wide range of $k$, the 
 attractor approximation works well ($k\gtrsim 10^{-3}$Mpc$^{-1}$ at $z\lesssim100$). Deviations between the numerical 
 and analytical solution arise when ${\rm K}\sim 1$.}
 \label{fig:DeltaTheta}
\end{figure*}

In Fig.~\ref{fig:DeltaTheta} we present $\Omega_{\rm ds}\Delta_{\rm ds}$ as a function of redshift for four different 
scales and for models 2--6. We see that the attractor solution is manifestly a very good approximation of the numerical 
solution for scales $k\gtrsim 10^{-3}$Mpc$^{-1}$ at $z\lesssim 100$. For smaller scales, the attractor solution 
works well at low redshifts but only approximately for $z\gtrsim 1$ since the assumption ${\rm K}\sim 1$ is violated. 
The existence of an attractor and our analytical approximations allow us to derive several analytical results 
describing the properties of the dark sector, as we show in the next subsection.

\subsection{New expressions for the Modified Gravity parameters in Horndeski models}\label{sect:MGparams}
A generic modified gravity model can either modify the Poisson equation or yield an anisotropic stress that leads to 
differences between the gravitational potentials (corresponding to $Z$ and $Y$ in our notation) or both at the same 
time. Hence, it is useful to recast the Einstein field equations in the following form
\begin{subequations}\label{eqn:MG}
 \begin{align}
  -\frac{2}{3}{\rm K}^2Z = &\, \mu_{Z}({\rm K},a)\Omega_{\rm m}\Delta_{\rm m}\,, \label{eqn:muZ}\\
  -\frac{2}{3}{\rm K}^2Y = &\, \mu_{Y}({\rm K},a)\Omega_{\rm m}\Delta_{\rm m}\,, \label{eqn:muY}\\
  -\frac{2}{3}{\rm K}^2\Psi = &\, \Sigma({\rm K},a)\Omega_{\rm m}\Delta_{\rm m}\,, \label{eqn:Sigma}\\
   \frac{Z}{Y} = &\, \eta({\rm K},a)\,. \label{eqn:eta}
 \end{align}
\end{subequations}

In~(\ref{eqn:muZ}) and (\ref{eqn:muY}) the functions $\mu_Z$ and $\mu_Y$ parameterise the modifications to the Poisson 
equations for the gravitational potentials. In the conformal Newtonian gauge, $Z=\phi$ is the space-space component of 
the metric while $Y=\psi$ is the time-time component of the metric. Since $\mu_Y$ (often simply called $\mu$) is 
related to the time-time component of the metric and can therefore be seen as an effective modification of the Newton 
constant, it has also been called $G_{\rm eff}$ (or $G_{\rm matter}$) in previous works \citep{Bean2010,Gleyzes2014}. 
Using Eqs.~(\ref{eqn:00}) and (\ref{eqn:ij}) together with Eqs.~(\ref{eqn:MG}), these two functions, $\mu_Z$ and 
$\mu_Y$, can be written in terms of the perturbed fluid variables as
\begin{align}
 \mu_Z = &\, 1 + \frac{\Omega_{\rm ds}\Delta_{\rm ds}}{\Omega_{\rm m}\Delta_{\rm m}}\,, \label{eqn:muZ_Delta}\\
 \mu_Y = &\, \mu_Z - 2\frac{\Omega_{\rm ds}w_{\rm ds}\Pi_{\rm ds}}{\Omega_{\rm m}\Delta_{\rm m}}\,. 
 \label{eqn:muY_Delta}
\end{align}
In~(\ref{eqn:Sigma}) the function $\Sigma$ parameterises the departure of the lensing (Weyl) potential 
$\Psi\equiv(Z+Y)/2$ from its GR equivalent. It is related to the other functions via
\begin{equation}
 \Sigma = \frac{1}{2}\left(\mu_Z+\mu_Y\right) = \frac{1}{2}\mu_Y\left(1+\eta\right)\,.
\end{equation}
In previous works, $\Sigma$ has also been called $G_{\rm light}$ \citep{Pogosian2016}. 
In~(\ref{eqn:eta}), the function $\eta$ (dubbed $\gamma$ in \cite{Gleyzes2014,Zucca2019}) is the gravitational slip, 
not to be confused with the metric variable ``$\eta$'' of the synchronous gauge. It is related to the other functions 
via
\begin{equation}
 \eta = \frac{\mu_Z}{\mu_Y}\,.
\end{equation}

This specific parametrisation of deviations from GR with the functions $\mu_Z$, $\mu_Y$, $\eta$ and $\Sigma$ has been 
widely used in recent research on dark energy and modified gravity. See for instance \cite{Planck2016_XIV} for current 
CMB constraints on these functions. Note that since there are only two metric potentials, only two of these four 
functions are necessary to characterise a particular model; the most common choices are ($\mu_Y, \eta$) or $(\mu_Y, 
\Sigma)$.

Here we present new analytical expressions for these four functions, based on the analytical approximation of the 
attractor solution of the previous section. 
By simply replacing $\Delta_{\rm ds}$ in Eqn.~(\ref{eqn:muZ_Delta}) by its attractor, see Eqn.~(\ref{eqn:dAIC}), we get 
an expression for $\mu_Z$. Then we insert this expression into Eqn.~(\ref{eqn:muY_Delta}) and write 
$w_{\rm ds}\Pi_{\rm ds}=C_{\Pi\Delta_{\rm ds}}\Delta_{\rm ds}+
\tfrac{\Omega_{\rm m}}{\Omega_{\rm ds}}C_{\Pi\Delta_{\rm m}}\Delta_{\rm m}$ where in accordance to our previous 
discussion we neglect the velocities. Again we replace $\Delta_{\rm ds}$ by its attractor and get the expression for 
$\mu_Y$. Finally, we use these expressions for $\mu_Z$ and $\mu_Y$ to write $\eta$ and $\Sigma$. All together, these 
expressions read
\begin{subequations}\label{eqn:MG_EoS}
 \begin{align}
  \mu_Z = &\, \frac{\mu_{Z,0} + \mu_{Z,\infty}\left({\rm K}/{\rm K}_{\ast}\right)^2}
                   {1 + \left({\rm K}/{\rm K}_{\ast}\right)^2}\,, \label{eqn:muZ_aK} \\
  \mu = \mu_Y = &\, \frac{\mu_{Y,0} + \mu_{Y,\infty}\left({\rm K}/{\rm K}_{\ast}\right)^2}
                   {1 + \left({\rm K}/{\rm K}_{\ast}\right)^2}\,, \label{eqn:muY_aK} \\
  \eta = &\, \frac{\mu_{Z,0} + \mu_{Z,\infty}\left({\rm K}/{\rm K}_{\ast}\right)^2}
                  {\mu_{Y,0} + \mu_{Y,\infty}\left({\rm K}/{\rm K}_{\ast}\right)^2}\,, \label{eqn:eta_aK} \\
  \Sigma = &\, \frac{\Sigma_0 + \Sigma_{\infty}\left({\rm K}/{\rm K}_{\ast}\right)^2}
                    {1 + \left({\rm K}/{\rm K}_{\ast}\right)^2}\,, \label{eqn:Sigma_aK}
 \end{align}
\end{subequations}
where
\begin{equation}\label{eqn:K2star}
 {{\rm K}_{\ast}}^2 \equiv \frac{\gamma_1(\gamma_2-\alpha_{\rm T}/3)}{\alpha_{\rm B}^2c_{\rm s}^2}\,,
\end{equation}
and where the functions $\mu_{Z,0}$, $\mu_{Y,0}$ and $\Sigma_{0}$ represent the values in the limit 
${\rm K}\rightarrow0$ and are given by
\begin{equation*}
 \mu_{Z,0} \equiv \frac{\gamma_2-\gamma_7}{\gamma_2-\alpha_{\rm T}/3}\frac{1}{\bar{M}^2}\,, \quad
 \mu_{Y,0} \equiv \mu_{Z,0}(1+\alpha_{\rm T})\,, \quad 
 \eta_0 \equiv \frac{\mu_{Z,0}}{\mu_{Y,0}} = \frac{1}{1+\alpha_{\rm T}}\,, \quad
 \Sigma_{0} \equiv \frac{1}{2}\left(\mu_{Y,0}+\mu_{Z,0}\right)\,,
\end{equation*}
with
\begin{equation*}
 \bar{M}^2\equiv \frac{M^2}{M_{\rm pl}^2}\,.
\end{equation*}
The functions $\gamma_1$, $\gamma_2$ and $\gamma_7$ are given in Appendix~\ref{sect:coefficientsEoS}.

The functions $\mu_{Z,\infty}$, $\mu_{Y,\infty}$, $\eta_{\infty}$ and $\Sigma_{\infty}$ represent the corresponding 
values in the limit ${\rm K}\rightarrow\infty$ and are given by
\begin{subequations}\label{eqn:mu_eta_Sigma_infty}
 \begin{align}
 & \mu_{Z,\infty} =  \frac{\alpha c_{\rm s}^2+2\alpha_{\rm B}[\alpha_{\rm B}(1+\alpha_{\rm T})+
                             \alpha_{\rm T}-\alpha_{\rm M}]}{\alpha c_{\rm s}^2\bar{M}^2}\,, \label{eqn:muZ_infty} \\
 & \mu_{Y,\infty} =  \frac{\alpha c_{\rm s}^2(1+\alpha_{\rm T})+2[\alpha_{\rm B}(1+\alpha_{\rm T})+
                             \alpha_{\rm T}-\alpha_{\rm M}]^2}{\alpha c_{\rm s}^2\bar{M}^2}\,, \label{eqn:muY_infty} \\
 & \eta_{\infty} = \frac{\mu_{Z,\infty}}{\mu_{Y,\infty}} = 
                    \frac{\alpha c_{\rm s}^2+2\alpha_{\rm B}[\alpha_{\rm B}(1+\alpha_{\rm T})+
                            \alpha_{\rm T}-\alpha_{\rm M}]}
                           {\alpha c_{\rm s}^2(1+\alpha_{\rm T})+2[\alpha_{\rm B}(1+\alpha_{\rm T})+
                            \alpha_{\rm T}-\alpha_{\rm M}]^2}\,, \label{eqn:eta_infty} \\
 & \Sigma_{\infty} = \frac{\mu_{Z,\infty}+\mu_{Y,\infty}}{2}
                  =  \frac{\alpha c_{\rm s}^2(2+\alpha_{\rm T})+
                           2[\alpha_{\rm B}(1+\alpha_{\rm T})+\alpha_{\rm T}-\alpha_{\rm M}]
                            [\alpha_{\rm B}(2+\alpha_{\rm T})+\alpha_{\rm T}-\alpha_{\rm M}]}
                             {2\alpha c_{\rm s}^2\bar{M}^2}\,, \label{eqn:Sigma_infty}
 \end{align}
\end{subequations}
where $\alpha=\alpha_{\rm K}+6\alpha_{\rm B}^2$. Note that $\alpha_{\rm K}$ does not appear in any of the expressions 
(\ref{eqn:mu_eta_Sigma_infty}).

\begin{table}[!ht]
 \flushright
 \begin{adjustbox}{width=1.1\textwidth}
  \small
  \begin{tabular}{ccccc|cc}
   \hline
   & $\bar{M}^2\mu_{Z,\infty}$ & $\bar{M}^2\mu_{Y,\infty}$ & $\eta_{\infty}$ & $\bar{M}^2\Sigma_{\infty}$ & 
     $\gamma_{\infty}$ & $g_{\infty}$ \\
   \hline
   $\Lambda$CDM & $1$ & $1$ & $1$ & $1$ & $0$ & $0$ \\
   & & & & & & \\
   Model 1 & $1$ & $1$ & $1$ & $1$ & $0$ & $0$ \\
   & & & & & & \\
   Model 2 & $1+\frac{2\alpha_{\rm B}(\alpha_{\rm B}-\alpha_{\rm M})}{\alpha c_{\rm s}^2}$
           & $1+\frac{2(\alpha_{\rm B}-\alpha_{\rm M})^2}{\alpha c_{\rm s}^2}$
           & $\frac{\alpha c_{\rm s}^2+2\alpha_{\rm B}(\alpha_{\rm B}-\alpha_{\rm M})}
                   {\alpha c_{\rm s}^2+2(\alpha_{\rm B}-\alpha_{\rm M})^2}$
           & $1+\frac{(\alpha_{\rm B}-\alpha_{\rm M})(2\alpha_{\rm B}-\alpha_{\rm M})}
                     {\alpha c_{\rm s}^2}$
           & $-\frac{2\alpha_{\rm M}(\alpha_{\rm B}-\alpha_{\rm M})}
                    {\alpha c_{\rm s}^2+2\alpha_{\rm B}(\alpha_{\rm B}-\alpha_{\rm M})}$
           & $\frac{\alpha_{\rm M}(\alpha_{\rm B}-\alpha_{\rm M})}
                   {\alpha c_{\rm s}^2+(\alpha_{\rm B}-\alpha_{\rm M})(2\alpha_{\rm B}-\alpha_{\rm M})}$ \\
   & & & & & & \\
   Model 3 & $1+\frac{2\alpha_{\rm B}^2}{\alpha c_{\rm s}^2}$
           & $1+\frac{2\alpha_{\rm B}^2}{\alpha c_{\rm s}^2}$
           & $1$ & $1+\frac{2\alpha_{\rm B}^2}{\alpha c_{\rm s}^2}$ & $0$ & $0$ \\
   & & & & & & \\
   Model 4 & $1$ & $1+\frac{2\alpha_{\rm M}^2}{\alpha c_{\rm s}^2}$
           & $\left(1+\frac{2\alpha_{\rm M}^2}{\alpha c_{\rm s}^2}\right)^{-1}$
           & $1+\frac{\alpha_{\rm M}^2}{\alpha c_{\rm s}^2}$
           & $\frac{2\alpha_{\rm M}^2}{\alpha c_{\rm s}^2}$
           & $-\frac{\alpha_{\rm M}^2}{\alpha c_{\rm s}^2+\alpha_{\rm M}^2}$\\
   & & & & & & \\
   Model 5 & $1+\frac{2\alpha_{\rm B}(\alpha_{\rm B}-\alpha_{\rm M})}{\alpha c_{\rm s}^2}$ 
           & $1+\frac{2(\alpha_{\rm B}-\alpha_{\rm M})^2}{\alpha c_{\rm s}^2}$
           & $\frac{\alpha c_{\rm s}^2+2\alpha_{\rm B}(\alpha_{\rm B}-\alpha_{\rm M})}
                   {\alpha c_{\rm s}^2+2(\alpha_{\rm B}-\alpha_{\rm M})^2}$
           & $1+\frac{(\alpha_{\rm B}-\alpha_{\rm M})(2\alpha_{\rm B}-\alpha_{\rm M})}
                     {\alpha c_{\rm s}^2}$
           & $-\frac{2\alpha_{\rm M}(\alpha_{\rm B}-\alpha_{\rm M})}
                    {\alpha c_{\rm s}^2+2\alpha_{\rm B}(\alpha_{\rm B}-\alpha_{\rm M})}$
           & $\frac{\alpha_{\rm M}(\alpha_{\rm B}-\alpha_{\rm M})}
                   {\alpha c_{\rm s}^2+(\alpha_{\rm B}-\alpha_{\rm M})(2\alpha_{\rm B}-\alpha_{\rm M})}$\\
   & & & & & & \\
   Model 6 & $\bar{M}^2\times$ Eqn.~(\ref{eqn:muZ_infty}) 
           & $\bar{M}^2\times$ Eqn.~(\ref{eqn:muY_infty}) & Eqn.~(\ref{eqn:eta_infty})
           & $\bar{M}^2\times$ Eqn.~(\ref{eqn:Sigma_infty})
           & Eqn.~(\ref{eqn:gamma_infty}) & Eqn.~(\ref{eqn:g_infty}) \\
   & & & & & & \\
   No slip model & $1$ & $1$ & $1$ & $1$ & $0$ & $0$ \\
   \hline 
  \end{tabular}
 \end{adjustbox}
 \caption{\label{tab:DEphenomenology_infinity} Time-evolution of the phenomenological MG functions in the limit 
 ${\rm K}\rightarrow\infty$. For completeness we also consider the no slip model \citep{Linder2018}.}
\end{table}

The expressions for $\mu_{Y,\infty}$ and $\eta_{\infty}$ have already been obtained in previous works using different 
formulations of the Effective Field Theory formalism. \cite{Gleyzes2014} (see Eqs.~(139) and (140)) and 
\cite{Lombriser2015a} (see Eqs.~(3.18) and (3.19)) used the $\alpha$s functions introduced in \cite{Bellini2014} while 
\cite{Gubitosi2013} (see Eq.~(70)) provided the expression for $\mu_{Y,\infty}$ based on a different set of functions 
which represent the coefficients of the perturbed operators and are constructed only with background quantities in the 
case of $\alpha_{\rm T}=0$. We refer to Table~2 of \cite{Bellini2014} for a translation between the perturbed variables 
$\alpha_{\rm x}$ and those used in \cite{Gubitosi2013}.

To our knowledge, a few other works have obtained expressions that attempt model the scale dependence of these 
functions in Horndeski theories. In \cite{DeFelice2011} (see Eqs.~(45) and (52)) the authors considered the equations 
of motion of the perturbed scalar field and applied the quasi-static approximation (i.e. neglected the time derivatives 
of the metric potentials and of the perturbed scalar field). The remaining terms were used to rewrite the resulting 
Poisson equation in terms of an effective gravitational constant $\mu$. \cite{Bloomfield2013a} (see Eqs.~(4.15) and 
(4.18)) used a similar approach to \cite{DeFelice2011} by considering only terms with ${\rm K}^2\gg 1$ and neglecting 
time derivatives of the fields. \cite{Silvestri2013} used the quasi-static approximation and arguments of locality and 
general covariance to provide a general expression for $\mu_{Y}$ and $\eta$ (see Eqs.~(24) and (25)). As stated in 
\cite{Silvestri2013}, these expressions have the same form as those in \cite{DeFelice2011}. 
See also Eqs.~(19) and (20) of \cite{Pogosian2016}. 
\cite{Lombriser2015a} (see Eqs.~(3.7) and (3.8)) used a so-called \textit{semi-dynamical} approach. To study the 
evolution of perturbations beyond the quasi-static regime on small scales, these authors introduce terms that take into 
account corrections from the perturbed velocities and time derivatives of the metric potentials determined at a pivot 
scale of choice.
All these approaches have two assumptions in common: they are valid on small scales and a variant of the 
quasi-static approximation is applied. The first assumption implies that only terms multiplied by ${\rm K}^2$ are 
considered with respect to the others; the second is equivalent to neglecting time derivatives of the relevant 
quantities appearing in the equations considered. More formally, these two assumptions imply 
${\rm K}^2|X|\gg\{|X^{\prime}|, |X^{\prime\prime}|\}$, where $X=\{Z, Y, \delta\phi, \Delta_{\rm ds}\}$, where the 
appropriate variables are considered according to the set of equations the quasi-static approximation is applied to. 
Here, $\delta\phi$ represents the perturbation of the scalar degree of freedom.

The large scale limit of the expressions in \cite{DeFelice2011,Bloomfield2013a,Silvestri2013,Lombriser2015a} differs 
from each other and ours due to the different assumptions. We will perform a detailed comparison of the different 
formulae for the modified gravity functions $\mu$ and $\eta$ in a forthcoming paper.

In Table~(\ref{tab:DEphenomenology_infinity}) we report the limiting values of the MG functions in the limit 
${\rm K}\rightarrow\infty$, in $\Lambda$CDM and in the six classes of models described in 
subsection~\ref{sect:Models}. We also added a line corresponding to the so-called ``no slip'' model of 
\cite{Linder2018}. The ``no slip'' model is a particular class of models that have $\alpha_{\rm M}=\alpha_{\rm B}$, 
$\alpha_{\rm K}$ free and $\alpha_{\rm T}=0$ (they are actually a subclass of the class of model 5). Hence one finds 
that $\gamma_7=0$ in the ``no slip'' model and, as a consequence, $\eta_{\infty}=1$ (``no slip'') while the other 
modified gravity functions are equal to $1/\bar{M}^2$. Moreover in the Table we have added two additional columns 
corresponding to the functions $\gamma_{\infty}\equiv1/\eta_{\infty}-1$ and 
$g_{\infty}\equiv\tfrac{\mu_{Z,\infty}-\mu_{Y,\infty}}{\mu_{Z,\infty}+\mu_{Y,\infty}}=
\tfrac{\eta_{\infty}-1}{\eta_{\infty}+1}$ which were used in \citep{Hu2007a}. 
This latter quantity is very well constrained in the Solar System by the Cassini mission \cite{Will2014}. 
Their general expressions are:
\begin{subequations}
 \begin{align}
  \gamma_{\infty} = &\, \frac{\alpha\alpha_{\rm T}c_{\rm s}^2+
                             2[\alpha_{\rm B}(1+\alpha_{\rm T})+\alpha_{\rm T}-\alpha_{\rm M}]
                              [\alpha_{\rm B}\alpha_{\rm T}+\alpha_{\rm T}-\alpha_{\rm M}]}
                             {\alpha c_{\rm s}^2+2\alpha_{\rm B}
                              [\alpha_{\rm B}(1+\alpha_{\rm T})+\alpha_{\rm T}-\alpha_{\rm M}]} \,,
                             \label{eqn:gamma_infty}\\
  g_{\infty} = &\, -\frac{\alpha\alpha_{\rm T}c_{\rm s}^2+
                          2[\alpha_{\rm B}(1+\alpha_{\rm T})+\alpha_{\rm T}-\alpha_{\rm M}]
                           [\alpha_{\rm B}\alpha_{\rm T}+\alpha_{\rm T}-\alpha_{\rm M}]}
                         {\alpha c_{\rm s}^2(2+\alpha_{\rm T})+
                          2[\alpha_{\rm B}(1+\alpha_{\rm T})+\alpha_{\rm T}-\alpha_{\rm M}]
                           [\alpha_{\rm B}(2+\alpha_{\rm T})+\alpha_{\rm T}-\alpha_{\rm M}]}\,. \label{eqn:g_infty}
 \end{align}
\end{subequations}

Finally we note that we chose to write our formulae for the modified gravity functions, 
Eqs.~(\ref{eqn:muZ_aK})--(\ref{eqn:Sigma_aK}) using ${\rm K}_{\ast}^2$ defined in Eqn.~(\ref{eqn:K2star}). 
Phenomenologically, ${\rm K}_{\ast}$ represents the transition scale between two regimes, large 
(${\rm K}\rightarrow 0$) and small (${\rm K}\rightarrow \infty$) scales. 
Indeed with this notation the expressions look relatively simple and compact. 
However it is true that in this form, if, for instance $\gamma_1$ or $\alpha_{\rm B}$ are zero, the expressions become 
ill-defined. In such cases, one can still use our expressions, but need to multiply both the numerator and denominator 
by ${\rm K}_{\ast}^2$ and potentially $\alpha_{\rm B}^2c_{\rm s}^2$. In this way, the expressions are always well 
defined.

\subsection{Simplified EoS for perturbations}\label{sect:simplified}
In this subsection we are interested in finding simplified expressions for the entropy perturbations and anisotropic 
stress of the dark sector, as these functions constitute a way of characterising a dark sector theory, independent and 
complementary to the modified gravity parameters of the previous section.

First we give the simplified expressions for the EoS that we obtained thanks to the attractor solution. Second we go 
through each class of models of subsection~\ref{sect:Models} and provide formulae for the EoS considering only the 
leading terms.

As we explained in the previous section for ${\rm K}^2\gg 1$, velocity perturbations are subdominant compared to 
density perturbations. With this assumption, we were able to express the dark sector density perturbation in terms of 
the matter perturbation, using the attractor solution of Eqn.~(\ref{eqn:dAIC}). Now we consider the EoS for 
perturbations for $w_{\rm ds}\Gamma_{\rm ds}$, $w_{\rm ds}\Pi_{\rm ds}$ and $w_{\rm ds}\zeta_{\rm ds}$ presented in 
Eqs.~(\ref{eqn:Delta}), (\ref{eqn:Theta}) and (\ref{eqn:zeta_ds}), respectively. 
In these expressions we neglect the velocity perturbations of both matter and dark sector and use the attractor 
solution to replace one density perturbation in terms of the other. This gives
\begin{subequations}
 \begin{align}
  w_{\rm ds}\Gamma_{\rm ds} & = \frac{\Omega_{\rm m}}{\Omega_{\rm ds}}\left\{
                                 c_{\rm a,ds}^2 + 
                                 \frac{\Gamma_0+\Gamma_{\infty}\left({\rm K}/{\rm K}_{\ast}\right)^2}
                                      {1 + \left({\rm K}/{\rm K}_{\ast}\right)^2}\right\}\Delta_{\rm m}\,,\\
                            & = -\frac{c_{\rm a,ds}^2+\Gamma_0 + 
                                      (c_{\rm a,ds}^2+\Gamma_{\infty})\left({\rm K}/{\rm K}_{\ast}\right)^2}
                                      {1-\mu_{Z,0} + 
                                      (1-\mu_{Z,\infty})\left({\rm K}/{\rm K}_{\ast}\right)^2}
                                 \Delta_{\rm ds}\,, \nonumber\\
  w_{\rm ds}\Pi_{\rm ds} & = \frac{\Omega_{\rm m}}{\Omega_{\rm ds}}\left\{
                             \frac{\Pi_0 + \Pi_{\infty}\left({\rm K}/{\rm K}_{\ast}\right)^2}
                                  {1 + \left({\rm K}/{\rm K}_{\ast}\right)^2}\right\}\Delta_{\rm m}\,,\\
                         & =  -\frac{\Pi_0 + \Pi_{\infty}\left({\rm K}/{\rm K}_{\ast}\right)^2}
                                    {1-\mu_{Z,0} + (1-\mu_{Z,\infty})\left({\rm K}/{\rm K}_{\ast}\right)^2}
                                \Delta_{\rm ds}\,, \nonumber\\
  w_{\rm ds}\zeta_{\rm ds} & = \frac{\Omega_{\rm m}}{\Omega_{\rm ds}}\left\{
                               c_{\rm a,ds}^2 + 
                               \frac{\zeta_0 + \zeta_{\infty}\left({\rm K}/{\rm K}_{\ast}\right)^2}
                                    {1 + \left({\rm K}/{\rm K}_{\ast}\right)^2}
                               \right\}\Delta_{\rm m}\,,\\
                           & = -\frac{c_{\rm a,ds}^2+\zeta_0 + 
                                      (c_{\rm a,ds}^2+\zeta_{\infty})\left({\rm K}/{\rm K}_{\ast}\right)^2}
                                     {1-\mu_{Z,0} + (1-\mu_{Z,\infty})\left({\rm K}/{\rm K}_{\ast}\right)^2}
                                     \Delta_{\rm ds}\,, \nonumber 
 \end{align}
\end{subequations}
where 
\begin{equation*}
 \Gamma_0 \equiv \mu_{Z,0}\left(\frac{\alpha_{\rm T}}{3}-c_{\rm a,ds}^2\right)\,, \quad 
 \Gamma_{\infty} \equiv \frac{1}{3}[\mu_{Y,\infty}-\mu_{Z,\infty}(1+3c_{\rm a,ds}^2)]\,, \quad 
 \Pi_0 \equiv \frac{1}{2}(\mu_{Z,0}-\mu_{Y,0})\,, \quad 
 \Pi_{\infty} \equiv \frac{1}{2}(\mu_{Z,\infty}-\mu_{Y,\infty})\,.
\end{equation*}
We also deduce that
\begin{equation*}
 \zeta_0 \equiv \frac{2}{3}\Pi_0 + \Gamma_0\,, \quad 
 \zeta_{\infty} \equiv \frac{2}{3}\Pi_{\infty} + \Gamma_{\infty}\,.
\end{equation*}

This enables us to define the equation-of-state parameter for perturbations in Horndeski theories, 
$w_{\Gamma} \equiv w_{\rm ds}\Gamma_{\rm ds}/\Delta_{\rm ds}$ and 
$w_{\Pi} \equiv w_{\rm ds}\Pi_{\rm ds}/\Delta_{\rm ds}$, and for future numerical implementation, 
$w_{\zeta} \equiv \tfrac{2}{3}w_{\Pi}+w_{\Gamma}$, given by
\begin{subequations}
 \begin{align}
  w_{\Gamma} & = -\frac{w_{\rm ds}+\Gamma_0 + 
                       (w_{\rm ds}+\Gamma_{\infty})\left({\rm K}/{\rm K}_{\ast}\right)^2}
                       {1-\mu_{Z,0} + (1-\mu_{Z,\infty})\left({\rm K}/{\rm K}_{\ast}\right)^2}\,,\\
  w_{\Pi} & = -\frac{\Pi_0 + \Pi_{\infty}\left({\rm K}/{\rm K}_{\ast}\right)^2}
                    {1-\mu_{Z,0} + (1-\mu_{Z,\infty})\left({\rm K}/{\rm K}_{\ast}\right)^2}\,,\\
  w_{\zeta} & = -\frac{w_{\rm ds}+\zeta_0 + 
                       (w_{\rm ds}+\zeta_{\infty})\left({\rm K}/{\rm K}_{\ast}\right)^2}
                      {1-\mu_{Z,0} + (1-\mu_{Z,\infty})\left({\rm K}/{\rm K}_{\ast}\right)^2}\,.
\end{align}
\end{subequations}
In a way similar to the new expressions for $\mu$ and $\eta$ one can use these expressions for $w_{\Pi}$ and 
$w_{\zeta}$ to solve the dynamics of perturbations in Horndeski models, as we will show in our next paper.

In each of the classes of models we can get simplified expressions for $w_{\rm ds}\Gamma_{\rm ds}$, 
$w_{\rm ds}\Pi_{\rm ds}$ and $w_{\rm ds}\zeta_{\rm ds}$ by neglecting the velocities (except model 4). Since in each 
class of models some $\alpha$s are zero (except in model 6), the expressions simplify considerably. In model 1 
($k$-essence), since only $\alpha_{\rm K}$ is different from zero, we obtain without any approximation
\begin{equation}\label{eqn:aKsimple}
 w_{\rm ds}\Gamma_{\rm ds} = \left(c_{\rm s}^2-c_{\rm a, ds}^2\right)\Delta_{\rm ds} \,, \quad 
 w_{\rm ds}\Pi_{\rm ds} = 0\,, \quad 
 w_{\rm ds}\zeta_{\rm ds} = w_{\rm ds}\Gamma_{\rm ds}\,,
\end{equation}
where $c_{\rm s}^2$ and $c_{\rm a, ds}^2$ are the perturbations and adiabatic sound speed. Note that for a 
$\Lambda$CDM background, $w_{\rm ds}\Gamma_{\rm ds} = \Delta_{\rm ds}$.

For model 2 ($f(R)$-like), we obtain
\begin{subequations}\label{eqn:aBaMsimple}
 \begin{align}
  w_{\rm ds}\Gamma_{\rm ds} & = \left\{c_{\rm s}^2-\frac{1}{3}\frac{\alpha_{\rm M}}{\alpha_{\rm B}}-c_{\rm a,ds}^2+
                                      \frac{{\rm M}_f^2}{{\rm K}^2}\right\}\Delta_{\rm ds} + \nonumber\\
                            & \quad  \frac{\Omega_{\rm m}}{\Omega_{\rm ds}}\left\{\left(
                                      c_{\rm s}^2-\frac{1}{3}\frac{\alpha_{\rm M}}{\alpha_{\rm B}}+
                                      \frac{{\rm M}_f^2}{{\rm K}^2}\right)\left(1-\frac{1}{\bar{M}^2}\right)+
                            \frac{1}{3}\left(\frac{\alpha_{\rm M}}{\alpha_{\rm B}}-1\right)\frac{1}{\bar{M}^2}
                                      \right\}\Delta_{\rm m}\,,\\
  w_{\rm ds}\Pi_{\rm ds} & = \frac{1}{2}\frac{\alpha_{\rm M}}{\alpha_{\rm B}}\left[
                             \Delta_{\rm ds}+
                             \frac{\Omega_{\rm m}}{\Omega_{\rm ds}}\left(1-\frac{1}{\bar{M}^2}\right)\Delta_{\rm m}
                             \right]\,,\\
  w_{\rm ds}\zeta_{\rm ds} & = \left\{c_{\rm s}^2-c_{\rm a,ds}^2+\frac{{\rm M}_f^2}{{\rm K}^2}\right\}\Delta_{\rm ds} 
                               + \frac{\Omega_{\rm m}}{\Omega_{\rm ds}}\left\{\left(
                                 c_{\rm s}^2+\frac{{\rm M}_f^2}{{\rm K}^2}\right)\left(1-\frac{1}{\bar{M}^2}\right)
                              +\frac{1}{3}\left(\frac{\alpha_{\rm M}}{\alpha_{\rm B}}-1\right)\frac{1}{\bar{M}^2}
                                      \right\}\Delta_{\rm m}\,,
 \end{align}
\end{subequations}
where ${\rm M}_f^2 \equiv -\epsilon_H(\bar{\epsilon}_H/\epsilon_H+\alpha_{\rm M}/\alpha_{\rm B}-2)/\alpha_{\rm B}$ 
generalises the mass term for $f(R)$ models, see e.g.\ \cite{Battye2018a}. Note that by setting 
$\alpha_{\rm M}=2\alpha_{\rm B}$, $c_{\rm s}^2=1$ and $\bar{M}^2=1$, one recovers the approximate $f(R)$ 
expressions already obtained in \cite{Battye2018a}.

For model 3, as well as models 4, 5 and 6, we further assume that the modes are subhorizon (${\rm K}^2\gg 1$) to 
derive the simplified EoS for perturbations. For model 3 we obtain
\begin{equation}\label{eqn:aKaBsimple}
 w_{\rm ds}\Gamma_{\rm ds} = \left(c_{\rm s}^2-c_{\rm a}^2\right)\Delta_{\rm ds} - 
                             2\frac{\alpha_{\rm B}^2}{\alpha}\frac{\Omega_{\rm m}}{\Omega_{\rm ds}}\Delta_{\rm m}\,, 
 \quad
 w_{\rm ds}\Pi_{\rm ds} = 0\,, \quad 
 w_{\rm ds}\zeta_{\rm ds} = w_{\rm ds}\Gamma_{\rm ds}\,.
\end{equation}
For $\alpha_{\rm B}=0$, we recover the expression in Eqn.~(\ref{eqn:aKsimple}).

For model 4, there is a subtlety because $C_{\Pi\Delta}=0$, so the leading terms in the anisotropic stress are 
proportional to the velocity perturbations. We obtain
\begin{subequations}\label{eqn:aKaMsimple}
 \begin{align}
  w_{\rm ds}\Gamma_{\rm ds} & = \left(c_{\rm s}^2-2\frac{\alpha_{\rm M}}{\alpha_{\rm K}}-c_{\rm a,ds}^2\right)
                                \Delta_{\rm ds}+
                                \left(c_{\rm s}^2-2\frac{\alpha_{\rm M}}{\alpha_{\rm K}}-c_{\rm a,m}^2\right)
                                \left(1-\frac{1}{\bar{M}^2}\right)\Delta_{\rm m} \\
  w_{\rm ds}\Pi_{\rm ds} & = \frac{1}{3}C_{\Pi\Theta}{\rm K}^2
                             \left[\Theta_{\rm ds}+\frac{\Omega_{\rm m}}{\Omega_{\rm ds}}
                             \left(1-\frac{1}{\bar{M}^2}\right)\Theta_{\rm m}\right]\,, \label{eqn:Pi4}\\
  w_{\rm ds}\zeta_{\rm ds} & = \left(c_{\rm s}^2-2\frac{\alpha_{\rm M}}{\alpha_{\rm K}}-c_{\rm a,ds}^2\right)
                               \Delta_{\rm ds}+
                               \left(c_{\rm s}^2-2\frac{\alpha_{\rm M}}{\alpha_{\rm K}}-c_{\rm a,m}^2\right)
                               \left(1-\frac{1}{\bar{M}^2}\right)\Delta_{\rm m}\,,
 \end{align}
\end{subequations}
where
\begin{equation}\label{eqn:CPiTheta}
 C_{\Pi\Theta} = \frac{\alpha_{\rm M}}{2\alpha_{\rm M}-\alpha_{\rm K}c_{\rm s}^2}\,.
\end{equation}
Even if $c^2_{\rm a,m}\approx 0$ for matter at late time, we kept this term for symmetry reasons with respect to the 
coefficient of $\Delta_{\rm ds}$.

For model 5 we obtain
\begin{subequations}\label{eqn:aKaBaMsimple}
 \begin{align}
  w_{\rm ds}\Gamma_{\rm ds} = & \left(c_{\rm s}^2-\frac{1}{3}\frac{\alpha_{\rm M}}{\alpha_{\rm B}}-
                                      c_{\rm a,ds}^2\right)\Delta_{\rm ds} + 
                                \frac{\Omega_{\rm m}}{\Omega_{\rm ds}}
                                  \left[\left(
                                   c_{\rm s}^2-\frac{1}{3}\frac{\alpha_{\rm M}}{\alpha_{\rm B}}\right)
                                  \left(1-\frac{1}{\bar{M}^2}\right)
                                  -\frac{2}{\bar{M}^2}\frac{\alpha_{\rm B}(\alpha_{\rm B}-\alpha_{\rm M})}{\alpha}
                                  \right]\Delta_{\rm m}\,,\\
  w_{\rm ds}\Pi_{\rm ds} = & \frac{1}{2}\frac{\alpha_{\rm M}}{\alpha_{\rm B}}\left[
                             \Delta_{\rm ds} + \frac{\Omega_{\rm m}}{\Omega_{\rm ds}}
                             \left(1-\frac{1}{\bar{M}^2}\right)\Delta_{\rm m}\right]\,,\\
  w_{\rm ds}\zeta_{\rm ds} = & \left(c_{\rm s}^2-c_{\rm a,ds}^2\right)\Delta_{\rm ds} + 
                               \frac{\Omega_{\rm m}}{\Omega_{\rm ds}}
                               \left[c_{\rm s}^2\left(1-\frac{1}{\bar{M}^2}\right)
                              -\frac{2}{\bar{M}^2}\frac{\alpha_{\rm B}(\alpha_{\rm B}-\alpha_{\rm M})}{\alpha}
                               \right]\Delta_{\rm m}\,.
 \end{align}
\end{subequations}
Note that when $\alpha_{\rm M}=0$, we recover model 3, while for $\alpha_{\rm B}=\alpha_{\rm M}=0$, we recover model 
1. We also note that the term $\alpha_{\rm M}/\alpha_{\rm B}=\alpha_{\rm M,0}/\alpha_{\rm B,0}$ becomes time 
independent for a particular functional form of the $\alpha$ functions, i.e.\ when they are all proportional to the 
same time-dependent function.

We finally consider model 6 for which none of the $\alpha$s are zero. Nevertheless the expressions simplify for 
${\rm K}^2\gg 1$. We obtain
\begin{subequations}\label{eqn:aKaBaMaTsimple}
 \begin{align}
  w_{\rm ds}\Gamma_{\rm ds} & = (\gamma_3-c_{\rm a,ds}^2)\Delta_{\rm ds}+
                                \frac{\Omega_{\rm m}}{\Omega_{\rm ds}}
                                \left[\gamma_3\left(1-\frac{1}{\bar{M}^2}\right)+
                                      \frac{\gamma_7}{\bar{M}^2}\right]\Delta_{\rm m}\,,\\
  w_{\rm ds}\Pi_{\rm ds} & = -\frac{1}{2}\left\{\gamma_8\Delta_{\rm ds}+
                                         \frac{\Omega_{\rm m}}{\Omega_{\rm ds}}
                                         \left[\frac{\alpha_{\rm T}}{\bar{M}^2}+
                                               \gamma_8\left(1-\frac{1}{\bar{M}^2}\right)\right]\Delta_{\rm m}
                                         \right\}\,,\\
  w_{\rm ds}\zeta_{\rm ds} & = (c_{\rm s}^2-c_{\rm a,ds}^2)\Delta_{\rm ds} + 
                               \frac{\Omega_{\rm m}}{\Omega_{\rm ds}}\left[
                               c_{\rm s}^2\left(1-\frac{1}{\bar{M}^2}\right)+
                               \frac{\gamma_7-\alpha_{\rm T}/3}{\bar{M}^2}
                               \right]\Delta_{\rm m}\,,
 \end{align}
\end{subequations}
where $\gamma_3$, $\gamma_7$ and $\gamma_8$ are defined in Appendix~\ref{sect:coefficientsEoS}.

We verified numerically that the simplified equations of state in Eqs.~(\ref{eqn:aBaMsimple}), (\ref{eqn:aKaBsimple}), 
(\ref{eqn:aKaMsimple}), (\ref{eqn:aKaBaMsimple}) and (\ref{eqn:aKaBaMaTsimple}) provide an excellent approximation to 
the full expressions. When we use the simplified EoS we obtain CMB anisotropy temperature power spectra that agree at 
the sub-percent level with the exact solutions down to $\ell\approx 10-20$. 
For the linear matter power spectrum we get sub-percent agreement for all scales with $k\gtrsim 10^{-2}$~Mpc$^{-1}$. 
We checked this for the numerically most challenging models, i.e.\ 2 and 4, with extreme values of $\alpha_{\rm x,0}$ 
such as $\alpha_{\rm B,0}=2.5$ and $\alpha_{\rm M,0}=4$. 
We leave a more detailed analysis of the simplified EoS to a forthcoming paper.

\section{Phenomenology of Horndeski theories}\label{sect:phenomenology}

\subsection{Understanding numerical results for several cosmological observables}\label{sect:numerical_results}
Using our novel code \verb|EoS_class| we have computed the dimensionless CMB angular temperature anisotropy power 
spectra $C_{\ell}^{\rm TT}$, the dimensionless angular power spectrum of the lensing potential $C_{\ell}^{\phi\phi}$ 
and the total linear matter power spectrum $P(k)$ in units (Mpc/$h$)$^3$, for the six classes of models of 
Section~\ref{sect:Models} and these are presented in the left hand panels of 
Figs.~\ref{fig:aBaM}--\ref{fig:aKaBaMaT}.

Let us first consider the global behaviour of $\alpha_{\rm K}$ across the models studied here. In model 2 
($f(R)$-like), $\alpha_{\rm K,0}=0$ while in all the other models $\alpha_{\rm K,0}=1$. 
More in detail, model 5 differs from model 2 only in terms of $\alpha_{\rm K}$ and we will be comparing these models 
and study the effect of the kineticity term on the observables. $\alpha_{\rm K}$ does not appear in the expressions 
for $\mu$ and $\gamma$ in the limit ${\rm K}\rightarrow\infty$, but it is present in the transition scale 
${\rm K}_{\ast}$ via the function $\gamma_1$. Comparing Fig.~\ref{fig:aBaM} with Fig.~\ref{fig:aKaBaM}, one realises 
that the $\alpha_{\rm K}$ is only important on very large scales ($\ell\lesssim20$ for the CMB temperature anisotropy 
and lensing potential power spectra and $k\lesssim 10^{-2}~h{\rm Mpc}^{-1}$ for the matter power spectrum), when 
analysing the right hand panels in Fig.\ref{fig:aBaM}.

The effects of modified gravity, for the particular models considered here, are typically not that visible in the 
matter power spectrum, they are more relevant in the CMB temperature anisotropy and lensing potential power spectra, 
therefore we will discuss them more in detail. In the following we will consider separately the effects of varying 
$\alpha_{\rm B}$ and $\alpha_{\rm M}$ to understand their global behaviour and translate this into the effects of 
varying $\mu=\mu_Y$.

The braiding term is responsible for the fifth force and for an enhancement of clustering. Increasing the value of 
$\alpha_{\rm B,0}$ leads to an excess of power on small scales for the matter power spectrum and on large scales for 
the angular power spectrum of the lensing potential. The temperature anisotropy power spectrum is affected on large 
scales via the Integrated Sachs-Wolfe (ISW) effect, which is usually reduced. An increase in the braiding translates 
into an increase of the effective gravitational constant $\mu$. 
Therefore, increasing $\mu$ leads to an increase of power in both $P(k)$ and $C_{\ell}^{\phi\phi}$.

The rate of running of the Planck mass $\alpha_{\rm M}$ introduces anisotropic stress. An increase in $\alpha_{\rm M}$ 
leads to an excess of power in the CMB temperature power spectrum on large scales (low multipoles) due to the ISW 
effect and at the same time to a deficit in power in the lensing potential power spectrum. An increase in 
$\alpha_{\rm M}$ leads to a decrease of $\mu$ and $\eta$ and as a consequence a decrease in $\Sigma$ and this explains 
the decrease of power in $C_{\ell}^{\phi\phi}$.

Model 2 is the class of $f(R)$-like models and has $\alpha_{\rm B}\neq 0$ and $\alpha_{\rm M} \neq 0$, while 
$\alpha_{\rm K}=\alpha_{\rm T}=0$. The condition $\alpha_{\rm B}\neq 0$ is characteristic of a modified Newton's law 
or an effective gravitational constant. For model 2, it can be written as
\begin{equation}
 \mu = \frac{G_{\rm eff}}{G_{\rm N}}
     = \frac{3{\rm M}_f^2 + [3c_{\rm s}^2+(\alpha_{\rm M}/\alpha_{\rm B}-1)^2]{\rm K}^2}
            {3{\rm M}_f^2+3c_{\rm s}^2{\rm K}^2}\frac{1}{\bar{M}^2}\,.
\end{equation}
The condition $\alpha_{\rm M}\neq0$ implies a non-vanishing slip (anisotropic stress)
\begin{equation}
 \eta = \frac{3{\rm M}_f^2 + [3c_{\rm s}^2-(\alpha_{\rm M}/\alpha_{\rm B}-1)]{\rm K}^2}
             {3{\rm M}_f^2 + [3c_{\rm s}^2+(\alpha_{\rm M}/\alpha_{\rm B}-1)^2]{\rm K}^2}\,.
\end{equation}
Their phenomenology is similar to $f(R)$ models as can be seen comparing the results of the panels of the left column 
of Fig.~\ref{fig:aBaM} with the top panels Fig.~3 of \cite{Battye2018a}.

In model 3 (KGB-like, $\alpha_{\rm K}\neq 0$ and $\alpha_{\rm B}\neq 0$) there is also a modification to Newton's law 
because the braiding is present ($\alpha_{\rm B}\neq 0$), but there is no anisotropic stress because 
$\alpha_{\rm M}=\alpha_{\rm T}=0$. One finds that
\begin{equation}\label{eqn:mu3}
 \mu_{\infty} = \frac{G_{\rm eff}}{G_{\rm N}}
              = 1+\frac{2\alpha_{\rm B}^2}{\alpha c_{\rm s}^2}\,, \quad 
 \eta_{\infty} = 1\,.
\end{equation}
Spectra for model 3 are presented in Fig.~\ref{fig:aKaB} for $\alpha_{\rm K,0}=1$ and four non-zero values of 
$\alpha_{\rm B,0}$ between 0.625 and 2.5. We see that the large-scale power increases monotonically with 
$\alpha_{\rm B,0}$ for all the spectra. 
When compared to model 2, we see that the effect of the variation of the braiding is stronger for the lensing potential 
and the temperature anisotropy, but smaller for the linear matter power spectrum.

In model 4 ($\alpha_{\rm K}\neq0$ and $\alpha_{\rm M}\neq 0$), although $\alpha_{\rm B}=0$, both $\mu$ and $\eta$ are 
non-trivial. One has
\begin{equation}
 \mu_{\infty} = \frac{G_{\rm eff}}{G_{\rm N}}
              = \left(1+\frac{2\alpha_{\rm M}^2}{\alpha_{\rm K}c_{\rm s}^2}\right)\frac{1}{\bar{M}^2}\,, \quad 
 \eta_{\infty} = \left(1+\frac{2\alpha_{\rm M}^2}{\alpha_{\rm K}c_{\rm s}^2}\right)^{-1}\,.
\end{equation}
Spectra for model 4 are presented in the left panels of Fig.~\ref{fig:aKaM} for $\alpha_{\rm K,0}=1$ and 
$\alpha_{\rm M,0}=1-4$. On large scales, we see that the CMB temperature power spectrum increases with 
$\alpha_{\rm M,0}$, however the lensing potential power spectrum decreases. 
As explained above, this can be linked to the decrease of the modified gravity parameter $\Sigma$ at late times, 
as shown on the bottom panel of Fig.~\ref{fig:MGparam_aKaM}.

For model 5, only $\alpha_{\rm T}=0$, and therefore neither $\mu$ nor $\eta$ are trivial. However, as before we can 
still obtain simplified expressions for the MG parameters in the small scale regimes (${\rm K}^2\gg 1$) and neglecting 
velocity perturbations:
\begin{align}
 \mu_{\infty} & = \frac{G_{\rm eff}}{G_{\rm N}} 
                = \frac{1+2(\alpha_{\rm B}-\alpha_{\rm M})^2}{\alpha c_{\rm s}^2}\frac{1}{\bar{M}^2}\,,\\
 \eta_{\infty} & = \frac{\alpha c_{\rm s}^2+2\alpha_{\rm B}(\alpha_{\rm B}-\alpha_{\rm M})}
                        {\alpha c_{\rm s}^2+2(\alpha_{\rm B}-\alpha_{\rm M})^2}\,,
\end{align}
with the slip the same as for model 2, as expected. 
Spectra for these models are presented on the left panel of Fig.~\ref{fig:aKaBaM} with $\alpha_{\rm K,0}=1$, and 
several values for the parameters $\alpha_{\rm B,0}$ and $\alpha_{\rm M,0}$.

Finally, model 6 is the most general Horndeski model with all the $\alpha$s different from zero. Therefore, one can not 
simplify further the modified gravity expressions beyond the forms presented in Eqn.~(\ref{eqn:MG_EoS}), or 
Eqs.~(\ref{eqn:mu_eta_Sigma_infty}) in the small scale regime. 
Spectra for these models are presented on the left panel of Fig.~\ref{fig:aKaBaMaT} with 
$\alpha_{\rm K,0}=\alpha_{\rm T,0}=1$, and several values for the parameters $\alpha_{\rm B,0}$ and 
$\alpha_{\rm M,0}$. By comparing Fig.~(\ref{fig:aKaBaM}) with Fig.~(\ref{fig:aKaBaMaT}), where the only difference is 
$\alpha_{\rm T}\neq 0$, we see that $\alpha_{\rm T}$ plays a minor role. The only noticeable difference is in the ISW 
effect for $\ell<5$ of the CMB temperature anisotropy power spectrum (upper left panels).

As previously stated, we are aware that models with $\alpha_{\rm T}\neq 0$ are now excluded by gravitational waves 
observations \citep{LigoVirgo2017,LigoVirgoIntegral2017,LigoVirgo2017a}. We decided to present these numerical results 
to show the robustness of our numerical implementation.

\subsection{The time evolution of the modified gravity parameters}
The commonly used modified gravity parameters, $\mu$ (or the effective modified gravitational constant $G_{\rm eff}$), 
$\eta$ (the slip) and $\Sigma$ (or $G_{\rm light}$), as well as the dark sector sound speed $c_{\rm s}^2$ and the 
effective Planck mass $\bar{M}^2$, are particularly useful to study the phenomenology of perturbations in dark sector 
theories. These are also the parameters that will be subject to observational constraints by forthcoming 
large-scale-structure and galaxy surveys, e.g.\ LSST \citep{LSST2009}, \textit{Euclid} \citep{Euclid2018}.

One can understand why these parameters are useful by considering the equation of motion for the gauge invariant matter 
perturbation. Indeed, following the same procedure as in \cite{Battye2018a} (see Eqn.~(19a)), one can obtain an 
evolution equation for the matter density perturbations sourced by the dark sector density perturbations (under the 
assumptions of negligible velocity perturbations that generally corresponds to the small scale regime, 
${\rm K}^2\gg 1$, and is valid for the modes of observational interest). Then with $\mu(=\mu_Y=G_{\rm eff}/G_{\rm N})$ 
as defined in Eqn.~(\ref{eqn:muY_Delta}) one can write:
\begin{equation}
 \Delta_{\rm m}^{\prime\prime} + (2-\epsilon_H)\Delta_{\rm m}^{\prime} - 
 \frac{3}{2}\frac{G_{\rm eff}}{G_{\rm N}}\Omega_{\rm m}\Delta_{\rm m} = 0\,,
\end{equation}
where in each class of Horndeski models the effective gravitational constant takes the forms presented in the previous 
section. 
This equation has been used to study the dark sector phenomenology in a number of works, see e.g.\ \cite{Battye2018a} 
for $f(R)$ gravity and \cite{Kimura2011,Pace2014,NazariPooya2016} for other classes of Horndeski theories. 
Here we present the redshift evolution of the modified gravity parameters in Horndeski theories. For simplicity we 
present the behaviour of these function for sub-horizon modes, ${\rm K}^2\gg 1$.

\begin{figure*}[tb]
 \centering
 \includegraphics[width=6.5cm,angle=0]{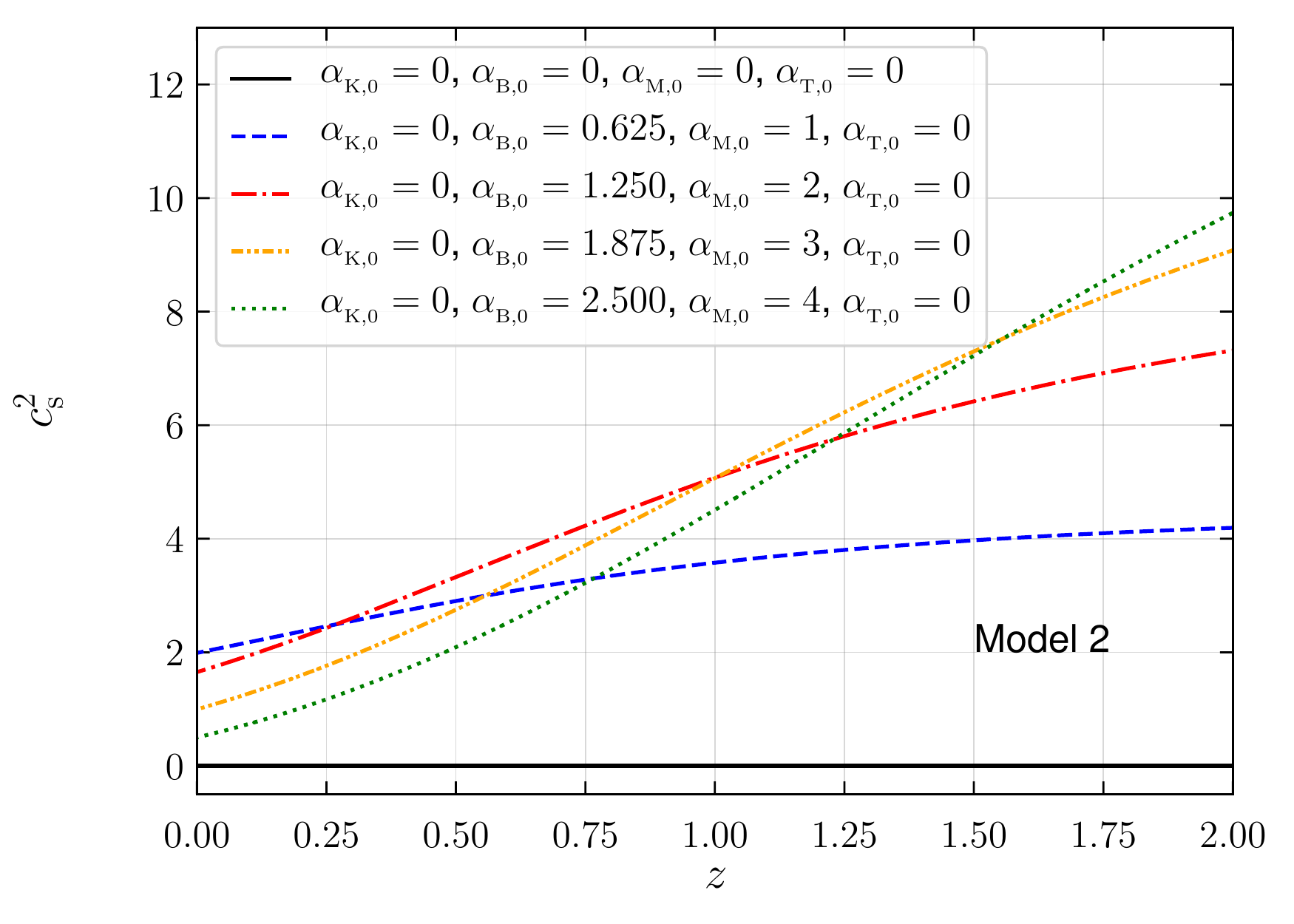}
 \includegraphics[width=6.5cm,angle=0]{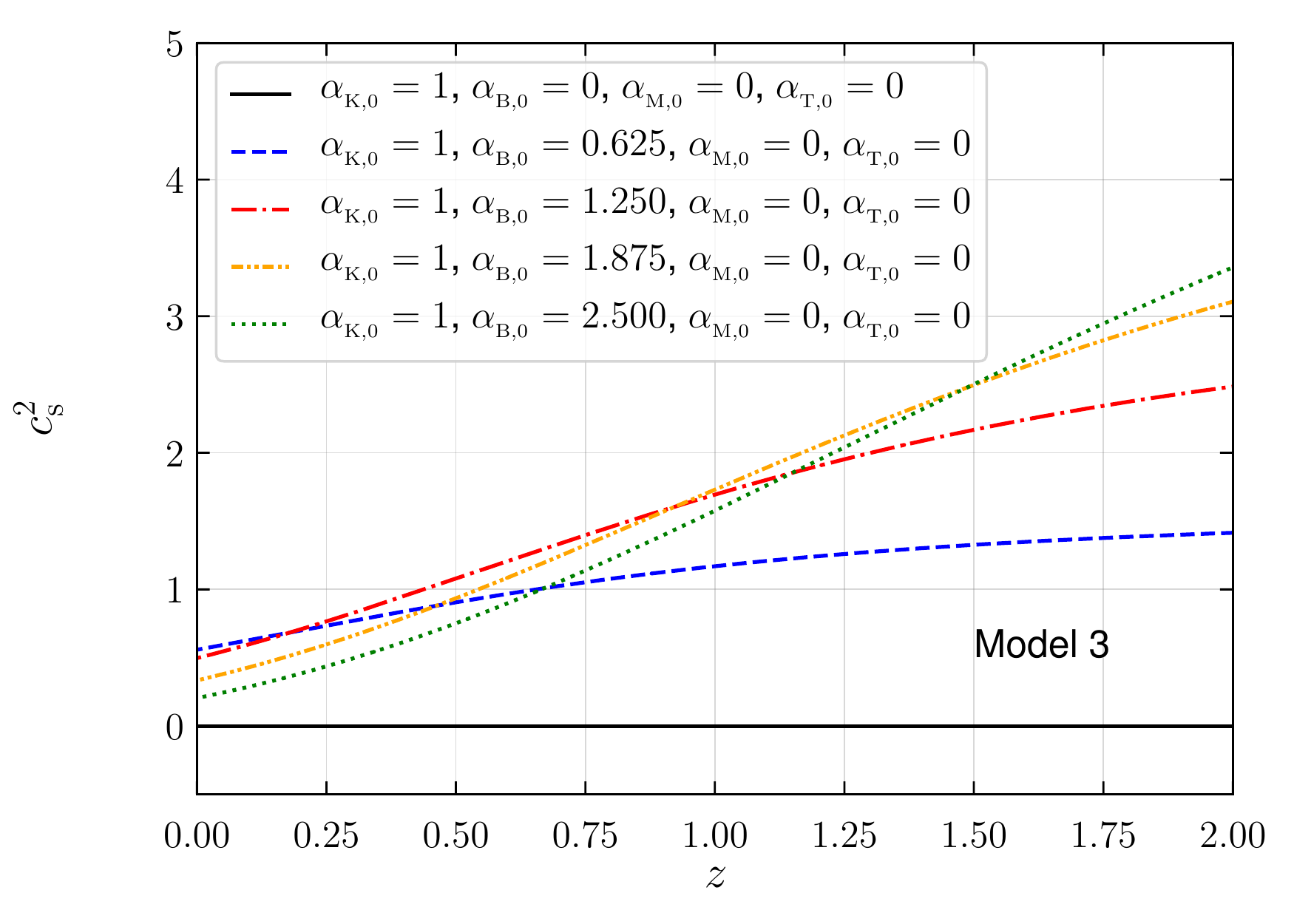}
 \includegraphics[width=6.5cm,angle=0]{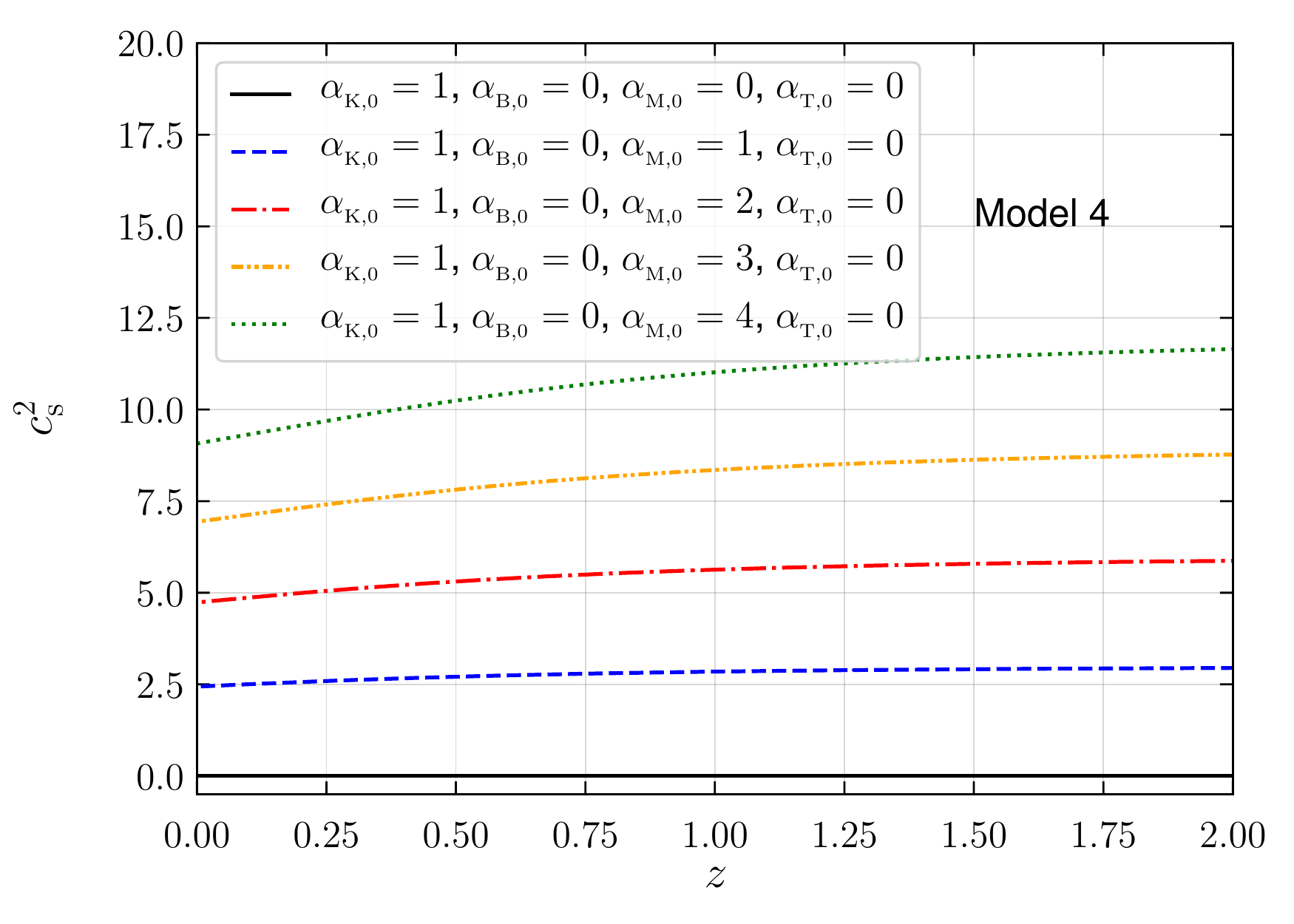}
 \includegraphics[width=6.5cm,angle=0]{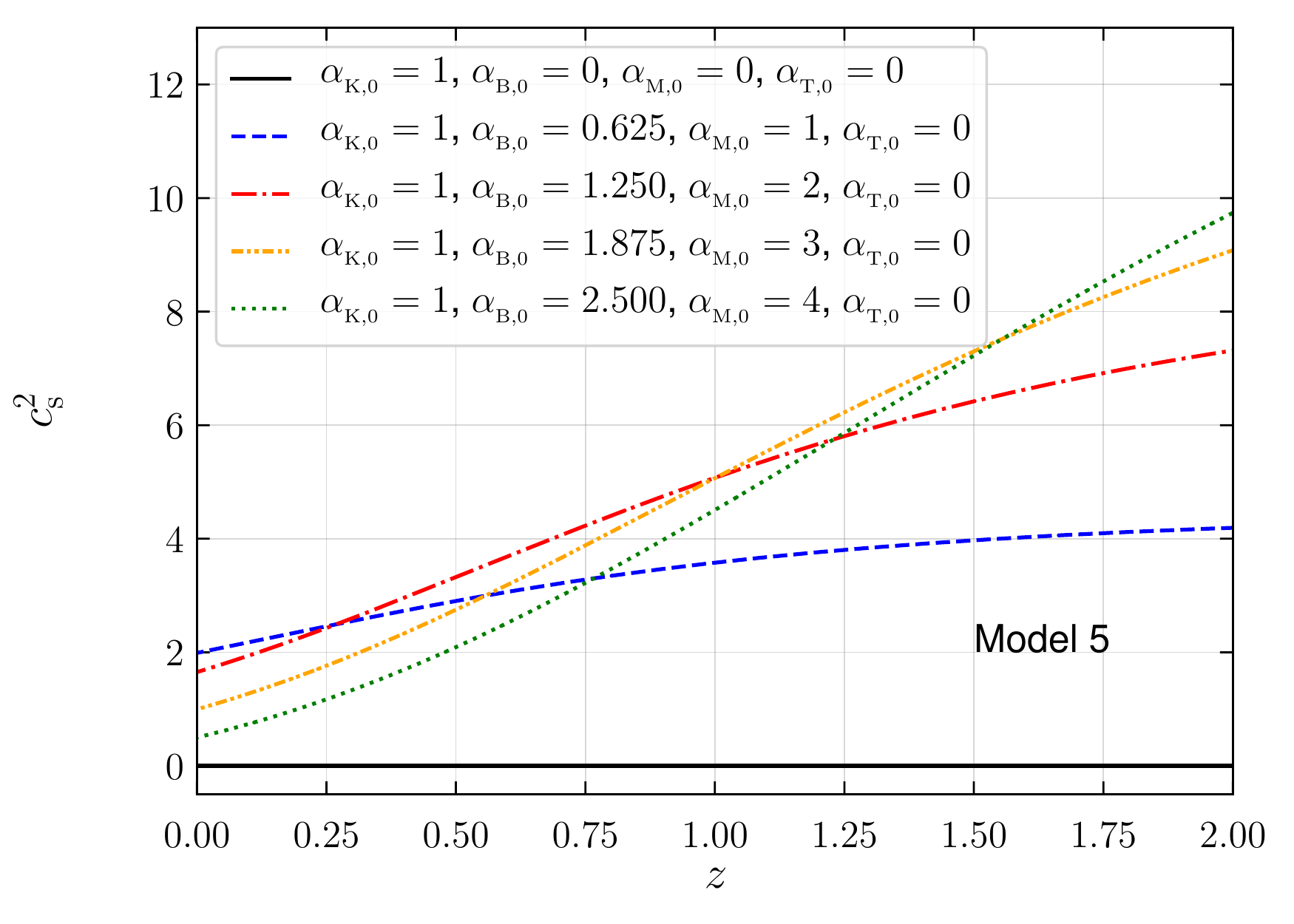}
 \includegraphics[width=6.5cm,angle=0]{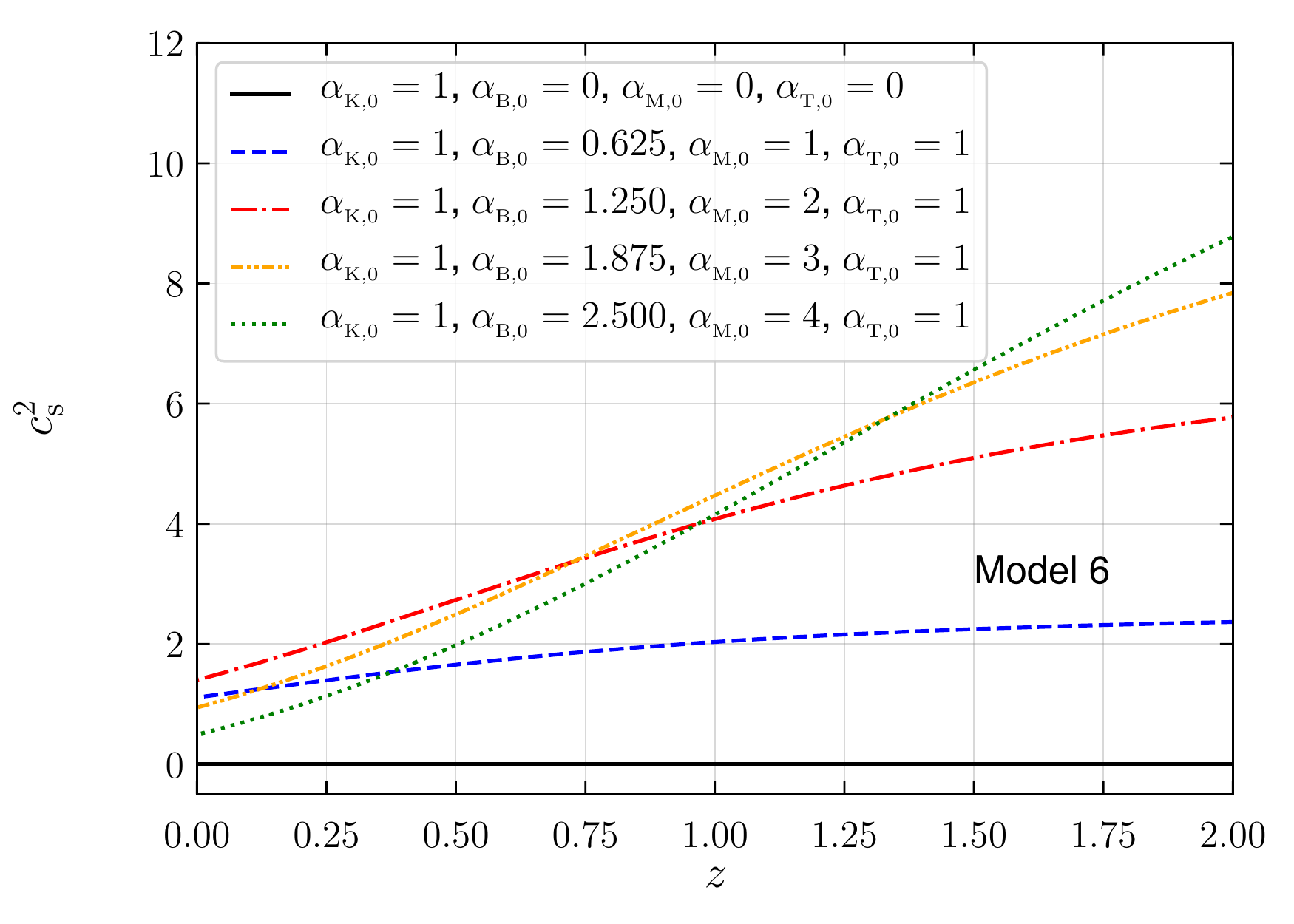}
 \includegraphics[width=6.5cm,angle=0]{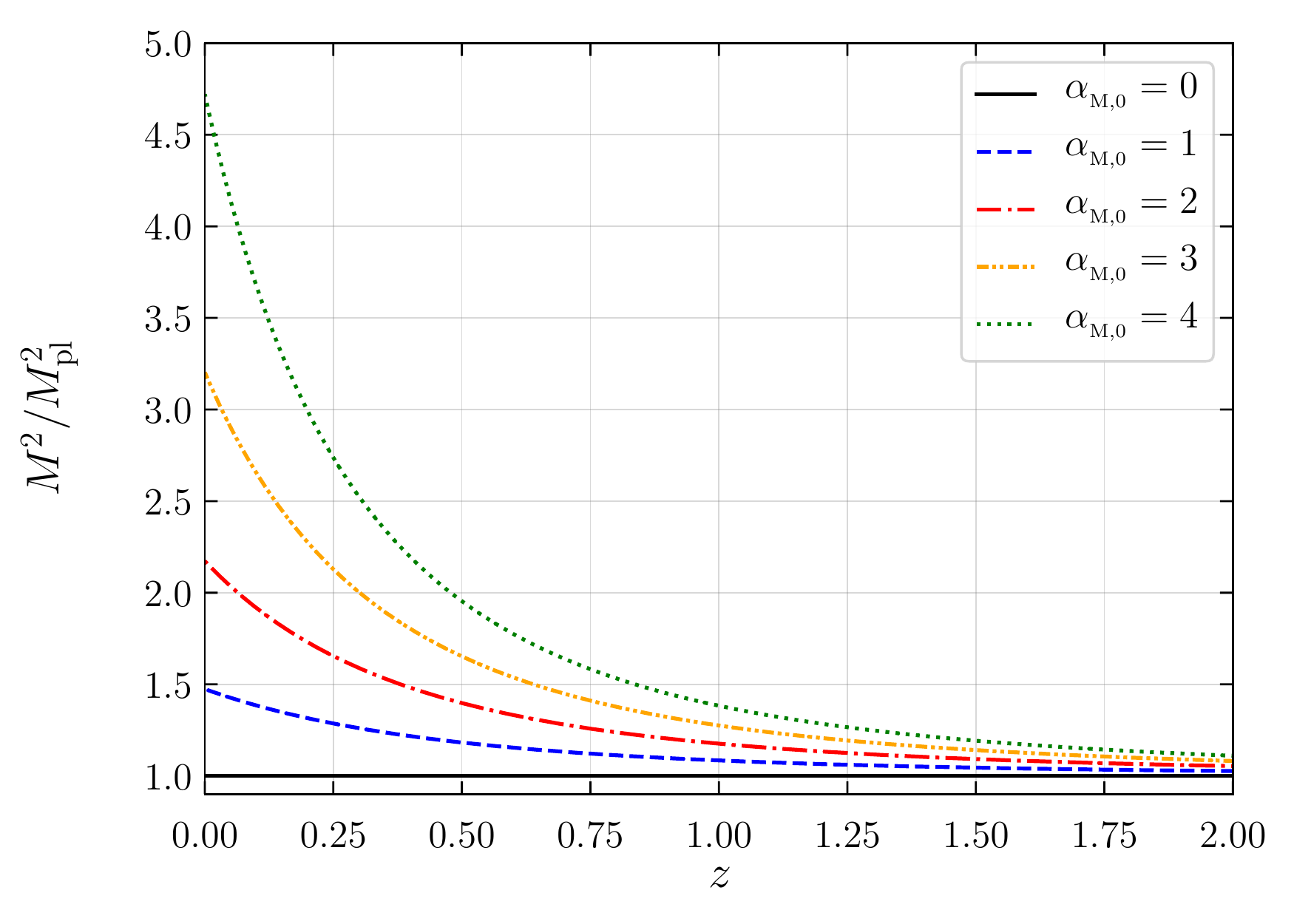}
 \cprotect\caption[justified]{Evolution of the perturbation sound speed $c_{\rm s}^2$ for models 2--6 and for 
 the effective Planck mass $\bar{M}^2=M^2/M_{\rm pl}^2$ for models 2 and 4--6. From left to right, in the top line we 
 present the effective sound speed for models 2 and 3, in the middle line for models 4 and 5 and in the bottom line 
 for model 6 and the effective Planck mass. 
 Line styles and colours are as in Figs.~\ref{fig:aBaM}--\ref{fig:aKaBaMaT} for models 2--6, respectively.}
 \label{fig:cs2_M2}
\end{figure*}

In Fig.~\ref{fig:cs2_M2} we present the sound speed of dark sector perturbations as a function of redshift for models 
2--6 and for different sets of values for $\alpha_{\rm x,0}$. 
The sound speed is important because it enters both the expressions of $G_{\rm eff}$ and the slip $\eta$, see 
Eqs.~(\ref{eqn:muY_infty}) and (\ref{eqn:eta_infty}). In particular we see that when $c_{\rm s}^2$ becomes large, 
$G_{\rm eff}$ and $\eta$ reduce to their $\Lambda$CDM limit and therefore one does not expect a \textit{modified 
gravity} behaviour of the matter and metric perturbations. We see in Fig.~\ref{fig:cs2_M2} that the sound speed is 
typically large at early times. This is because the sound speed $c_{\rm s}^2$ is proportional to $1/\alpha$ and 
$\alpha\propto\Omega_{\rm ds}$ in these models.

We note that for $k$-essence-like models (Model 1), the sound speed is $c_{\rm s}^2=3(1+w_{\rm ds})/\alpha_{\rm K}$ 
which becomes zero for the $\Lambda$CDM background that we are assuming. This corresponds to the solid black line on 
Fig.~\ref{fig:cs2_M2}. In this case, the correct procedure to obtain the modified gravity parameters is to first 
evaluate their expressions by equating $\alpha_{\rm x}=0$ and then by setting $c_{\rm s}^2$. Therefore one obtains 
$\mu=\eta=\Sigma=1$.

In the bottom right panel of Fig.~\ref{fig:cs2_M2} we also show the evolution with respect to redshift of the effective 
Planck mass $\bar{M}^2$. Since $\alpha_{\rm M}\equiv d\ln{\bar{M}^2}/d\ln{a}$, $\bar{M}^2$ is larger at late time for 
larger values of $\alpha_{\rm M,0}$.

The effective gravitational constant $\mu_Y$ affects clustering and peculiar motion of galaxies while $\Sigma$ affects 
light geodesics. The first is constrained by galaxy clustering and redshift-space distortions 
\citep{Song2011,Simpson2013,Asaba2013}, while the latter by the CMB, weak lensing and galaxy number counts 
\citep{Asaba2013,Moessner1998}. The parameters $\mu_Y$ and $\eta$ correspond to the parameters $Q$ and $R$ introduced 
in \citep{Bean2010}, respectively. We refer the reader to \cite{Bean2010} for a comparison of different notations 
used in the literature.

\begin{figure}[!t]
 \centering
 \includegraphics[width=6.5cm,angle=0]{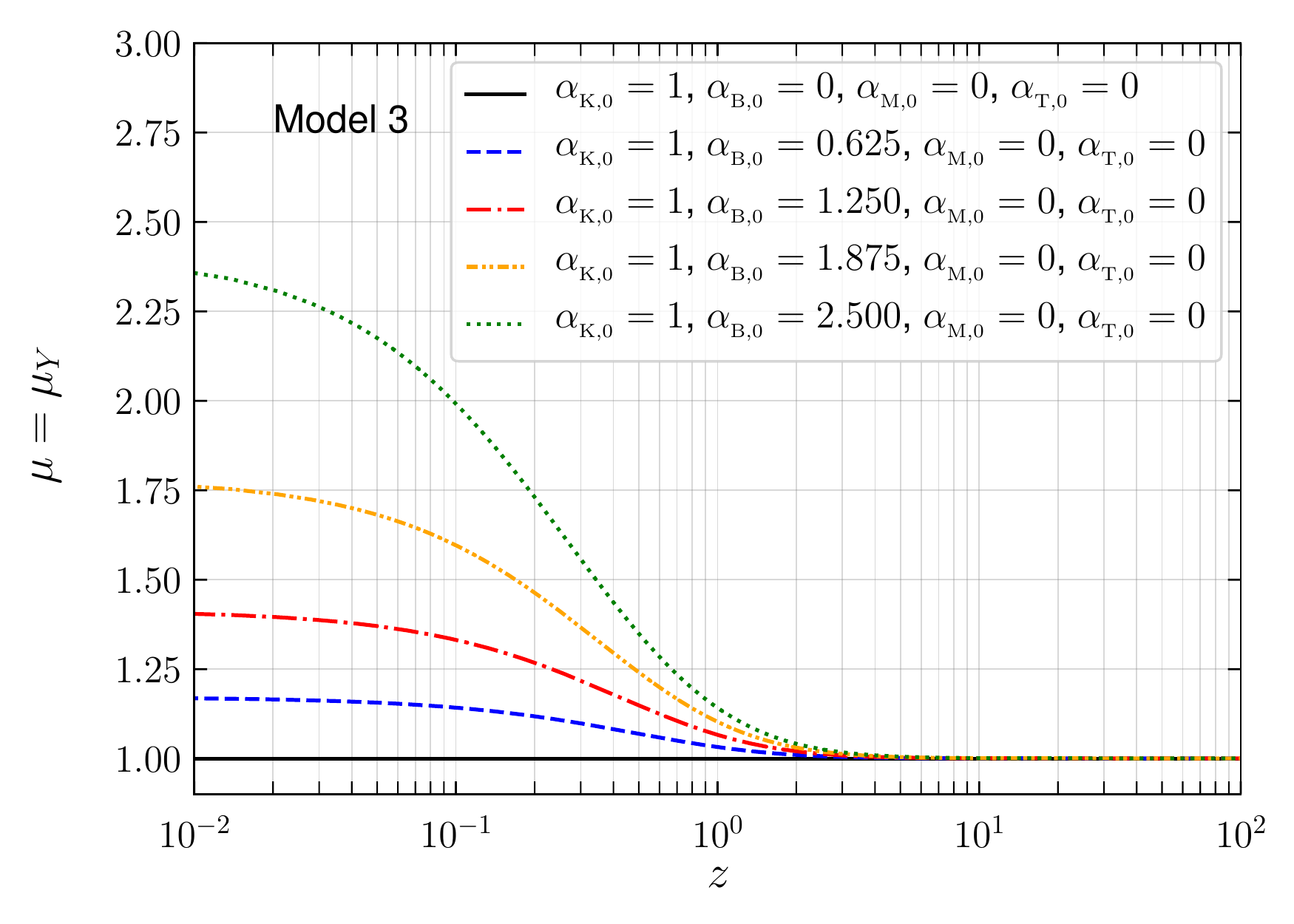}
 \caption[justified]{Evolution of the effective gravitational constant $\mu$ parameter for ${\rm K}^2\gg 1$ as a 
 function of redshift $z$ for different values of the normalisation $\alpha_{\rm B,0}$ for model 3. Line styles and 
 colours are as in Fig.~\ref{fig:aKaB}. For this model, $\mu_{Z}=\mu_{Y}=\Sigma$ as $\eta=1$.}
 \label{fig:MGparam_aKaB}
\end{figure}

\begin{figure}[!t]
 \centering
 \includegraphics[width=6.5cm,angle=0]{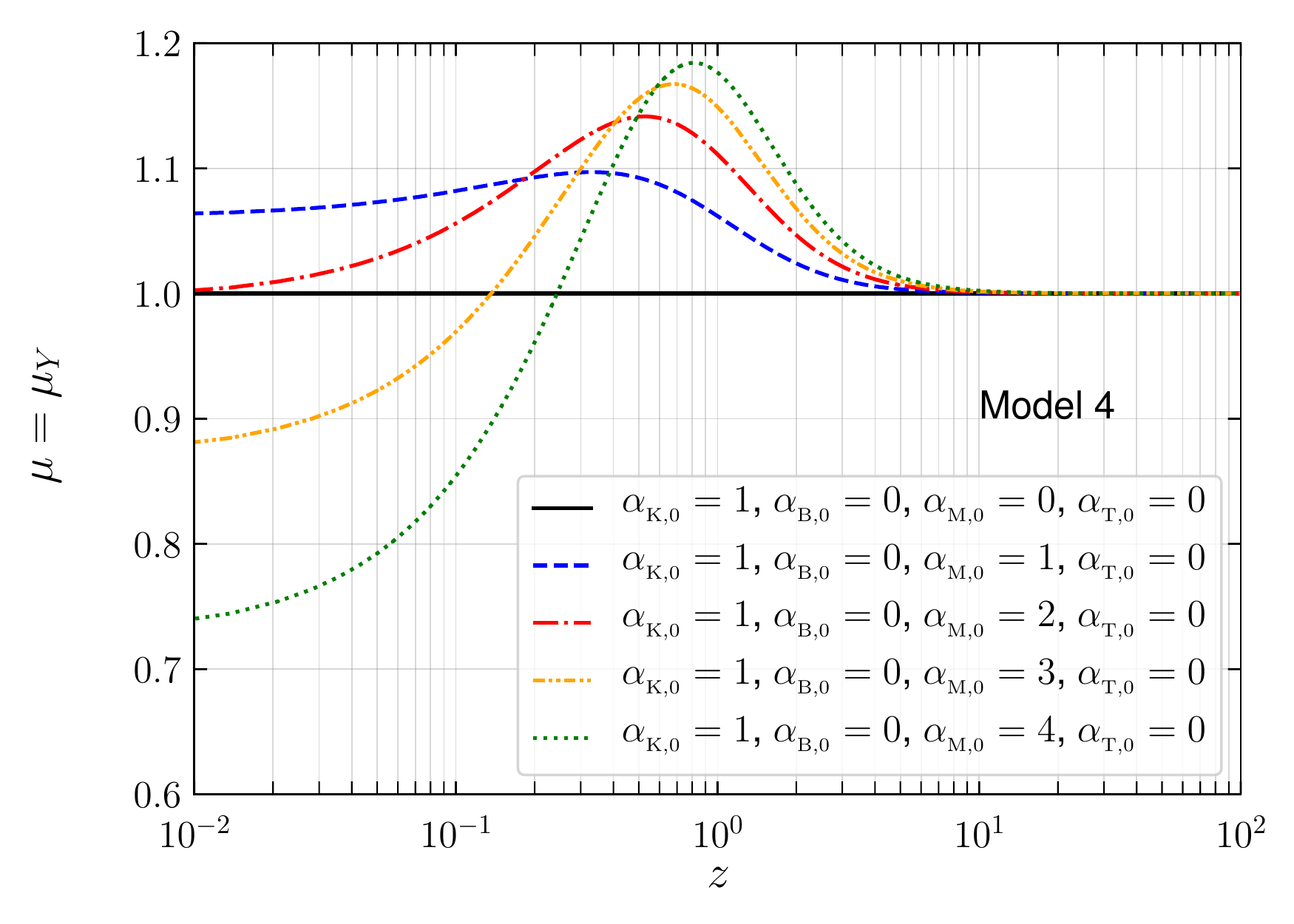}
 \includegraphics[width=6.5cm,angle=0]{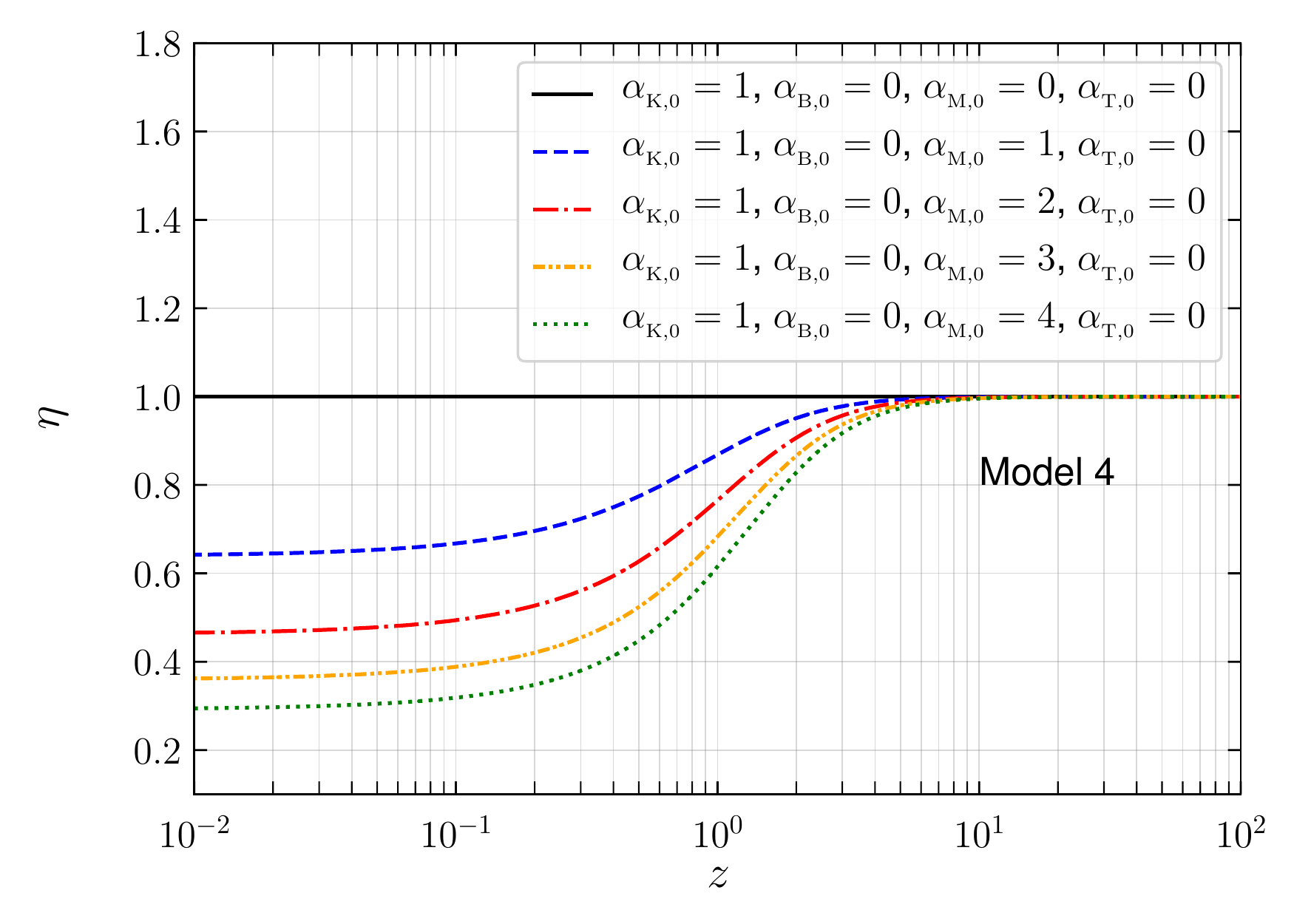}
 \includegraphics[width=6.5cm,angle=0]{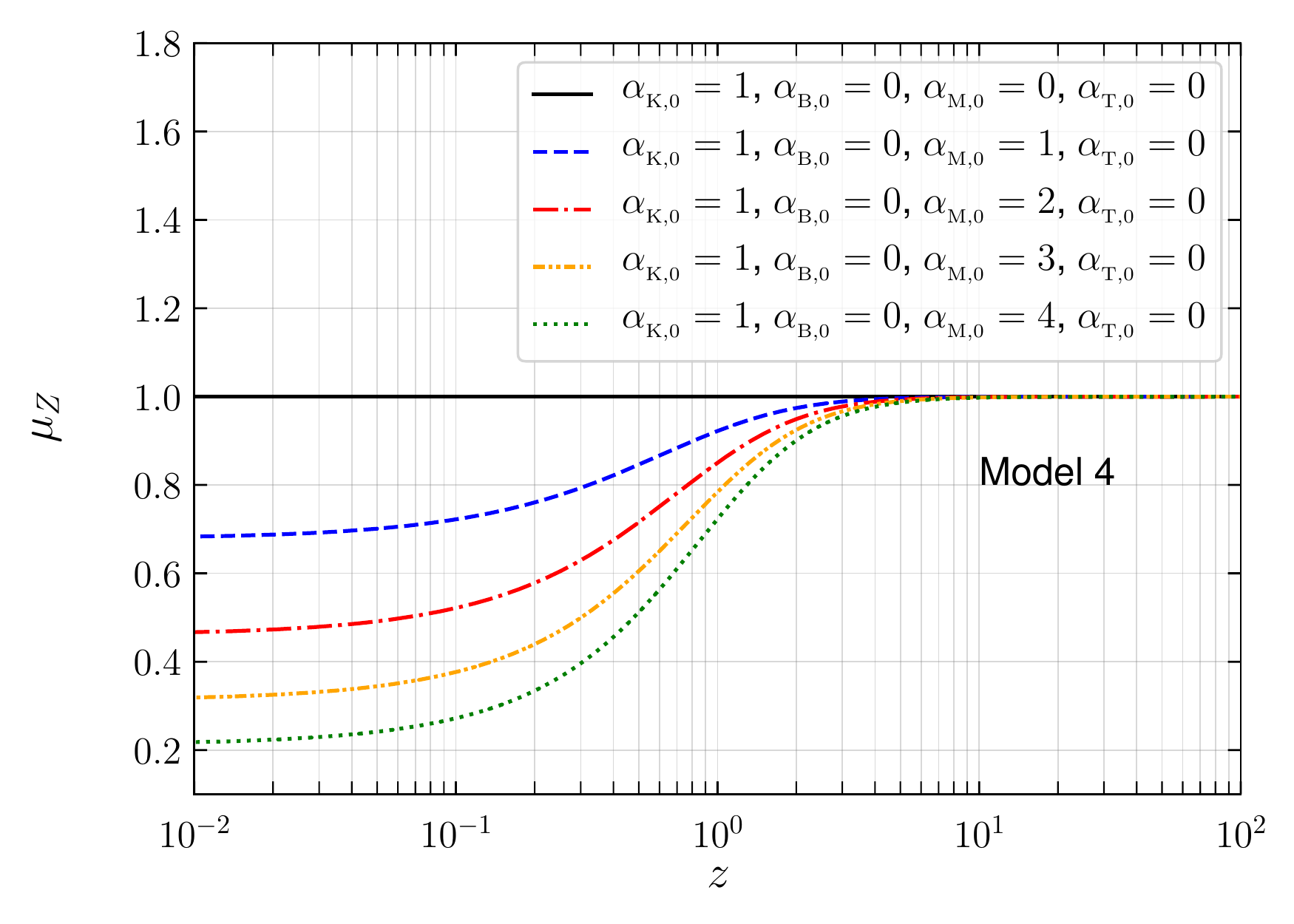}
 \includegraphics[width=6.5cm,angle=0]{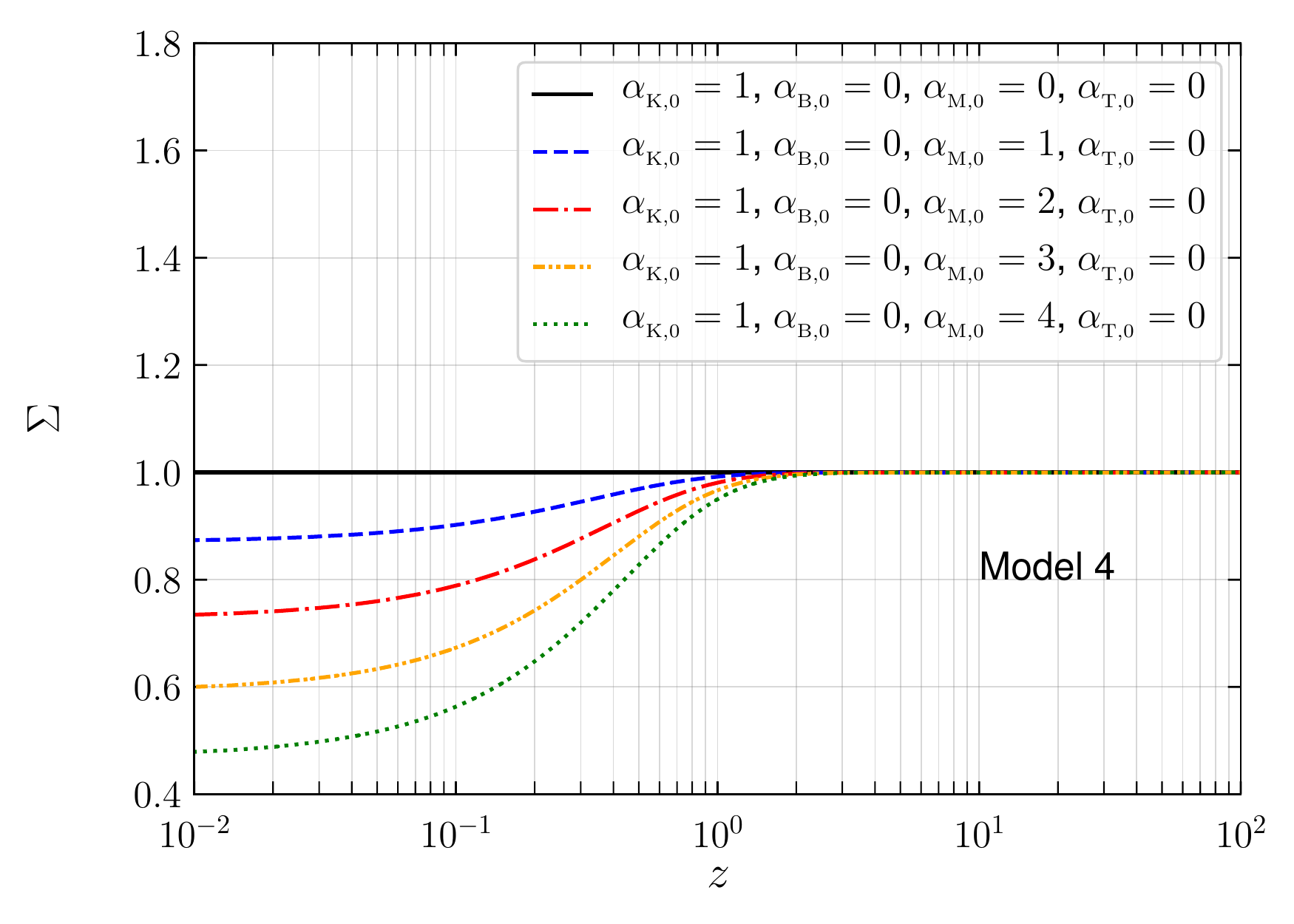}
 \caption[justified]{Evolution of the modified gravity parameters for ${\rm K}^2\gg 1$ as a function of redshift $z$ 
 for the different 
 values of $\alpha_{\rm M,0}$ for model 4. From top left in clockwise order we show $\mu$, $\eta$, $\Sigma$ and 
 $\mu_Z$, respectively. Line styles and colours are as in Fig.~\ref{fig:aKaM}.}
 \label{fig:MGparam_aKaM}
\end{figure}

\begin{figure}[!t]
 \centering
 \includegraphics[width=6.5cm,angle=0]{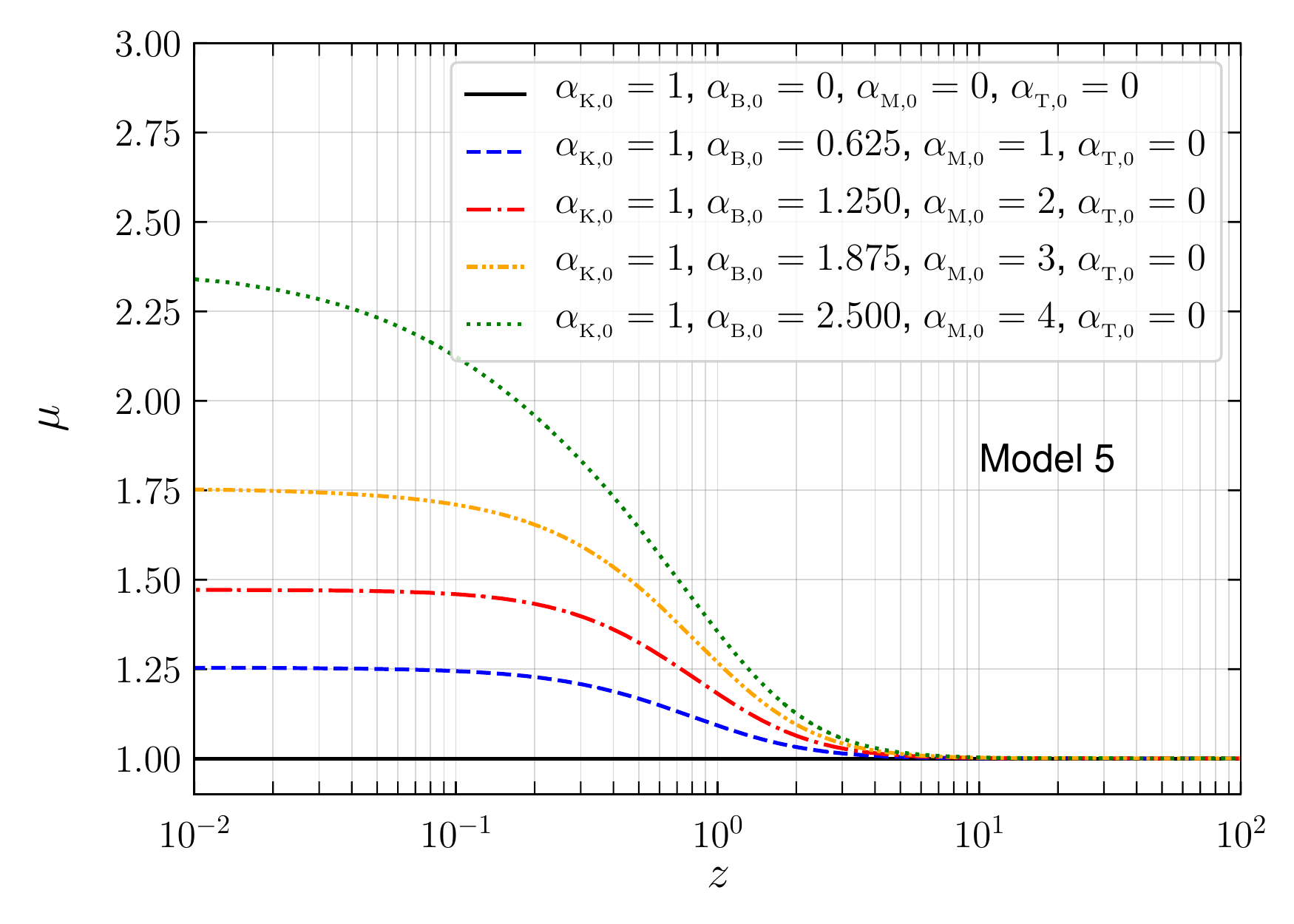}
 \includegraphics[width=6.5cm,angle=0]{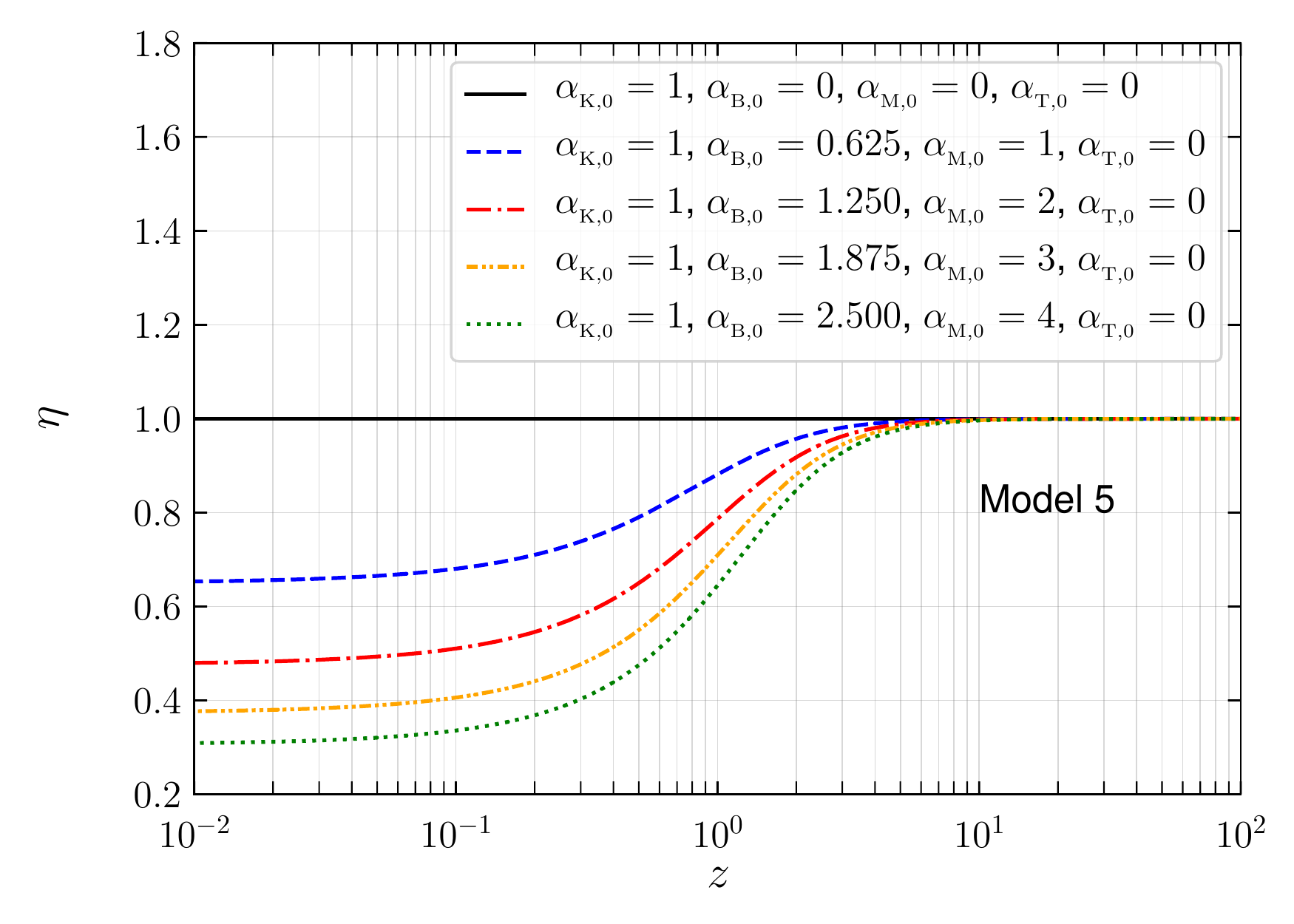}
 \includegraphics[width=6.5cm,angle=0]{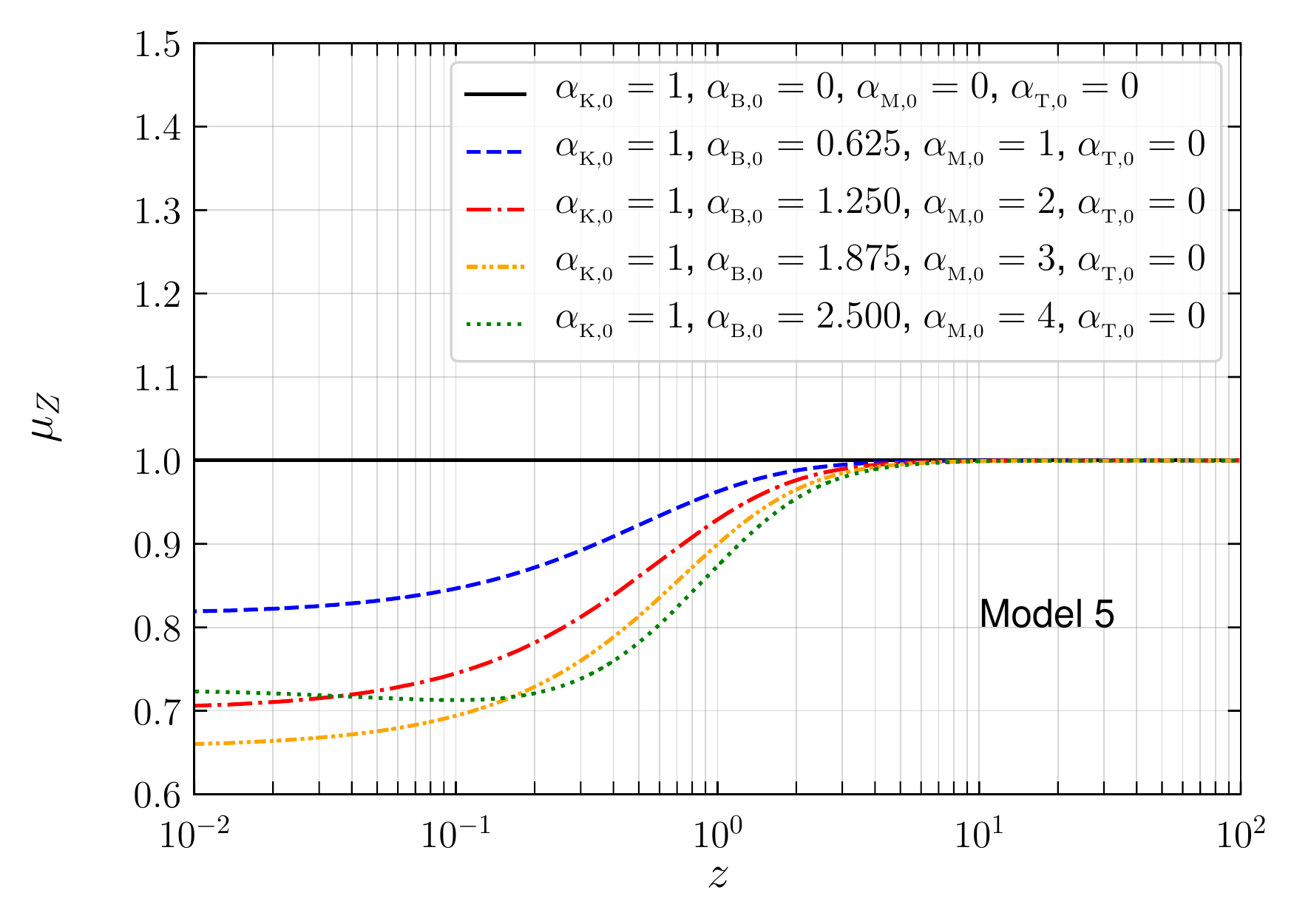}
 \includegraphics[width=6.5cm,angle=0]{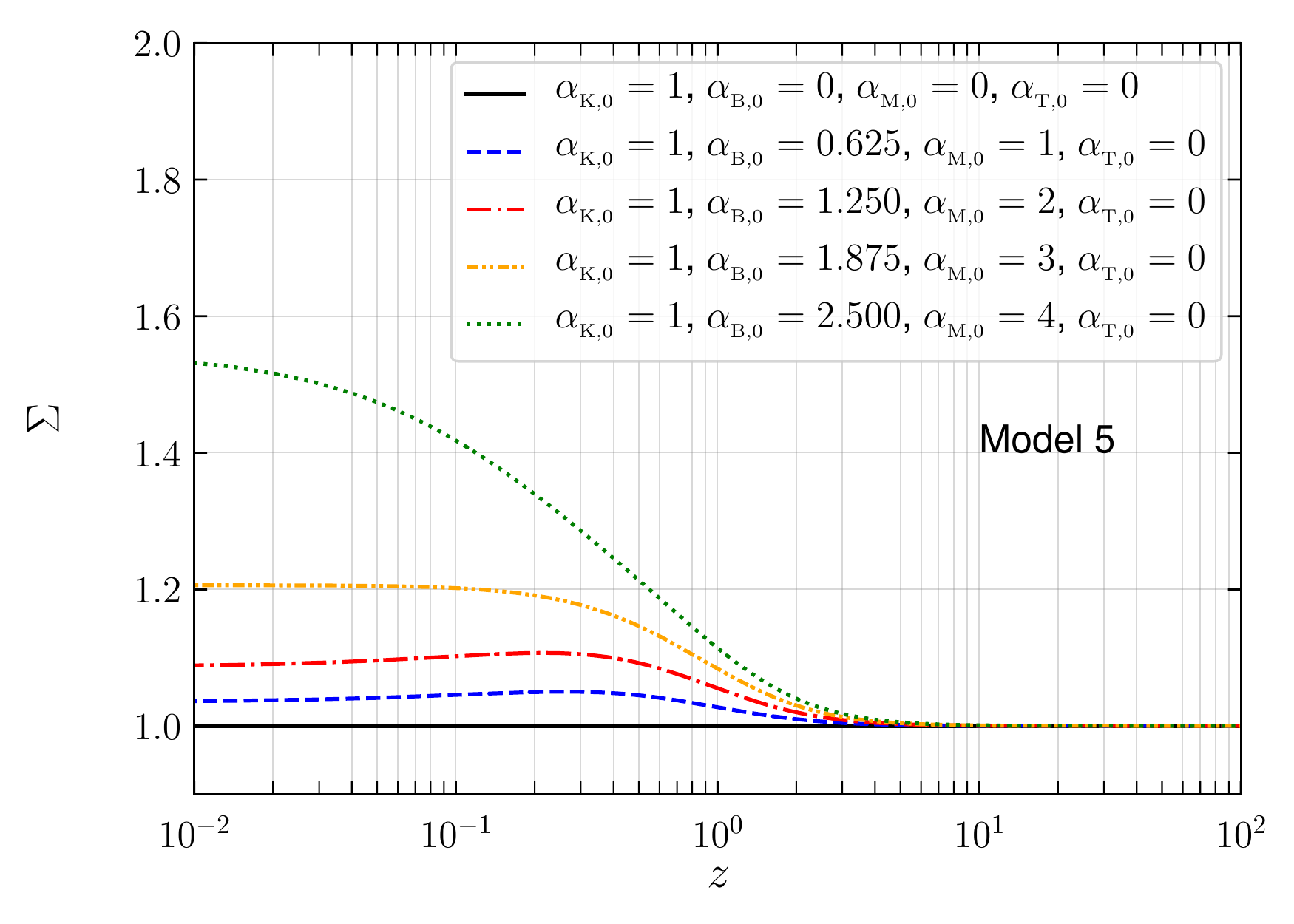}
 \caption[justified]{Evolution of the modified gravity parameters for ${\rm K}^2\gg 1$ as a function of redshift $z$ 
 for the different 
 values of $\alpha_{\rm B,0}$ and $\alpha_{\rm M,0}$ for model 5. From top left in clockwise order we show $\mu$, 
 $\eta$, $\Sigma$ and $\mu_Z$, respectively. Line styles and colours are as in Fig.~\ref{fig:aKaBaM}.}
 \label{fig:MGparam_aKaBaM}
\end{figure}

\begin{figure}[!t]
 \centering
 \includegraphics[width=6.5cm,angle=0]{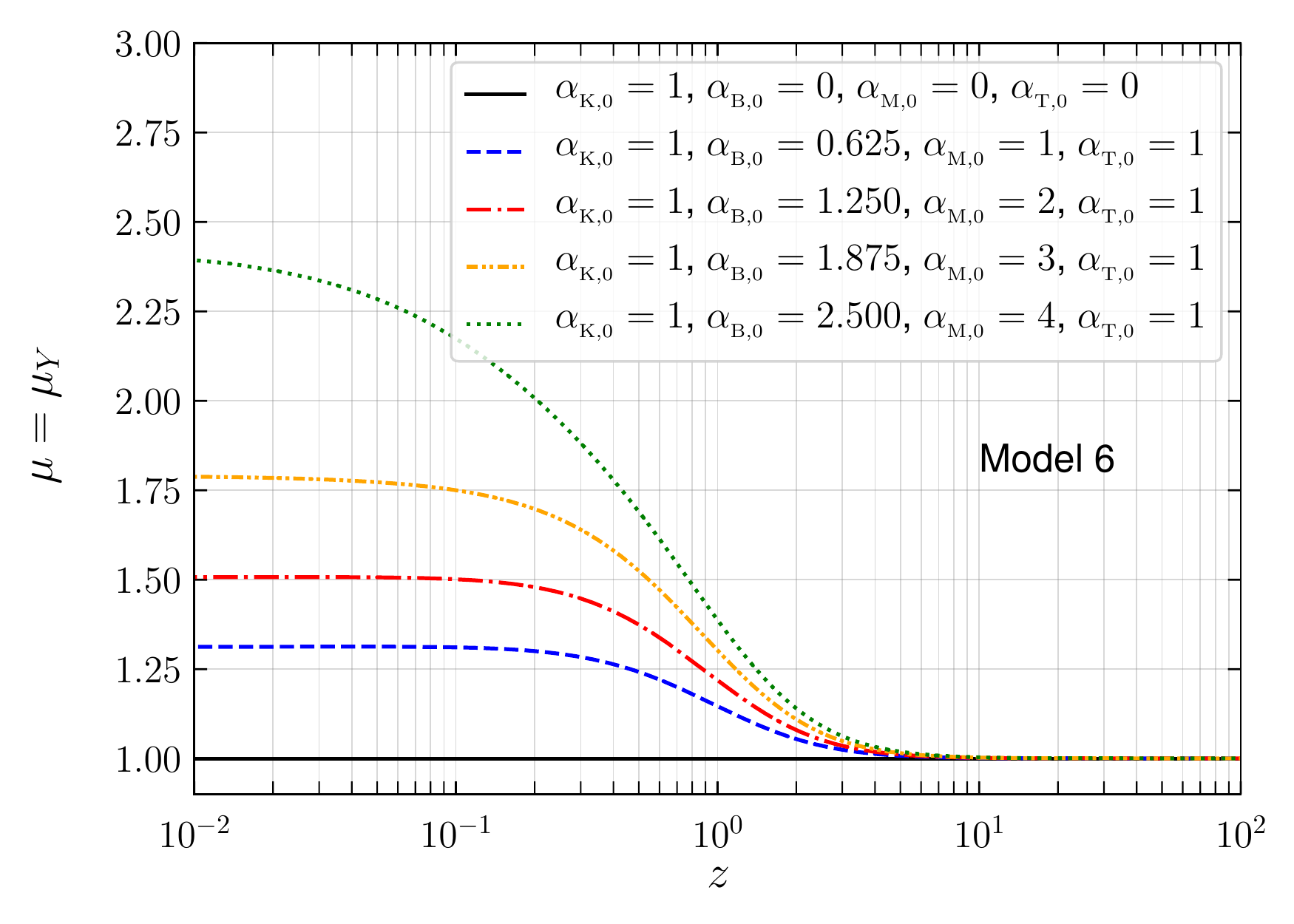}
 \includegraphics[width=6.5cm,angle=0]{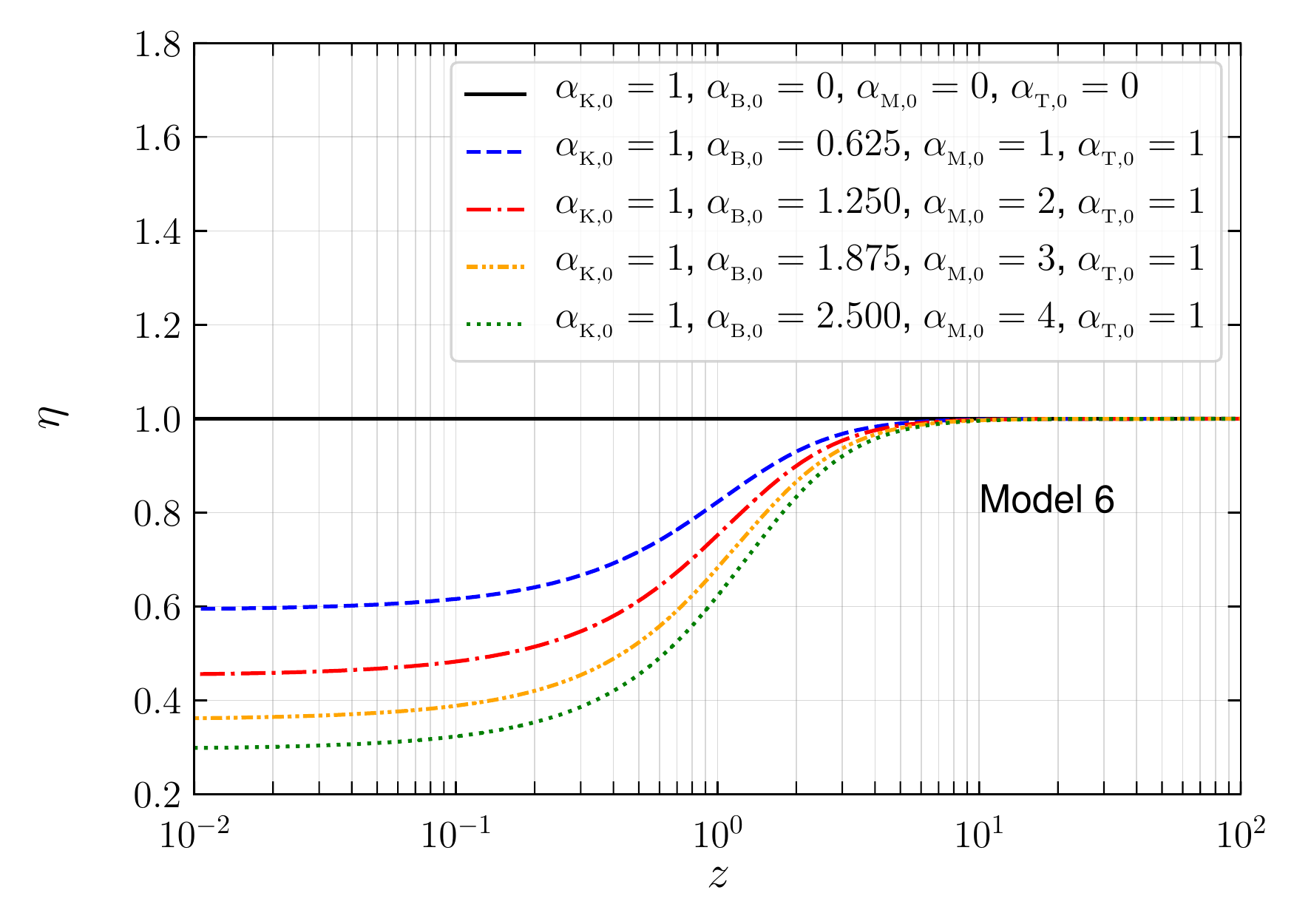}
 \includegraphics[width=6.5cm,angle=0]{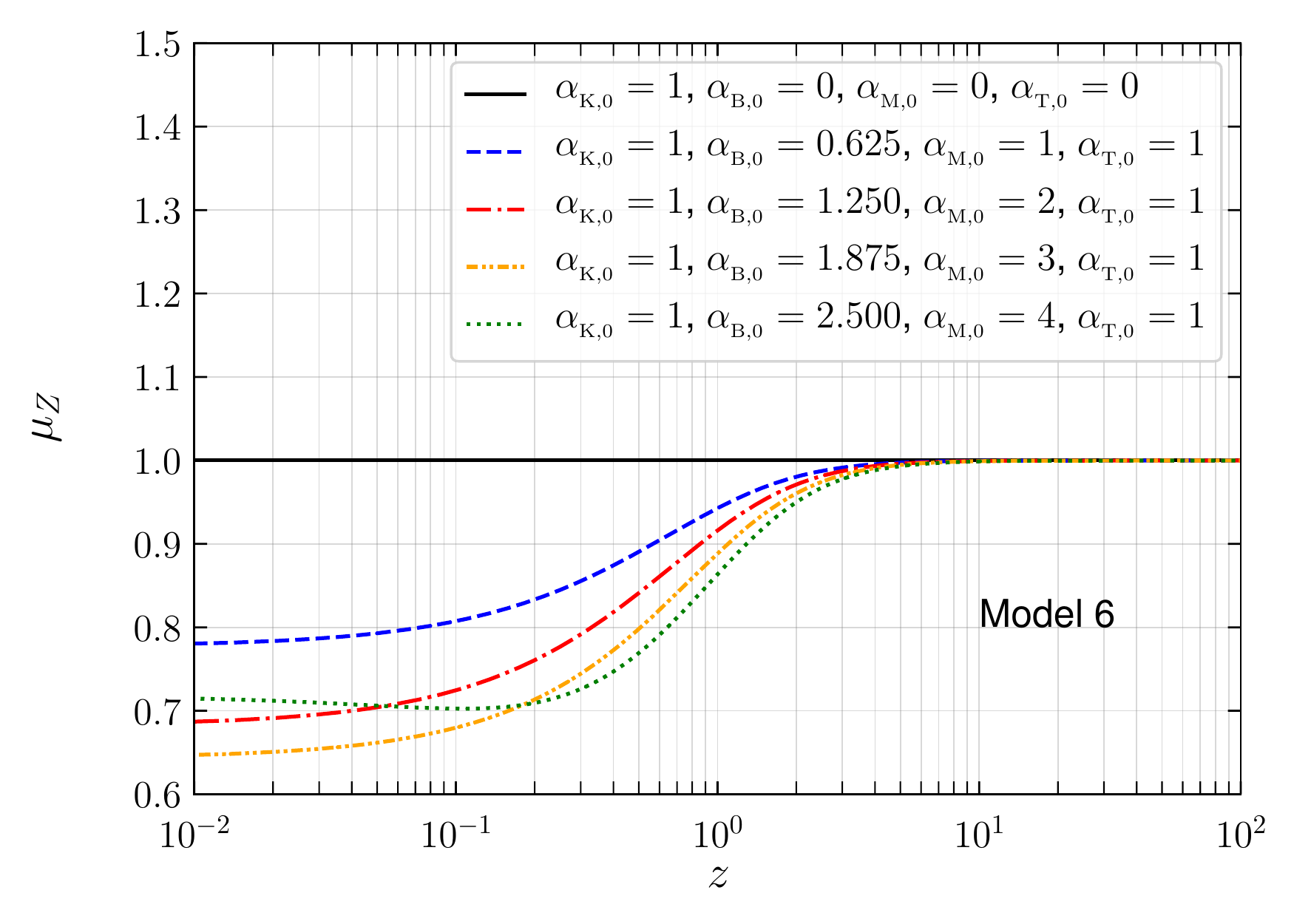}
 \includegraphics[width=6.5cm,angle=0]{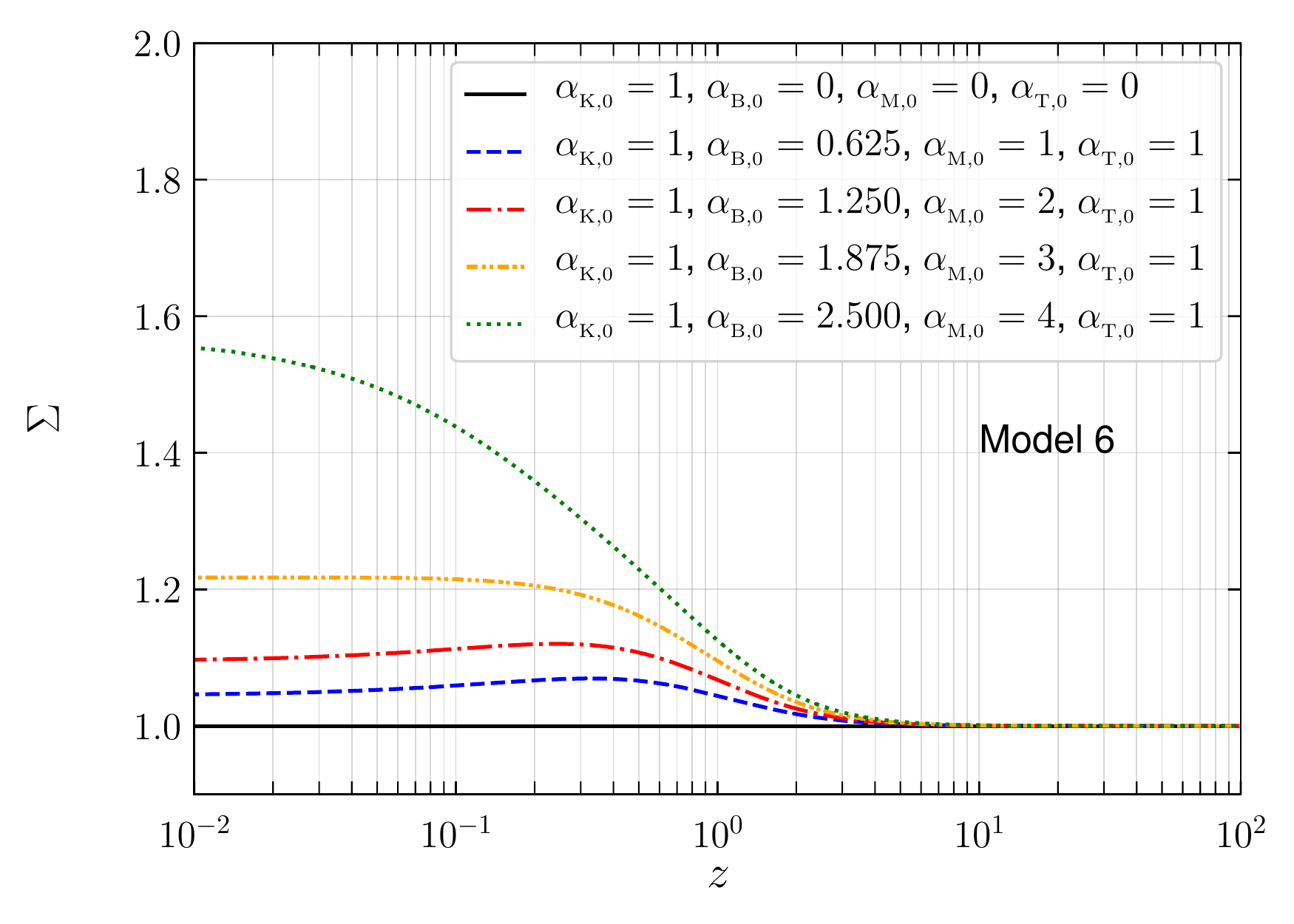}
 \caption[justified]{Evolution of the modified gravity parameters for ${\rm K}^2\gg 1$ as a function of redshift $z$ 
 for the different values of $\alpha_{\rm B,0}$ and $\alpha_{\rm M,0}$ at fixed $\alpha_{\rm T,0}=1$ for model 6. 
 From top left in clockwise order we show $\mu$, $\eta$, $\Sigma$ and $\mu_Z$, respectively. Line styles and colours 
 are as in Fig.~\ref{fig:aKaBaMaT}.}
 \label{fig:MGparam_aKaBaMaT}
\end{figure}

In Figs.~\ref{fig:MGparam_aKaB}--\ref{fig:MGparam_aKaBaMaT} we present the evolution of the modified gravity 
parameters $\mu_{Z}$, $\mu_{Y}$, $\eta$ and $\Sigma$ for the models 3--6. For simplicity we will consider only the 
values for ${\rm K}^2\gg1$ and therefore we will not consider the scale dependence of this functions. Because of this 
choice, model 2 evolves identically to model 5 and we will not consider it further. We compared the full numerical 
solution obtained with our code \verb|EoS_class| with the analytical expectation and found that they agree for 
$k\gtrsim 2\times 10^{-3}~h{\rm Mpc}^{-1}$. On larger scales, we see a departure of the analytical result from the 
full numerical solution, as the condition ${\rm K}^2\gg 1$ is violated. We will leave a detailed comparison of the 
analytical expressions with the exact numerical results in a forthcoming paper.

In Fig.~\ref{fig:MGparam_aKaB} we present the time evolution of $\mu$ for model 3. In this class of models, the other 
modified gravity parameters are either trivial or identical to $\mu$. That is because $w_{\rm ds}\Pi_{\rm ds}=0$, 
i.e.\ $\eta=1$. We understand that $\mu$ increases at late times with increasing $\alpha_{\rm B}$ by looking at 
Eqn.~(\ref{eqn:mu3}) and by comparing with Fig.~\ref{fig:aKaB} we see that this increase of $\mu$ is associated with an 
amplification of matter clustering and ISW.

In Fig.~\ref{fig:MGparam_aKaM} we present the time evolution of the modified gravity parameters for model 4 
($\mu_Z$, $\mu$, $\eta$ and $\Sigma$), for $\alpha_{\rm K,0}=1$ and different values of $\alpha_{\rm M,0}$. The 
functions $\mu_Z$, $\eta$ and $\Sigma$ become smaller than one and decrease at late times. The effect is more 
pronounced for larger values of $\alpha_{\rm M,0}$. For $\mu$, the effect is also more pronounced for larger values of 
$\alpha_{\rm M,0}$. Nevertheless the effect has a non-trivial time-dependence. In particular for all the values of 
$\alpha_{\rm M,0}$ considered, $\mu$ peaks at around $z\approx 0.7$, between 10 and 20\% above the corresponding 
$\Lambda$CDM value ($\mu=1$). This different behaviour can be understood looking at the formula in 
Table~\ref{tab:DEphenomenology_infinity}.

Since $\alpha_{\rm T}$ plays a very minor role (at least for the value we considered, see discussion in 
Sect.~\ref{sect:numerical_results}), we discuss models 5 and 6 together. These models are the most general with 
$\alpha_{\rm K}$, $\alpha_{\rm B}$ and $\alpha_{\rm M}$ different from zero. The modified gravity functions for models 
5 and 6 are shown in Figs.~\ref{fig:MGparam_aKaBaM} and \ref{fig:MGparam_aKaBaMaT}, respectively. We show the modified 
gravity functions for a different set of values of the $\alpha$s, with $\alpha_{\rm K,0}=1$ and four different values 
of $\alpha_{\rm B,0}$ between 0.625 and 2.5 and four different values of $\alpha_{\rm M,0}$ between 1 and 4. We see 
that $\mu$ and $\Sigma$ depart from unity and increase at late times. Meanwhile, $\mu_Z$ and $\eta$ decrease in a 
similar way as model 4. Hence, comparing with model 4, we can say that $\alpha_{\rm M}$ plays a more important role 
than $\alpha_{\rm B}$ for $\Sigma$ and $\mu_Z$. Moreover, we can say that $\alpha_{\rm B}$ plays a more important role 
than $\alpha_{\rm M}$ for $\mu$ and $\Sigma$. Again, this can be understood looking at the expressions in 
Table~\ref{tab:DEphenomenology_infinity}. In particular, by noticing the effect of the effective Planck mass 
$\bar{M}^2$ on these functions. The increase of $\Sigma$ can be linked with the increase of the lensing potential in 
Fig.~\ref{fig:aKaBaM} and \ref{fig:aKaBaMaT}.

\subsection{The growth index}
The growth index is defined as $\gamma\equiv\ln{f}/\ln{\Omega_{\rm m}}$ where 
$f=\Delta_{\rm m}^{\prime}/\Delta_{\rm m}$ is the growth rate \citep[see][]{Peebles1980}. 
These quantities (or more precisely $\gamma$ and $f\sigma_8$) have a clear $\Lambda$CDM expectation and have already 
been measured from galaxy surveys and redshift space distortions (RSD) experiment \citep{GilMarin2018}. For instance, 
in $\Lambda$CDM one expects a scale-independent $f\sigma_8$ and a constant growth index $\gamma=6/11$. 
With future surveys we will be able to determine with high significance whether these quantities match or depart from 
their $\Lambda$CDM expectation. It is therefore crucial to have clear understanding and predictions of the properties 
of the growth index and growth rate in modified gravity models and this can be done using expressions deduced in 
previous sections.

Here, we present the time evolution of the growth index in Horndeski theories. 
Taking the derivative of the growth rate, the equation describing the time evolution of the growth index is, e.g.\ 
\citep{Pogosian2010,Battye2018a}
\begin{equation}\label{eqn:gamma}
 \gamma^{\prime} + \frac{3w_{\rm ds}\Omega_{\rm ds}}{\ln{\Omega_{\rm m}}}\gamma + 
 \frac{\Omega_{\rm m}^{\gamma}}{\ln{\Omega_{\rm m}}} - 
 \frac{3}{2}\frac{\Omega_{\rm m}^{1-\gamma}}{\ln{\Omega_{\rm m}}}\frac{G_{\rm eff}}{G_{\rm N}} = 
 \frac{3w_{\rm ds}\Omega_{\rm ds}-1}{2\ln{\Omega_{\rm m}}}\;.
\end{equation}
Since it is not possible to find an analytical solution to this equation, the standard procedure is to linearise it 
and assume that the early time solution holds also at later time when matter is no longer the dominant component, see 
\cite{Battye2018a} for details of the derivation of Eqn.~(\ref{eqn:gamma}) and its linearization. (Although 
\cite{Battye2018a} focuses on $f(R)$ gravity, the equations for the growth index, Eqs.~(22)-(23) in that work 
apply to Horndeski models provided that $\epsilon$ is replaced by $G_{\rm eff}/G_{\rm N}$.) The solution can be found 
analytically only when $G_{\rm eff}/G_{\rm N}$ is approximately constant, which was the case for $f(R)$ models in 
\cite{Battye2018a} but is not generally true for the models we investigate here. Hence, we have to rely on 
numerical solutions.

\begin{figure}
 \centering
 \includegraphics[width=6.5cm,angle=0]{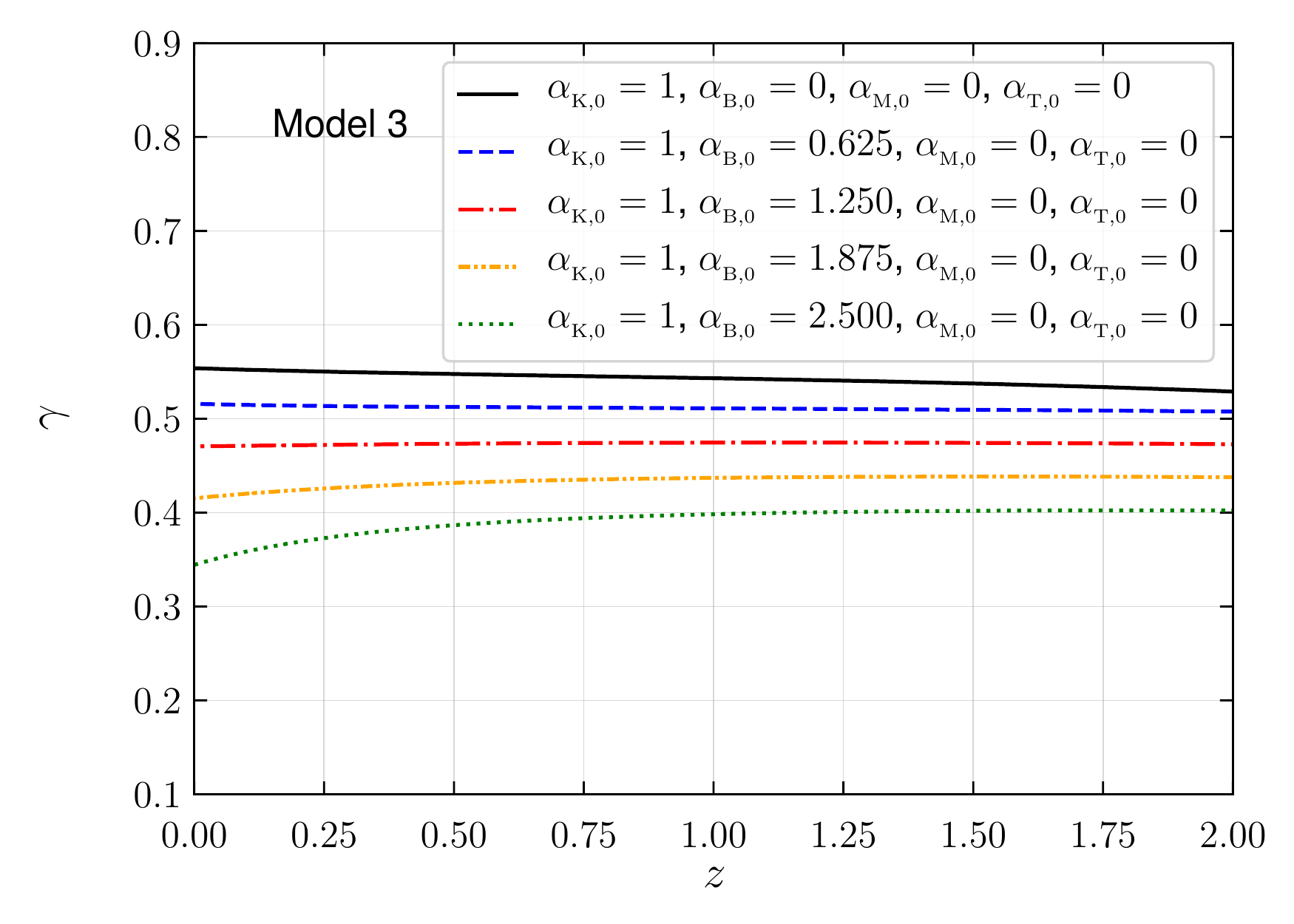}
 \includegraphics[width=6.5cm,angle=0]{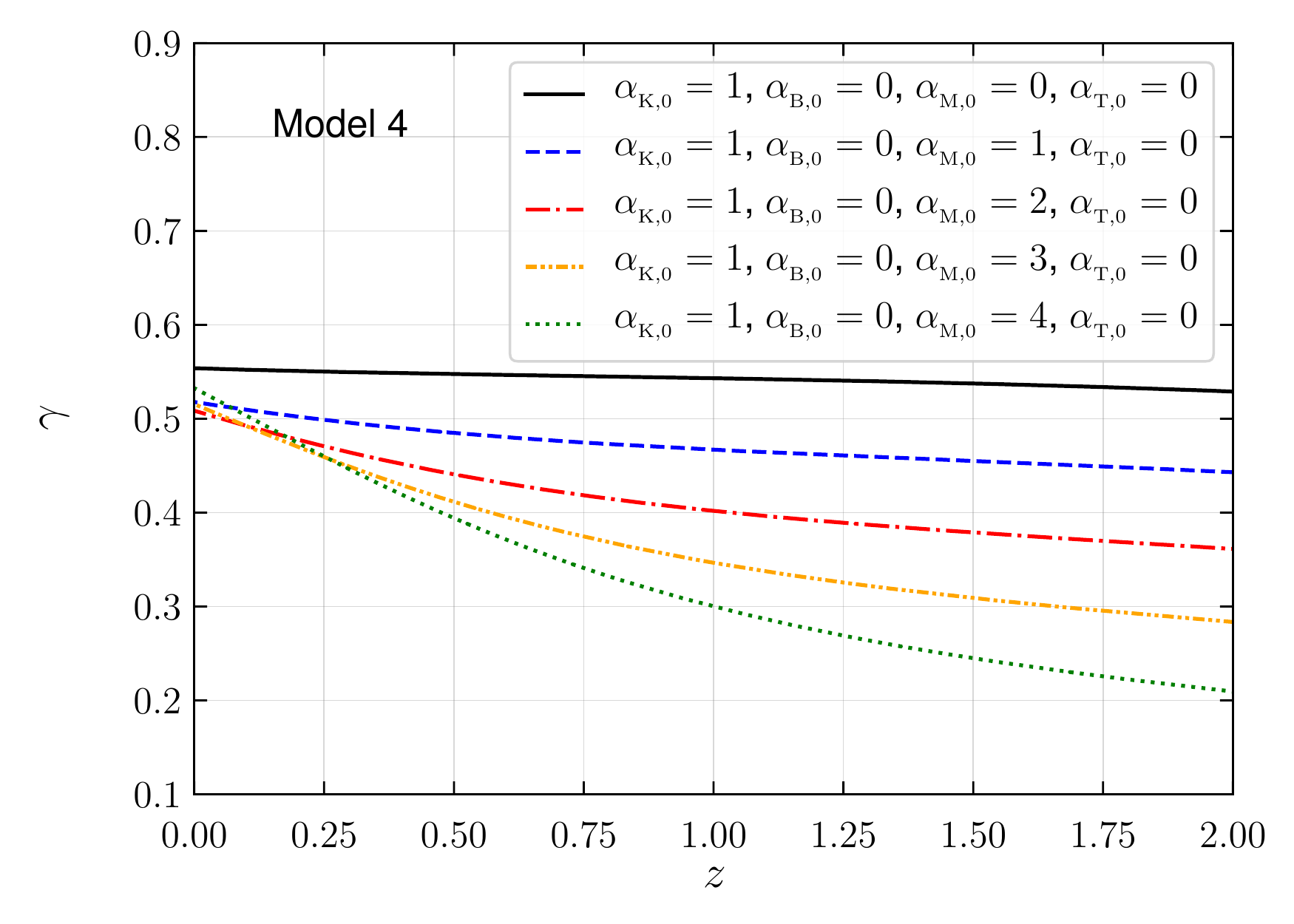}
 \includegraphics[width=6.5cm,angle=0]{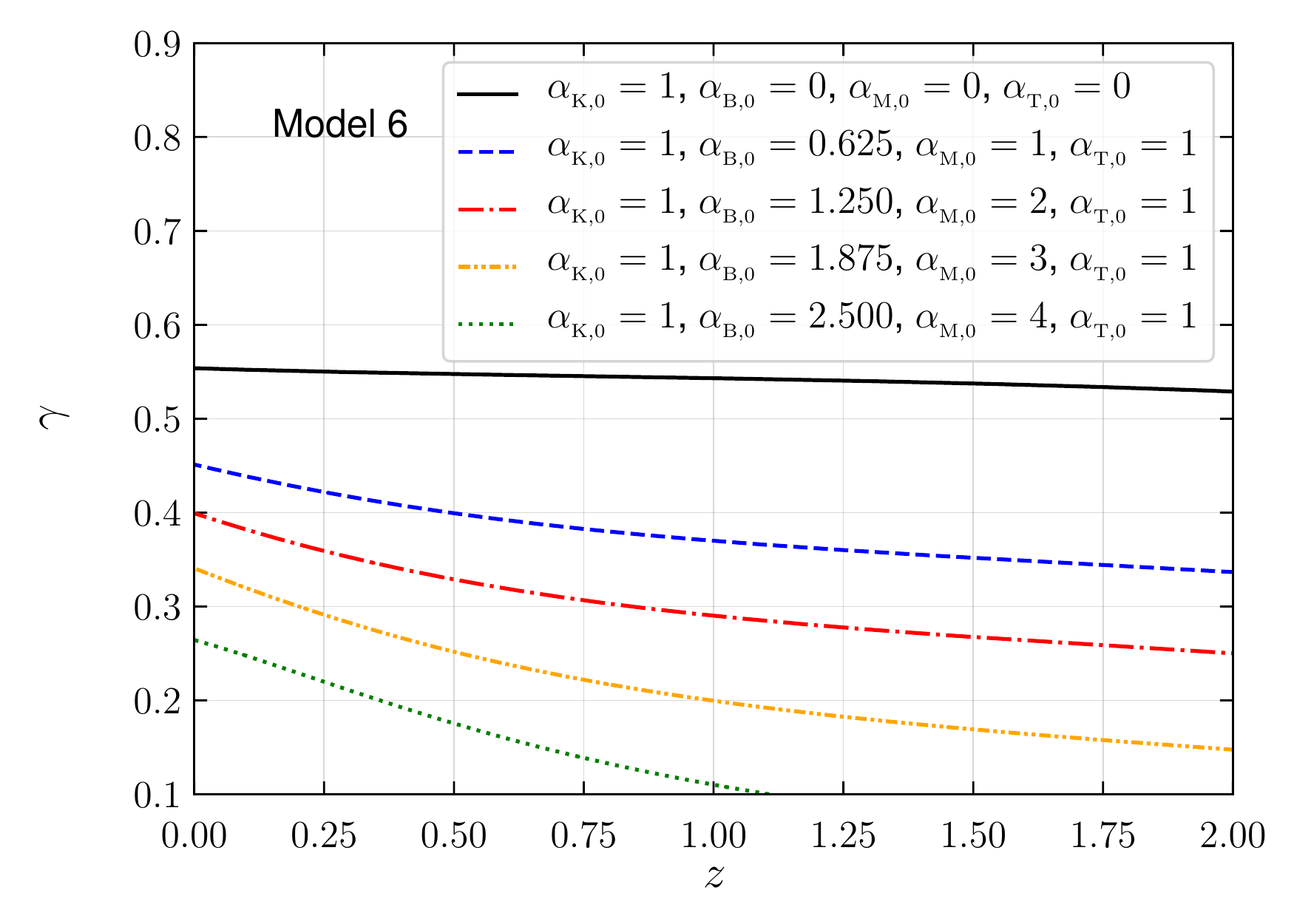}
 \includegraphics[width=6.5cm,angle=0]{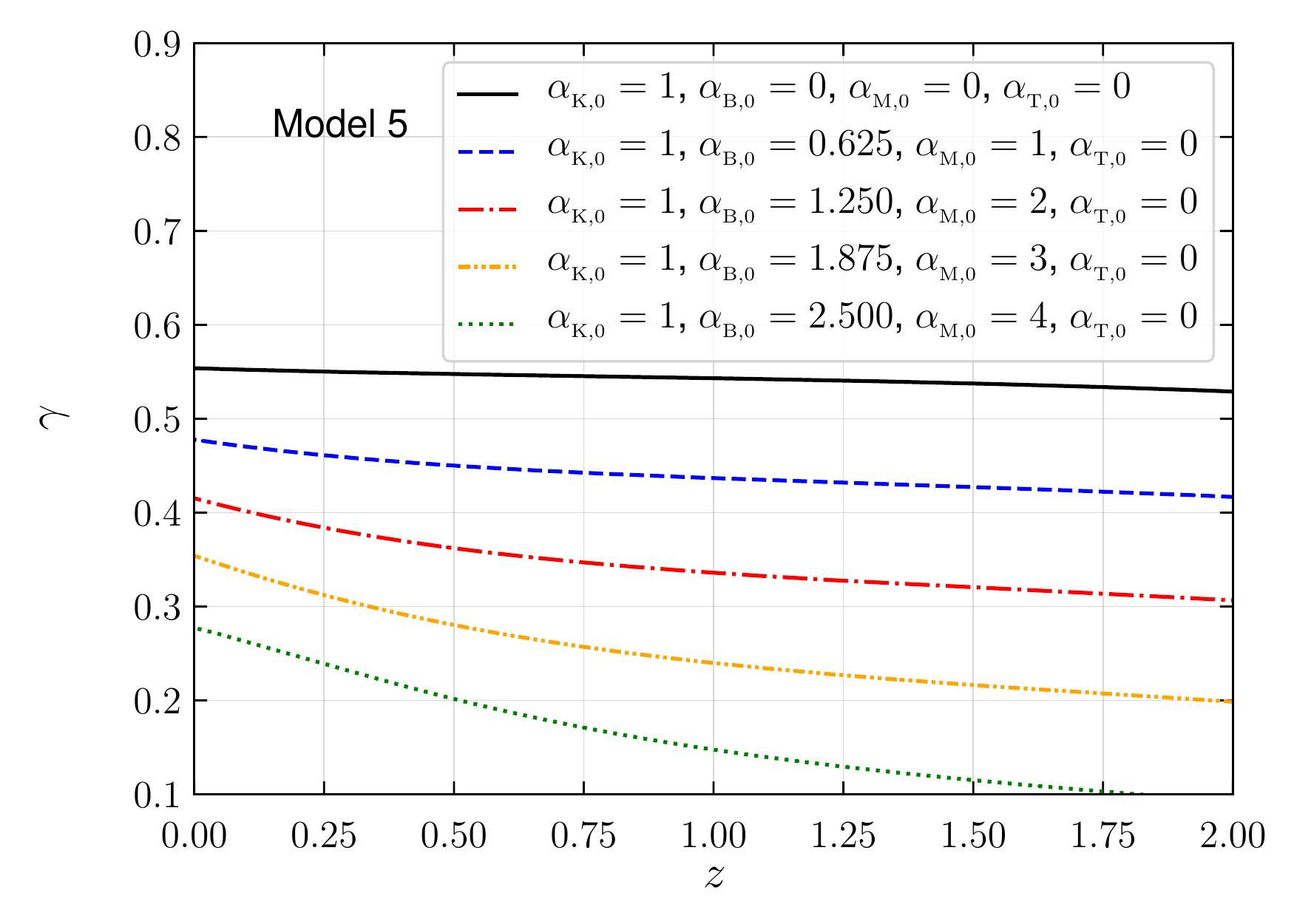}
 \caption[justified]{Time evolution for the growth index $\gamma$ for ${\rm K}^2\gg 1$. From top left in clockwise 
 order we show models 3, 4, 5 and 6, respectively, Line styles and colours are as in Fig.~\ref{fig:aKaB}, 
 \ref{fig:aKaM}, \ref{fig:aKaBaM}, and \ref{fig:aKaBaMaT}, respectively.}
 \label{fig:gamma}
\end{figure}

We present our results for models 3--6 in Fig.~\ref{fig:gamma}. As in the previous section, we focus on the regime 
${\rm K}^2\gg 1$. 
In all cases presented here, for all the different models, the effect of modified gravity decreases the growth index 
compared to its $\Lambda$CDM expectation (solid black lines in the figure). We note this is also the case in $f(R)$ 
gravity \citep{Battye2018a}. If the growth index is lower, it means that the growth rate, 
$f=\Omega_{\rm m}^{\gamma}$, is larger and therefore matter clustering is amplified. Hence in general one can expect 
an excess of power on small scales for the matter power spectrum $P(k)$. Nevertheless, one should also take into 
account another important aspect, namely the amount of time during which $\gamma$ remains significantly lower than the 
$\Lambda$CDM expectation value, as at early time all the models recover this value.

\subsection{The \texorpdfstring{\boldmath{$E_G$}}{EG} function}
Zhang et~al. (2007) \citep{Zhang2007} introduced the $E_G$ parameter as test for modified gravity and dark energy 
models. This quantity is defined as the ratio of the Laplacian of the Newtonian potentials to the peculiar velocity 
divergence. In our notation, using Einstein field equations, this quantity can be written as
\begin{equation}
 E_G = \frac{\Omega_{\rm m,0}\Sigma}{\Delta_{\rm m}^{\prime}/\Delta_{\rm m}}\,,
\end{equation}
where $\Sigma$, defined in Eqn.~(\ref{eqn:Sigma}), is proportional to the lensing (Weyl) potential and 
$\Delta_{\rm m}^{\prime}/\Delta_{\rm m}=\Omega_{\rm m}^{\gamma}$ is the growth rate discussed in the previous section. 
See also \citep{Lombriser2017,Ghosh2018} for recent updates on the measurability of the $E_G$ parameter.

Reyes et~al. (2010) \cite{Reyes2010} were the first to measure $E_G$. They used galaxy-velocity cross-correlation and 
galaxy-galaxy lensing data from SDSS \citep{York2000}, and obtained a mean $E_G=0.39\pm0.06$ (68\%), at $z=0.32$ on 
small scales (${\rm K}^2\gg 1$). In $\Lambda$CDM, with $\Omega_{\rm m,0}=0.31$, we calculate 
$E_G^{\Lambda{\rm CDM}}=0.452\pm0.029$ (68\%) at $z=0.32$, where we quote the same uncertainty as \cite{Reyes2010}. 
The theoretical uncertainty is obtained by taking into account the error on the measure of $\Omega_{\rm m,0}$ and 
since we do not repeat the analysis of \cite{Reyes2010}, we assume the same error bars. 
We note that our central $\Lambda$CDM value is substantially larger than the one quoted in \cite{Reyes2010}. This is 
because they assumed a lower value for $\Omega_{\rm m,0}$, namely $\Omega_{\rm m,0}=0.256\pm0.018$ from WMAP 5-year 
data \citep{Dunkley2009}. With our value of $\Omega_{\rm m,0}$, consistent with recent measurements 
\citep{Planck2018_VI}, the value of $E_G$ measured by \cite{Reyes2010} is $\approx 1\sigma$ lower than the current 
$\Lambda$CDM value, although they are comparable within errors.

\begin{figure}
 \centering
 \includegraphics[width=6.5cm,angle=0]{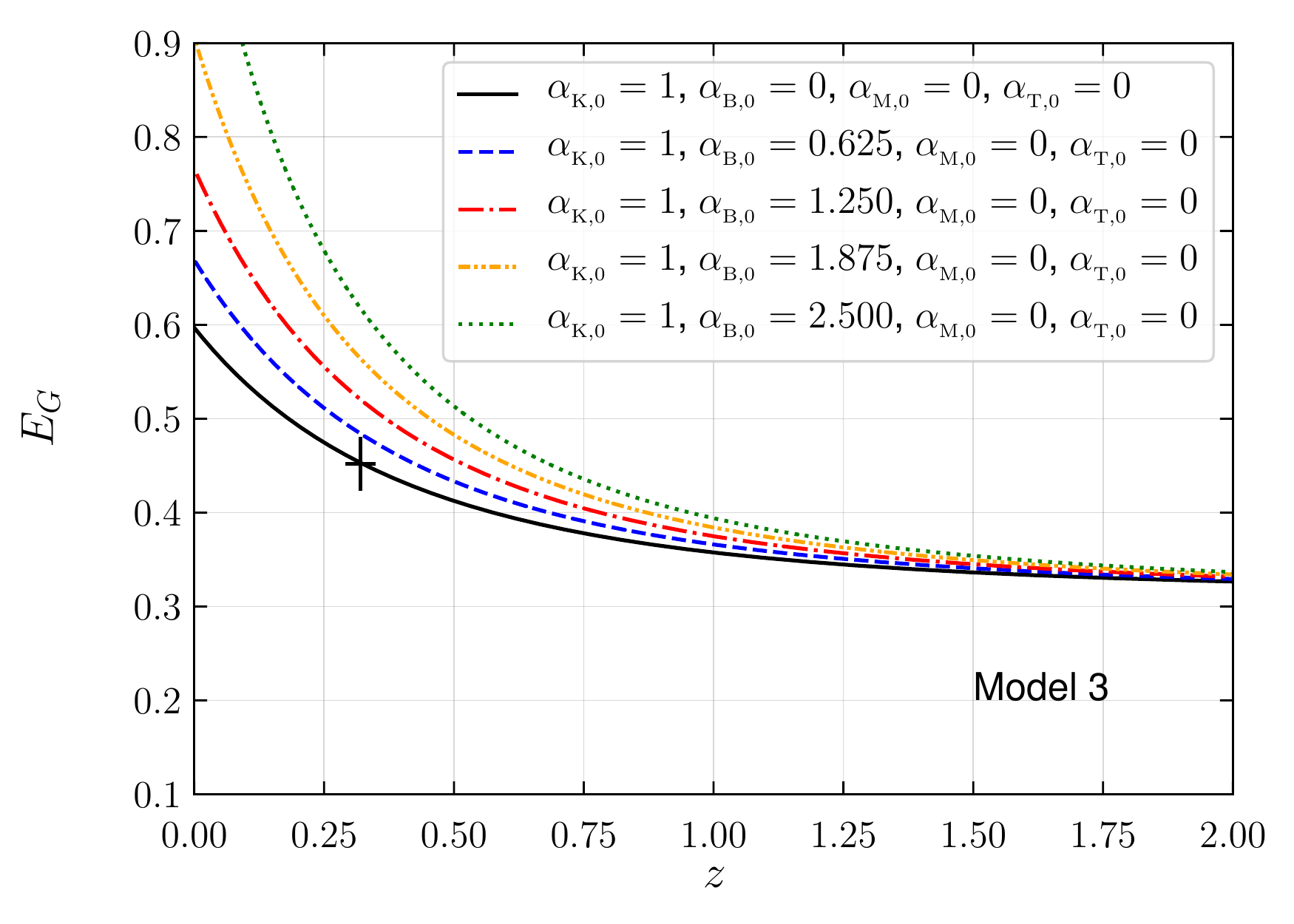}
 \includegraphics[width=6.5cm,angle=0]{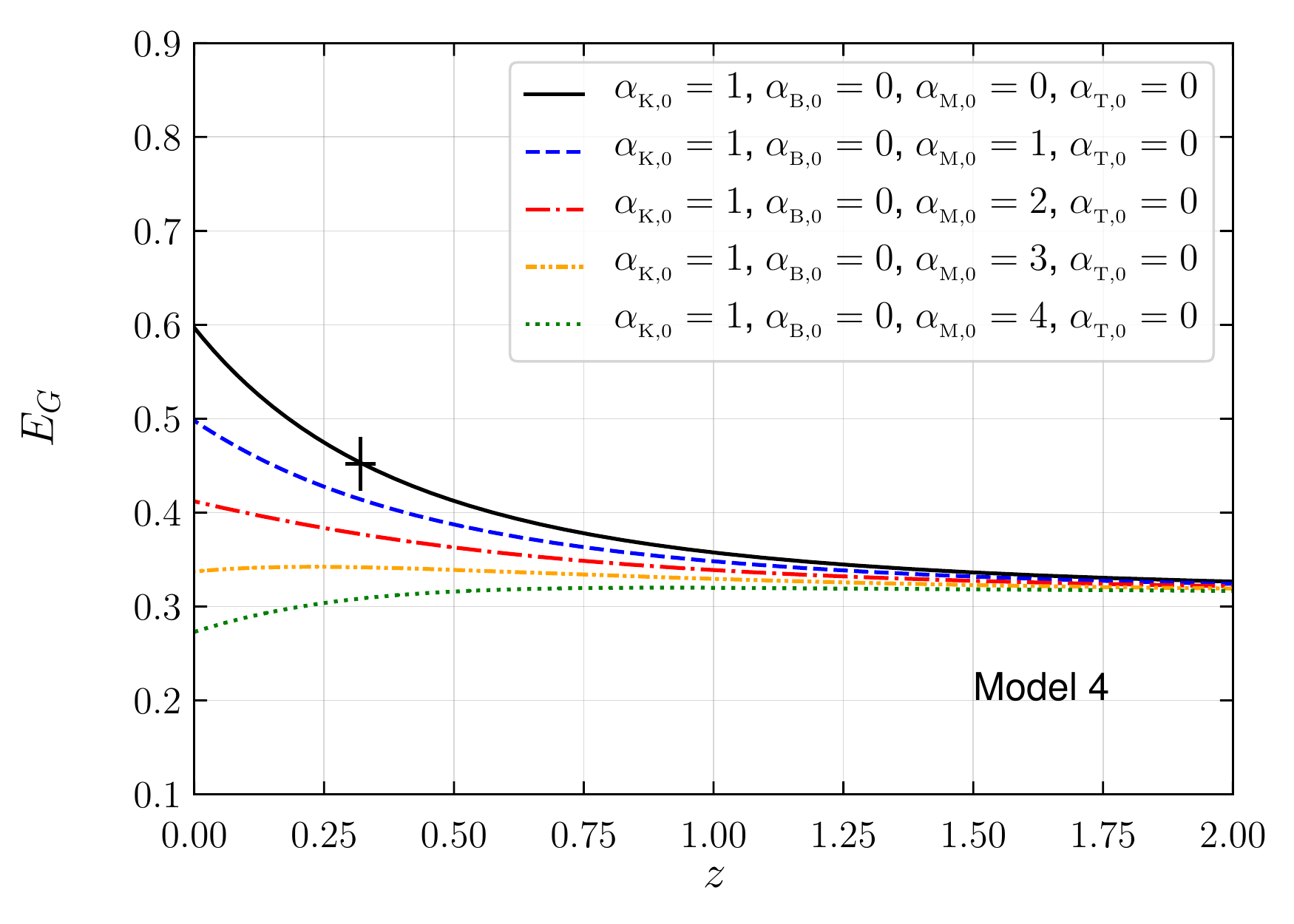}
 \includegraphics[width=6.5cm,angle=0]{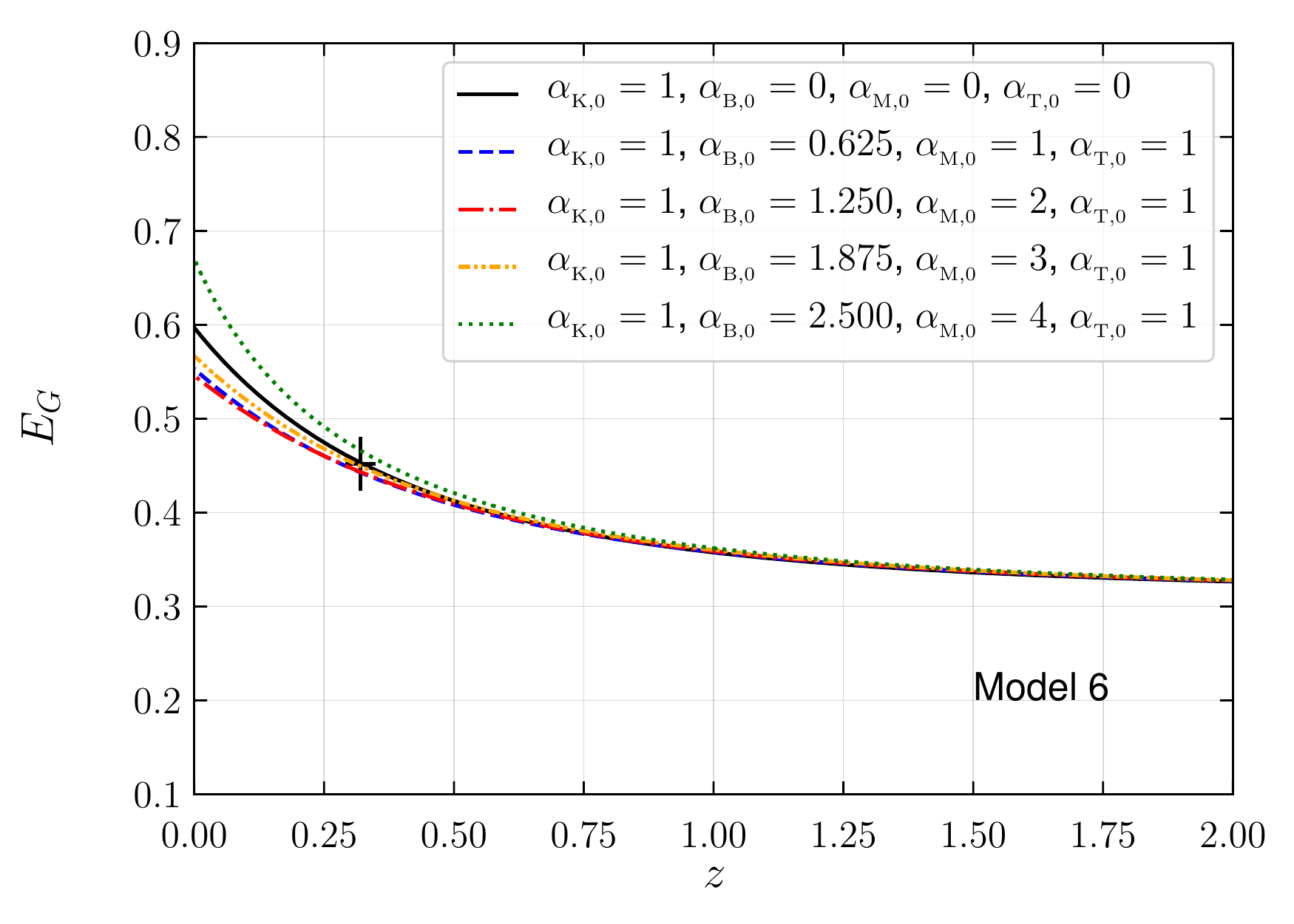}
 \includegraphics[width=6.5cm,angle=0]{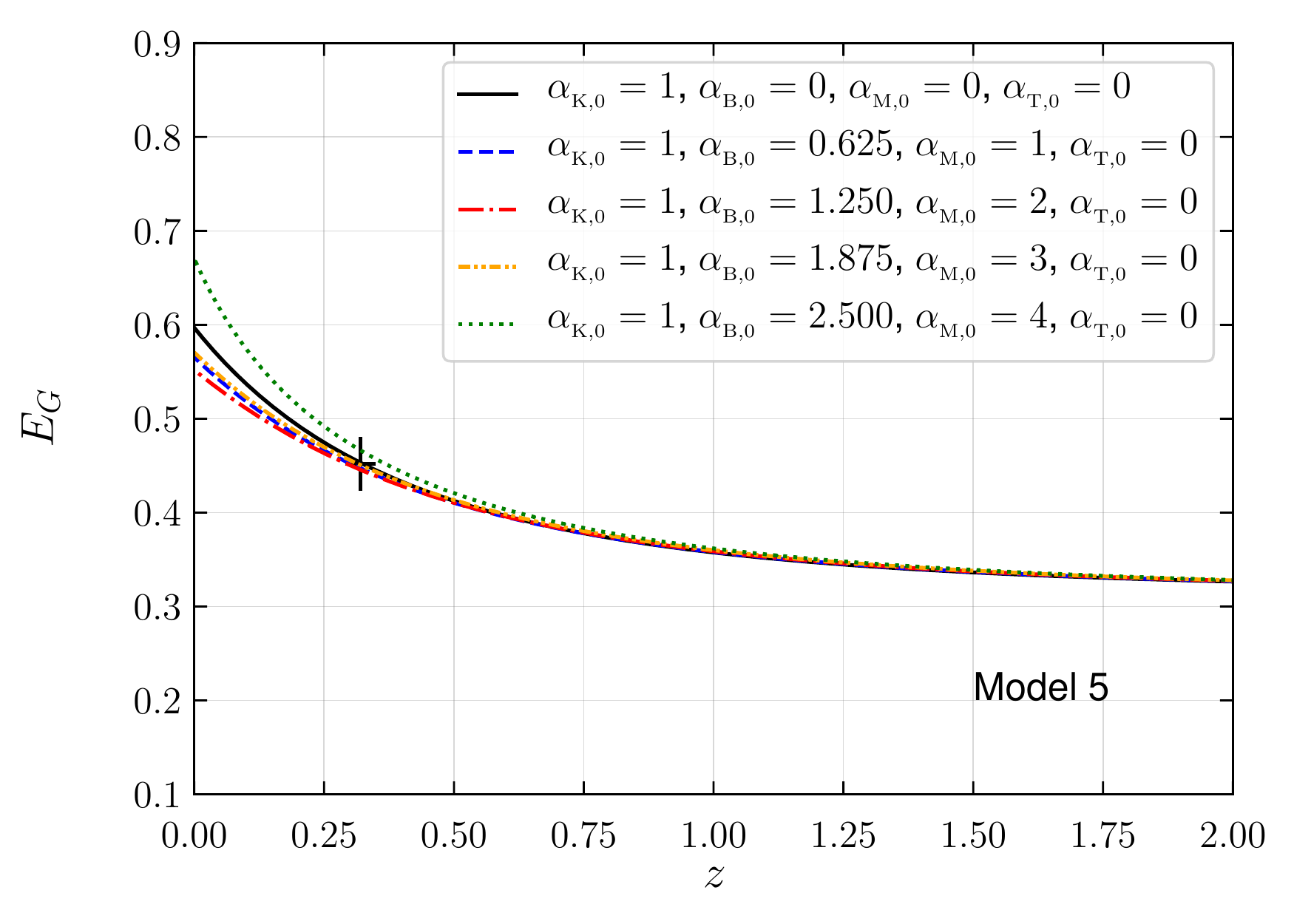}
 \caption[justified]{Time evolution for the quantity $E_G$. From top to bottom, we show models 3 to 6, respectively. 
 Line style and colours are as in Fig.~\ref{fig:aKaB}--\ref{fig:aKaBaMaT}. The point with errorbars represent the 
 expected value $E_G=0.452\pm0.029$ for a $\Lambda$CDM model at $z=0.32$ with the cosmology assumed in this work. 
 In agreement with the measurements of \cite{Reyes2010}, for all the models we consider the value of 
 $\Sigma_{\infty}$.}
 \label{fig:EG}
\end{figure}

In Fig.~\ref{fig:EG} we present the redshift evolution of the $E_G$ parameter for Horndeski models 3--6 and the same 
values of the $\alpha_{\rm x,0}$ and cosmological parameters of Fig.~\ref{fig:gamma} as used before in this work. 
We have included the $\Lambda$CDM expectation, namely $E_G^{\Lambda{\rm CDM}}=0.452\pm0.029$ (68\%), as the black point 
at $z=0.32$. 
Moreover, the solid black line shows the time evolution of $E_G$ for model 1 (Quintessence/$k$-essence) with only 
$\alpha_{\rm K}\neq 0$. For the value considered here ($\alpha_{\rm K,0}=1$), this model is indistinguishable from 
$\Lambda$CDM. In all the panels, we see that at early time ($z\gtrsim1$), all the models give similar results as 
$\Lambda$CDM. Differences are most important for models 3 and 4 over a range of redshifts $z\lesssim 1.5$, 
while for models 5 and 6 differences are noticeable only at very low redshift $z\lesssim 0.3$.

We see that for the models we have considered in our analysis, a measurement of $E_G$ with the uncertainty that we 
used (7\% relative uncertainty) would not discriminate between the parameters used for models 5 and 6 and 
$\Lambda$CDM. However, it would exclude models 3 and 4 for $\alpha_{\rm B,0}\gtrsim0.625$ and 
$\alpha_{\rm M,0}\gtrsim 1$. 
It would be interesting to update the measurements of $E_G$ with the latest galaxy survey data to check or revise 
these conclusions, with current and forecasted uncertainties.

We also note that this situation (the fact that some models are not discriminated by the $E_G$ statistics but are with 
the CMB temperature anisotropy power spectrum) is similar to what happens for $f(R)$ models, as shown in Fig.~4 (left 
panel) of \cite{Lombriser2012}. Indeed, when $B_0\simeq 10$, the value of $E_G$ for $f(R)$ models is very close to the 
expectation value of the $\Lambda$CDM cosmology.

\section{Conclusions}\label{sect:conclusions}
In this work we studied modifications to General Relativity corresponding to Horndeski theories 
\citep{Horndeski1974,Deffayet2011,Kobayashi2011}. Rather than using a scalar field or an Effective Field Theory 
approach, we have recast the Horndeski modifications as a dark energy fluid for both the background and the linear 
cosmological perturbations. Thus, we have extended the works of \cite{Battye2016a,Battye2018a,Battye2019}, and 
applied the EoS approach \citep{Battye2012,Battye2013,Battye2013a,Battye2014} to the full Horndeski theories for the 
first time.

We have implemented the EoS approach for Horndeski into a modified version of the \verb|CLASS| code 
\citep{Lesgourgues2011a,Blas2011}. 
Our code, \verb|EoS_class|, will be made publicly available online upon acceptance of this 
manuscript.\footnote{Website:\href{https://github.com/fpace}
{https://github.com/fpace}} 
Our code is as fast as \verb|hi_class| \citep{Zumalacarregui2017} (an independent code for Horndeski theories based on 
the EFT approach \citep{Bellini2014}) for the models we studied. If we account for the different versions of 
\verb|CLASS| used in \verb|EoS_class| and \verb|hi_class|, we get agreement with the latter within 0.1\% relative 
error. Moreover, unlike \verb|hi_class|, our code can solve the dynamics of cosmological perturbations in Horndeski 
models with $\alpha_{\rm K}=0$.

For sub-horizon modes, following a procedure similar to the quasi-static approximation where time derivatives are 
neglected, we have obtained an analytical approximation for the attractor solution of the gauge-invariant dark sector 
density perturbation. Using our analytical result, we derived new expressions for the modified gravity parameters, 
e.g.\ $\mu$, $\eta$ and $\Sigma$, that feature a scale dependence. 
Similar formulae had been obtained in previous works 
\citep{DeFelice2011,Lombriser2015a,Bloomfield2013a,Silvestri2013,Pogosian2016}. 
Although all the formulae for the modified gravity parameters agree in the small scale limit, they all differ in the 
large scale limit. Therefore, in a forthcoming paper, we will explicitly compare these formulae between them and with 
the exact numerical solution. The open question is: which one of the scale-dependent formulae for the modified gravity 
parameters has the widest range of validity when ${\rm K}^2=k^2/(aH)^2\ll 1$? Having an accurate analytical 
approximation of the modified gravity parameters is useful because it enables a simple implementation of modified 
gravity dynamics, which can be observationally tested, as well as a relatively simple phenomenological discussion.

With our exact numerical solutions computed with \verb|EoS_class| and our analytical approximations of the modified 
gravity parameters we have studied the phenomenology of Horndeski models at the level of CMB temperature and lensing 
and matter power spectra as well as the growth index and the $E_G$ function of \cite{Zhang2007}.

\appendix

\section{Background expressions for Horndeski models}\label{sect:background}
In this section we present explicit expressions for the background quantities describing Horndeski models.

The density $\rho_{\rm ds}$ and the pressure $P_{\rm ds}$ are, respectively \cite{DeFelice2011,Nishi2014}
\begin{align}
 \rho_{\rm ds} = &\, 2XG_{2,X} - G_2 - XG_{3,\phi} - 6H\dot{\phi}XG_{3,X} - 
                     6H^2(G_4 -4XG_{4,X}-4X^2G_{4,XX}) - \nonumber\\
                 &\, 6H\dot{\phi}(G_{4,\phi}+2XG_{4,\phi X}) + 
                     2H^3\dot{\phi}X(5G_{5,X}+2XG_{5,XX}) + 
                     3H^2X(3G_{5,\phi}+2XG_{5,\phi X}) + \nonumber\\
                 &\, 3H^2M_{\rm pl}^2\,, \label{eqn:rho} \\
 P_{\rm ds} = &\, G_2 + X\left(2\ddot{\phi}G_{3,X}-G_{3,\phi}\right) + 2\left(3H^2+2\dot{H}\right)G_4 - 
                  4H\left(\dot{X}+3HX\right)G_{4,X} - \nonumber\\
              &\, 8X\left(\dot{H}G_{4,X}+H\dot{X}G_{4,XX}\right) + 
                  2\left(\ddot{\phi}+2H\dot{\phi}\right)G_{4,\phi} - 2XG_{4,\phi\phi} + 
                  4\left(\ddot{\phi}-2H\dot{\phi}\right)XG_{4,\phi X} - \nonumber\\
              &\, 2X\left(2H^3\dot{\phi}+2H\dot{H}\dot{\phi}+3H^2\ddot{\phi}\right)G_{5,X} - 
                  4H^2\ddot{\phi}X^2G_{5,XX} - 2HX\left(\dot{X}-HX\right)G_{5,\phi X} - \nonumber\\
              &\, \left[2\left(HX\right)^{\hbox{$\cdot$}}+3H^2X\right]G_{5,\phi} - 2H\dot{\phi}XG_{5,\phi\phi} -
                  \left(3H^2+2\dot{H}\right)M_{\rm pl}^2\,. \label{eqn:P}
\end{align}

The expressions relevant to the equation of motion of the scalar field are
\begin{align}
 J \equiv &\, -2\dot{\phi}G_{2,X} - 6HXG_{3,X} + 2\dot{\phi}G_{3,\phi} - 
               12H^2\dot{\phi}\left(G_{4,X}+2XG_{4,XX}\right) - 12HXG_{4,\phi X} + \nonumber\\
          &\, \phantom{-}
               2H^3X\left(3G_{5,X}+2XG_{5,XX}\right) - 
               6H^2\dot{\phi}\left(G_{5,\phi}+XG_{5,\phi X}\right)\,,\label{eqn:J}\\
 P_{\phi} \equiv &\, G_{2,\phi} - X\left(G_{3,\phi\phi}-2\ddot{\phi}G_{3,\phi X}\right) + 
                     6\left(2H^2+\dot{H}\right)G_{4,\phi} + 6H\left(\dot{X}+2HX\right)G_{4,\phi X} + \nonumber \\
                 &\, 3H^2XG_{5,\phi\phi} + 2H^3\dot{\phi}XG_{5,\phi X}\;. \label{eqn:Pphi}
\end{align}

\section{Perturbation coefficients for Horndeski models}\label{sect:perturbations}
The perturbation coefficients, $\alpha_{\rm K}$, $\alpha_{\rm B}$, $\alpha_{\rm M}$ and $\alpha_{\rm T}$ in terms of 
the Horndeski functions $G_i$ are

\begin{eqnarray}
 M^2 & \equiv & 2G_4-4XG_{4,X}-XG_{5,\phi}-2H\dot{\phi}XG_{5,X}\,,\\
 H\alpha_{\rm M} & \equiv & \frac{d\ln{M^2}}{dt}\,,\\
 M^2H^2\alpha_{\rm K} & \equiv & 2X\left(G_{2,X}+2XG_{2,XX}-G_{3,\phi}-XG_{3,\phi X}\right)\nonumber\\
  && -12H\dot{\phi}X\left(G_{3,X}+XG_{3,XX}+3G_{4,\phi X}+2XG_{4,\phi XX}\right)\nonumber \\
  && +12H^2X\left(G_{4,X}+8XG_{4,XX}+4X^2G_{4,XXX}\right)\nonumber\\
  && +6H^2X\left(G_{5,\phi}+5XG_{5,\phi X}+2X^2G_{5,\phi XX}\right)\nonumber\\
  && +4H^3\dot{\phi}X\left(3G_{5,X}+7XG_{5,XX}+2X^2G_{5,XXX}\right)\,,\\
  2M^2H\alpha_{\rm B} & \equiv & 2\dot{\phi}X\left(G_{3,X}+2G_{4,\phi X}\right)
    +2\left(\frac{dG_4}{dt}-\frac{dX}{dt}G_{4,X}\right)\nonumber \\
  && -4HX\left(2G_{4,X}+4XG_{4,XX}+G_{5,\phi}+XG_{5,\phi X}\right)\nonumber\\
  && -2H^2\dot{\phi}X\left(3G_{5,X}+2XG_{5,XX}\right)\,,\nonumber\\
  & \equiv & 2\dot{\phi}\left(XG_{3,X}+G_{4,\phi}+2XG_{4,\phi X}\right)\nonumber \\
  && -4HX\left(2G_{4,X}+4XG_{4,XX}+G_{5,\phi}+XG_{5,\phi X}\right)\nonumber\\
  && -2H^2\dot{\phi}X\left(3G_{5,X}+2XG_{5,XX}\right)\,,\\
 M^2\alpha_{\rm T} & \equiv & 2X\left[2G_{4,X}+G_{5,\phi}-\left(\ddot{\phi}-H\dot{\phi}\right)G_{5,X}\right]\,.
\end{eqnarray}
Note that evaluating $\alpha_{\rm M}$ as time derivative of $M^2$, rather than in terms of the derivatives of $G_4$ and 
$G_5$ with respect to $\phi$ and $X$ and $\alpha_{\rm B}$ in terms of the time derivative $G_4$ is more general
since they allow to recover also results for $f(R)$ models.

\section{Coefficients for the EoS approach}\label{sect:coefficientsEoS}
In this section we provide the full expressions for the coefficients used in the EoS formalism. Having shown in 
\cite{Battye2018a} that the coefficients for the two gauge-invariant quantities $w_{\rm ds}\Gamma_{\rm ds}$ and 
$w_{\rm ds}\Pi_{\rm ds}$ lead to the same results for $f(R)$ models, here we generalise these expressions with the help 
of those provided by \cite{Gleyzes2014} in their Appendix~C, Eqs.~(184)-(195), for generic Horndeski models. 
As in \cite{Battye2018a}, we used a standard continuity equation for the dark sector fluid which implies, for the 
background,
\begin{align}
 \rho_{\rm de}^{\rm GLV} & = \rho_{\rm ds} + 3(M^2-M_{\rm pl}^2)H^2\;,\nonumber\\
 P_{\rm de}^{\rm GLV} & = P_{\rm ds} - \left(3-2\epsilon_{H}\right)H^2(M^2-M_{\rm pl}^2)\;,\nonumber
\end{align}
where $\rho_{\rm de}^{\rm GLV}$ and $P_{\rm de}^{\rm GLV}$ are the variables used in \cite{Gleyzes2014} while 
$\rho_{\rm ds}$ and $P_{\rm ds}$ those used in this work; $M_{\rm pl}^2$ and $M^2$ are the standard and the effective 
Planck mass, respectively. Note that this also leads to a different background equation of state $w$ for the two 
conditions:
\begin{equation}
 w_{\rm de}^{\rm GLV} = w_{\rm ds} + 
                        \frac{1}{3}\frac{2\epsilon_H-3(1+w_{\rm ds})}{\Omega_{\rm ds}+M^2/M_{\rm pl}^2-1}
                        \left(\frac{M^2}{M_{\rm pl}^2}-1\right)\;,
\end{equation}
where $\Omega_{\rm ds}$ represents the dark sector density parameter. 
The two agree only for minimally coupled models, i.e.\ when $M^2=M_{\rm pl}^2$. The background expressions are found by 
taking into account that $H$ is the same in both formalisms.

From now on, for compactness of the notation, we will define $\bar{M}^2\equiv M^2/M_{\rm pl}^2$.

For the perturbed variables, the relation between the variables in \cite{Gleyzes2014} and the ones adopted in our 
numerical implementation are:
\begin{align}
 \delta\rho_{\rm de}^{\rm GLV} & = \bar{M}^2\delta\rho_{\rm ds} + 
                                   \left(\bar{M}^2-1\right)\delta\rho_{\rm m}\;, \nonumber\\
 \delta P_{\rm de}^{\rm GLV} & = \bar{M}^2\delta P_{\rm ds} + 
                                 \left(\bar{M}^2-1\right)\delta P_{\rm m}\;, \nonumber\\
 q_{\rm m}^{\rm GLV}+q_{\rm de}^{\rm GLV} & = -\bar{M}^2
                                               \frac{
                                               \bar{\rho}_{\rm ds}\hat{\Theta}_{\rm ds}+
                                               \bar{\rho}_{\rm m}\hat{\Theta}_{\rm m}}
                                               {3H}\;, \nonumber\\
 q_{\rm de}^{\rm GLV} & = -\frac{1}{3H}\left[\bar{M}^2
                           \bar{\rho}_{\rm ds}\hat{\Theta}_{\rm ds}+
                           \left(\bar{M}^2-1\right)\bar{\rho}_{\rm m}\hat{\Theta}_{\rm m}
                           \right]\;, \nonumber\\
 \sigma_{\rm m}^{\rm GLV}+\sigma_{\rm de}^{\rm GLV} & = -\frac{a^2}{k^2}\bar{M}^2
                                                         (\bar{P}_{\rm ds}\Pi_{\rm ds}
                                                        +\bar{P}_{\rm m}\Pi_{\rm m})\;, \nonumber\\
 \sigma_{\rm de}^{\rm GLV} & = -\frac{a^2}{k^2}\left[\bar{M}^2\bar{P}_{\rm ds}\Pi_{\rm ds}
                               +\left(\bar{M}^2-1\right)\bar{P}_{\rm m}\Pi_{\rm m}\right]\;. \nonumber
\end{align}
For $\bar{M}^2=1+f_{R}$, as it is the case in $f(R)$ models, we recover the expressions presented in 
\cite{Battye2018a}. 
We also made the following identifications: $q_{\rm m}^{\rm GLV}=-\frac{\rho_{\rm m}\Theta_{\rm m}}{3H}$ and 
$\sigma_{\rm m}^{\rm GLV}=-\tfrac{a^2}{k^2}P_{\rm m}\Pi_{\rm m}$.
The relations between the perturbed expressions in the two formalisms are obtained by comparing the perturbed Einstein 
field equations.

With the help of the conversions reported above, we can now write the full expressions for the coefficients of the 
entropy perturbation $C_{\Gamma X}$ and anisotropic stress $C_{\Pi X}$, with $X=\Delta_{\rm ds,m}$, 
$\Theta_{\rm ds,m}$, $\Gamma_{\rm m}$ and $\Pi_{\rm m}$:

\begin{align}
 C_{\Gamma\Delta_{\rm ds}} & \equiv \frac{\gamma_1 \gamma_2 + \tilde{\gamma}_3 {\rm K}^2}
                                    {\gamma_1 + \alpha_{\rm B}^2 {\rm K}^2}-
                                    \frac{\mathrm{d}\bar{P}_{\rm ds}}{\mathrm{d}\bar{\rho}_{\rm ds}}\,,\\
 C_{\Gamma\Theta_{\rm ds}} & \equiv -\frac{1}{3}\frac{\gamma_1 \gamma_4 + \tilde{\gamma}_5 {\rm K}^2}
                                     {\gamma_1 + \alpha_{\rm B}^2 {\rm K}^2}+
                                     \frac{\mathrm{d}\bar{P}_{\rm ds}}{\mathrm{d}\bar{\rho}_{\rm ds}}\,,\\
 C_{\Gamma\Delta_{\rm m}} & \equiv \frac{\gamma_1 \gamma_2 + \tilde{\gamma}_3 {\rm K}^2}
                                   {\gamma_1 + \alpha_{\rm B}^2 {\rm K}^2}\left(1-\frac{1}{\bar{M}^2}\right)
                                   +\frac{\gamma_7}{\bar{M}^2}
                                   +\left(\frac{1}{\bar{M}^2}\frac{\alpha_{\rm K}}{\alpha}-1\right)
                                   \frac{\mathrm{d}\bar{P}_{\rm m}}{\mathrm{d}\bar{\rho}_{\rm m}} \,,\\
 C_{\Gamma\Theta_{\rm m}} & \equiv -\frac{1}{3}
                                    \left[\frac{\gamma_1 \gamma_4 + \tilde{\gamma}_5 {\rm K}^2}
                                    {\gamma_1 + \alpha_{\rm B}^2 {\rm K}^2}\left(1-\frac{1}{\bar{M}^2}\right)
                                   +\frac{\gamma_1\gamma_6 + 3\tilde{\gamma}_7 {\rm K}^2}
                                    {\gamma_1 + \alpha_{\rm B}^2 {\rm K}^2}\frac{1}{\bar{M}^2}
                                   +3\left(\frac{1}{\bar{M}^2}\frac{\alpha_{\rm K}}{\alpha}-1\right)
                                    \frac{\mathrm{d}\bar{P}_{\rm m}}{\mathrm{d}\bar{\rho}_{\rm m}}\right]\,,\\
 C_{\Gamma\Gamma_{\rm m}} & \equiv \frac{1}{\bar{M}^2}\frac{\alpha_{\rm K}}{\alpha}-1\,,\\
 C_{\Pi\Delta_{\rm ds}} & \equiv -\frac{1}{2}\frac{\gamma_1 \alpha_{\rm T} + \tilde{\gamma}_8 {\rm K}^2}
                                  {\gamma_1 + \alpha_{\rm B}^2 {\rm K}^2}\,,\\
 C_{\Pi\Theta_{\rm ds}} & \equiv \frac{1}{6}\frac{\gamma_9 {\rm K}^2}{\gamma_1 + \alpha_{\rm B}^2 {\rm K}^2}\,,\\
 C_{\Pi\Delta_{\rm m}} & \equiv -\frac{1}{2}
                                 \left[\frac{1}{\bar{M}^2}\alpha_{\rm T} + 
                                 \frac{\gamma_1 \alpha_{\rm T} + \tilde{\gamma}_8 {\rm K}^2}
                                 {\gamma_1 + \alpha_{\rm B}^2 {\rm K}^2}
                                 \left(1-\frac{1}{\bar{M}^2}\right)\right]\,,\\
 C_{\Pi\Theta_{\rm m}} & \equiv \frac{1}{6}
                                \left[\frac{\gamma_9 {\rm K}^2}{\gamma_1 + \alpha_{\rm B}^2 {\rm K}^2}
                                \left(1-\frac{1}{\bar{M}^2}\right) + 
                                \frac{\gamma_{10}{\rm K}^2}{\gamma_1 + \alpha_{\rm B}^2 {\rm K}^2}
                                \frac{1}{\bar{M}^2}\right]\,,\\
 C_{\Pi\Pi_{\rm m}} & \equiv -\left(1-\frac{1}{\bar{M}^2}\right)\,,
\end{align}
where, as before, ${\rm K}=\tfrac{k}{aH}$. Note that these coefficients refer to the dimensionful density and velocity 
perturbations and for clarity of notation we have used the labels $\Delta$ and $\Theta$ also for them. While this does 
not affect the coefficients for the dark sector variables, for the matter sector the coefficients for the 
dimensionless quantities differ by a factor $\Omega_{\rm m}(a)/\Omega_{\rm ds}(a)$ from the expressions above.

The $\gamma_i$ functions are given by:
\begin{align}
 \gamma_1 & \equiv \alpha_{\rm K} \frac{3\Omega_{\rm ds}(1+w_{\rm ds})-2\epsilon_H}{4\bar{M^2}}
                   +\frac{1}{2}\alpha\epsilon_H\,,\\
 \gamma_2 & \equiv c_s^2+\frac{\alpha_{\rm T}}{3} - 
                   2\frac{2\alpha_{\rm B}+\tilde{\Gamma}+(1+\alpha_{\rm B})(\alpha_{\rm M}-\alpha_{\rm T})}{\alpha}\,,
                   \\
 \gamma_3 & \equiv c_{\rm s}^2 + \frac{\gamma_8}{3}\,,\\
 \gamma_4 & \equiv \frac{1}{3\Omega_{\rm ds}(1+w_{\rm ds})+2\epsilon_H(\bar{M}^2-1)}\times\nonumber\\
          & \quad  \left\{9\Omega_{\rm ds}(1+w_{\rm ds})\frac{\alpha_{\rm K}}{\alpha}
                   \frac{\mathrm{d}\bar{P}_{\rm ds}}{\mathrm{d}\bar{\rho}_{\rm ds}}+
                   2(\epsilon_H-\bar{\epsilon}_H)\left(\bar{M}^2-\frac{\alpha_{\rm K}}{\alpha}\right)-
                   \right.\nonumber\\
          & \qquad 3\alpha_{\rm M}(\Omega_{\rm m}+w_{\rm m}\Omega_{\rm m})+
           \left. \frac{18\alpha_{\rm B}}{\alpha}\left[\alpha_{\rm B}(3+\alpha_{\rm M})+\tilde{\Gamma}\right]
                   (\Omega_{\rm m}+w_{\rm m}\Omega_{\rm m})\right\}\,,\\
 \gamma_5 & \equiv -1 - \frac{(6\alpha_{\rm B}-\alpha_{\rm K})(\alpha_{\rm T}-\alpha_{\rm M})}{6\alpha_{\rm B}^2} 
                   + \frac{\alpha_{\rm B}^2}{\alpha}\left(\frac{\alpha_{\rm K}}{\alpha_{\rm B}^2}\right)^{\prime}\,,\\
 \gamma_6 & \equiv \frac{\alpha_{\rm K}\alpha_{\rm M}-6\alpha_{\rm B}(3\alpha_{\rm B}+\tilde{\Gamma})}{\alpha}\,,\\
 \gamma_7 & \equiv \frac{\alpha_{\rm K}\alpha_{\rm T}-6\alpha_{\rm B}(\alpha_{\rm B}+\alpha_{\rm T}-\alpha_{\rm M})}
                        {3\alpha}\,,\\
 \gamma_8 & \equiv \alpha_{\rm T} + \frac{\alpha_{\rm T}-\alpha_{\rm M}}{\alpha_{\rm B}}\,,\\
 \gamma_9 & \equiv \frac{1}{2}\alpha(\alpha_{\rm T}-\alpha_{\rm M})\,,\\
 \gamma_{10} & \equiv 3\alpha_{\rm B}^2(\alpha_{\rm T}-\alpha_{\rm M})\,,
\end{align}
where the sound speed for the dark sector perturbations is given by
\begin{equation}\label{eqn:cs2}
 c_{\rm s}^2 = 
             -\frac{2(1+\alpha_{\rm B})[\alpha_{\rm B}(1+\alpha_{\rm T})-(\alpha_{\rm M}-\alpha_{\rm T})-\epsilon_H]
             +2\alpha_{\rm B}^{\prime}}{\alpha}
             -\frac{3(\Omega_{\rm m}+w_{\rm m}\Omega_{\rm m})}{\alpha \bar{M}^2}\,,
\end{equation}
where $\Omega_{\rm m}=1-\Omega_{\rm ds}$ and $w_{\rm m}\Omega_{\rm m}=2\epsilon_H/3-1-w_{\rm ds}\Omega_{\rm ds}$.

For numerical convenience we also defined $\tilde{\gamma}_i=\alpha_{\rm B}^2\gamma_i$:
\begin{align}
 \tilde{\gamma}_3 & \equiv \alpha_{\rm B}^2c_{\rm s}^2 + \frac{\tilde{\gamma}_8}{3}\,,\\
 \tilde{\gamma}_5 & \equiv -\alpha_{\rm B}^2 - \frac{(6\alpha_{\rm B}-\alpha_{\rm K})(\alpha_{\rm T}-\alpha_{\rm M})}
                                                    {6}
                           + \alpha_{\rm B}
                             \frac{\alpha_{\rm K}^{\prime}\alpha_{\rm B}-2\alpha_{\rm K}\alpha_{\rm B}^{\prime}}
                                  {\alpha}
                             \,,\\
 \tilde{\gamma}_8 & \equiv \alpha_{\rm B}^2\alpha_{\rm T} + \alpha_{\rm B}(\alpha_{\rm T}-\alpha_{\rm M})\,.
\end{align}
These expressions are useful when $\alpha_{\rm B}=0$.

We further define $\gamma_1\tilde{\Gamma}=\alpha_{\rm B}\gamma_1\Gamma$\footnote{Note that our expression in 
Eqn.~(\ref{eqn:g1Gamma}) corrects a typo in Eqn.~(194) of \cite{Gleyzes2014}.}
\begin{eqnarray}\label{eqn:g1Gamma}
 \gamma_1\tilde{\Gamma} & \equiv & \frac{\alpha_{\rm K}\alpha_{\rm B}}{4\bar{M}^2}
                           \left[3(3+\alpha_{\rm M})(\Omega_{\rm m}+w_{\rm m}\Omega_{\rm m})+
                                 2(\epsilon_H-\bar{\epsilon}_H)
                         -9\Omega_{\rm ds}(1+w_{\rm ds})
                         \frac{\mathrm{d}\bar{P}_{\rm ds}}{\mathrm{d}\bar{\rho}_{\rm ds}}\right]+
                           \nonumber\\
                        && \frac{3\Omega_{\rm ds}(1+w_{\rm ds})+2\epsilon_H(\bar{M}^2-1)}{4\bar{M}^2}
                           (\alpha_{\rm K}^{\prime}\alpha_{\rm B}-2\alpha_{\rm K}\alpha_{\rm B}^{\prime})-
                           \frac{1}{2}\alpha\alpha_{\rm B}(4\epsilon_H-\bar{\epsilon}_H)\,,
\end{eqnarray}
where $\epsilon_H=-\dot{H}/H^2=-H^{\prime}/H$ and $\bar{\epsilon}_H=-(\ddot{H}/H^3+4\dot{H}/H^2)$. We further have 
$\bar{\epsilon}_H=\epsilon_H^{\prime}+4\epsilon_H-2\epsilon_H^2$.

It is also useful to consider the following coefficients
\begin{align}
 \gamma_{12} & = \gamma_1\left[c_{\rm s}^2+\frac{\alpha_{\rm T}}{3} - 
                   2\frac{2\alpha_{\rm B}+(1+\alpha_{\rm B})(\alpha_{\rm M}-\alpha_{\rm T})}{\alpha}\right]
                -2\frac{\gamma_1\tilde{\Gamma}}{\alpha}\,,\\
 \gamma_{14} & = \frac{9\alpha_{\rm K}\Omega_{\rm ds}(1+w_{\rm ds})}{4\bar{M}^2}
                 \frac{d\bar{P}_{\rm ds}}{d\bar{\rho}_{\rm ds}}-
                 \frac{3\alpha_{\rm K}\alpha_{\rm M}(\Omega_{\rm m}+w_{\rm m}\Omega_{\rm m})}{4\bar{M}^2}+
               \nonumber\\
             & \quad \frac{9(\Omega_{\rm m}+w_{\rm m}\Omega_{\rm m})}{2\alpha\bar{M}^2}
               \left(\alpha_{\rm K}^{\prime}\alpha_{\rm B}^2-
               2\alpha_{\rm K}\alpha_{\rm B}\alpha_{\rm B}^{\prime}\right)+
               \frac{1}{2}(\epsilon_H-\bar{\epsilon}_H)\left(\alpha-\frac{\alpha_{\rm K}}{\bar{M}^2}\right)\,,\\
 \gamma_{16} & = \gamma_1\frac{\alpha_{\rm K}\alpha_{\rm M}-18\alpha_{\rm B}^2}{\alpha}-
                                         6\frac{\alpha_{\rm B}}{\alpha}\gamma_1\tilde{\Gamma}\,,\\
 \gamma_{109} & = \frac{1}{2}\alpha_{\rm K}(\alpha_{\rm M}-\alpha_{\rm T})\,,
\end{align}
where $\gamma_{12} \equiv \gamma_1\gamma_2$, $\gamma_{14} \equiv \gamma_1\gamma_4$, 
$\gamma_{16} \equiv \gamma_1\gamma_6$ and $\gamma_{109} \equiv \gamma_{10}-\gamma_9$. 
The first three are used in the code for models with $\alpha_{\rm B}\neq0$.

\section{Precision parameters for the numerical solutions}\label{sect:parameters}
In this section we report the precision parameters used in this work. For the comparison with results from the 
\verb|hi_class| code, we adopted the same values used in \cite{Bellini2018}, which we report here for 
completeness (see also their Appendix~C). We note that one can generally get accurate spectra for less demanding values 
of the precision parameters.

\begin{verbatim}
 l_max_scalars = 5000
 P_k_max_h/Mpc = 12
 perturb_sampling_stepsize = 0.010
 tol_perturb_integration = 1e-10
 l_logstep = 1.045
 l_linstep = 25
 l_switch_limber = 20
 k_per_decade_for_pk = 200
 accurate_lensing = 1
 delta_l_max = 1000
 k_max_tau0_over_l_max = 8
\end{verbatim}

\section*{Acknowledgements}
\noindent RAB and FP acknowledge support from Science and Technology Facilities Council (STFC) grant ST/P000649/1. DT 
is supported by an STFC studentship. 
BB acknowledges financial support from the European Research Council (ERC) Consolidator Grant 725456. 
FP thanks Miguel Zumalac{\'a}rregui for useful discussions about the design of the public version of the 
\verb|hi_class| code and Filippo Vernizzi for the comparison of the Horndeski coefficients. RAB and FP further 
acknowledge discussions with Emilio Bellini, Pedro Ferreira and Lucas Lombriser. We also thank an anonymous referee 
whose comments helped us to improve the scientific content of this work.

\bibliographystyle{JHEP}
\bibliography{EoS_Horndeski.bbl}

\label{lastpage}

\end{document}